\documentclass[10pt]{article} 

\usepackage[english]{babel}
\usepackage[utf8]{inputenc}

\usepackage{amssymb,dsfont,amsmath,amsfonts,mathrsfs,stmaryrd,bbold,xfrac}

\usepackage[enableskew]{youngtab}
\usepackage{slashed}
\usepackage{xcolor}
\usepackage{comment}
\usepackage[
    colorlinks=true,
    linkcolor=black,
    citecolor=black,
    urlcolor=black
]{hyperref}
\advance \textheight by \topskip
\DeclareMathAlphabet{\mathpzc}{OT1}{pzc}{m}{it}
\usepackage{xcolor}

\usepackage[normalem]{ulem}
\usepackage{graphicx}
\usepackage{adjustbox}
\usepackage{tikz}
\usepackage{soul}
\usepackage{authblk}
\usepackage[square,numbers,sort&compress]{natbib}
\usepackage{doi}

\usepackage{mathtools}

\usepackage{array}
\usepackage{longtable}

\usepackage{cancel}

\newcommand{\I}{\mathbf{1}_2}
\newcommand{\vecsigma}{\vec{\sigma}}

\def\be{\begin{equation}}
\def\ee{\end{equation}}
\def\bea{\begin{eqnarray}}
\def\eea{\end{eqnarray}}
\def\Epsilon{E} 

\usepackage{authblk}

\newcommand{\nablahat}{\hat{\nabla}}

\begin{document}

\title{\textbf{Interacting Galilean and Finite-Energy Carroll Fermions}}

\author[1,2]{Pulastya Parekh%
\thanks{\href{mailto:parekh@cecs.cl}{parekh@cecs.cl}}}

\author[1,2]{Aditya Sharma%
\thanks{\href{mailto:aditya.sharma@uss.cl}{ext.aditya.sharma@uss.cl}}}

\author[1,2]{Mauricio Valenzuela%
\thanks{\href{mailto:mauricio.valenzuela@uss.cl}{mauricio.valenzuela@uss.cl}}}

\affil[1]{Centro de Estudios Cient\'ificos (CECS), Avenida Arturo Prat 514, Valdivia, Chile}
\affil[2]{Facultad de Ingenier\'ia, Universidad San Sebasti\'an, sede Valdivia, General Lagos 1163, Valdivia
5110693, Chile}

\date{}

\maketitle

\begin{abstract}
We develop a common similarity-transformation framework for Galilean
and Carrollian Dirac contractions, encompassing L\'evy--Leblond and
magnetic Carroll spin-half systems. Explicit field maps relate free formulations
with different mass and energy parametrizations. We obtain the complete
five-parameter family of derivative-free Hermitian quartic potentials
for a pair of Grassmann-odd Pauli spinors, invariant under rotations, a common $U(1)$,
and triangular boosts. Higher-order and color-mixed interactions admit
exact classical solutions. An independent pairing channel generates
derivative interactions upon exact auxiliary elimination and sources
otherwise absent Carroll spinor components. Quartic deformations of a
Galilean fermionic gauge theory yield algebraic or spatial differential
equations while preserving a local phase symmetry. Nambu--Jona-Lasinio
contractions provide relativistic origins for selected channels.
\end{abstract}


\newpage

{
\hypersetup{linkcolor=black}
\setlength{\parskip}{0pt}
\tableofcontents
}
\clearpage

\section{Introduction}
\label{sec:intro}

The Galilean limit $(c\to\infty)$ and the
Carrollian limit $(c\to0)$ organize relativistic field theories into two distinct non-Lorentzian kinematic regimes. 
Galilean theories underlie non-relativistic
approximations in condensed matter physics, hydrodynamics, and nuclear
and collider physics; see
\cite{Caswell:1985ui,Brambilla:2004jw,Son:2005rv,Bedaque:2002mn,Geracie:2014nka,Petkou:2022bmz,Baiguera:2023fus}
and references therein.
Carrollian symmetry
\cite{Levy-Leblond:1965dsc,SenGupta:1966qer,Bacry:1968zf} arises in the
geometry of null hypersurfaces \cite{Duval:2014uoa,Ciambelli:2019lap} and
has implications for flat-space holography
\cite{Duval:2014uva,Alday:2024yyj,Donnay:2022aba,Ruzziconi:2026bix,Mason:2023mti},
black-hole horizons
\cite{Gray:2022svz,Donnay:2019jiz,Ecker:2023uwm,Ecker:2026noz},
hydrodynamics
\cite{Husnugil:2025edm,Marsot:2022imf,Arenas-Henriquez:2025rpt},
and tensionless (null) strings
\cite{Schild:1976vq,Isberg:1993av,Bagchi:2015nca,
Bagchi:2019cay,Bagchi:2020fpr}. These applications motivate field
theories with the corresponding kinematic symmetries.
Galilean-invariant field theory has a long history
\cite{Hagen:1972pd,Hagen:1971dk,Hagen:1970wn,Gomis:1985uf,
Levy-Leblond:1967,Duval:1990hj,Duval:1993pe,Niederer:1972zz,Duval:2024eod},
whereas the early ultralocal theories of Klauder
\cite{Klauder:1971zz,Klauder:2000ud,Klauder:1973nc} have only recently been
rediscovered as Carrollian field theories. Considerable recent effort has
therefore been devoted to constructing models with Galilean or Carrollian
symmetry
\cite{Biedenharn:1974rb,Santos:2004pq,Barnich:2014cwa,
Bergshoeff:2015sic,Bagchi:2016bcd,Bergshoeff:2017btm,
Hansen:2019svu,Chapman:2020vtn,Hansen:2020pqs,
Banerjee:2020qjj,deBoer:2021jej,Hansen:2021fxi,
Banerjee:2022uqj,Rivera-Betancour:2022lkc,
Banerjee:2022ocj,Chen:2023pqf,Ecker:2023uwm,
Mehra:2023rmm,Parekh:2023xms,Sharma:2023chs,
Cerdeira:2023ztm,Ara:2024fbr,Aviles:2025ygw,
Bal:2026xup,Campoleoni:2026wja,Majumdar:2025juk,
Majumdar:2026io,Saha:2025wkr,Sharma:2025rug,Zorba:2026ghe}.

To realize these kinematic symmetries, several complementary approaches for constructing non-Lorentzian field theories have been developed. Field rescalings
\cite{Bergshoeff:2015sic,Bagchi:2016bcd,Bergshoeff:2017btm,Hansen:2020pqs}
and expansions
\cite{Hansen:2019svu,Hansen:2021fxi,deBoer:2021jej,Rivera-Betancour:2022lkc,Cerdeira:2023ztm}
organize limits by powers of $c$. Null reduction obtains Galilean
\cite{Biedenharn:1974rb,Santos:2004pq,Elizalde:1975zd,Elizalde:1976vi,Julia:1994bs,Banerjee:2022uqj,Sharma:2023chs}
or Carrollian \cite{Duval:1984cj,Duval:1985vt,Chen:2023pqf,Saha:2025wkr,Majumdar:2025juk,Majumdar:2026io,Sharma:2025rug}
covariance from Lorentzian structures in one higher dimension.
The seed Lagrangian method extends a non-Lorentzian Lagrangian
block using the symmetry algebra
\cite{Bergshoeff:2022qkx,Koutrolikos:2023evq}, while intrinsic constructions
start directly from the non-Lorentzian geometry
\cite{Banerjee:2020qjj,Ciambelli:2023xqk,Gupta:2020dtl,Bagchi:2022eav}. Without assuming a Lagrangian, bootstrap approaches have also been used to constrain correlation
functions through the symmetry algebra; see, for example,
\cite{Bagchi:2017cpu,Chen:2020vvn,Chen:2023naw,Chetia:2025xeh,Chen:2022jhx,Marotta:2025qjh}. Distinct limiting theories are often termed “electric” and “magnetic”, following the terminology of Galilean electromagnetism 
\cite{Levy-Leblond:1965dsc,LeBellac:1973,Rousseaux:2013}. Furthermore, these methodologies have been successfully adapted to construct non-Lorentzian theories of gravity, including Galilean
\cite{Andringa:2010it,Andringa:2013mma,Bergshoeff:2014uea,Hartong:2015zia,Afshar:2015aku,Bergshoeff:2015uaa,Hansen:2019svu,Bergshoeff:2019ctr,Hansen:2020pqs,Bergshoeff:2022eog,Hartong:2022lsy}
and Carrollian
\cite{Bergshoeff:2017btm,Campoleoni:2022ebj,Hartong:2015xda,Bergshoeff:2022eog,Hartong:2022lsy}.

For fermions, known Galilean models include the massive
L\'evy--Leblond theory \cite{Levy-Leblond:1967eic}
and massless fermion theories \cite{Banerjee:2022uqj,Sharma:2023chs}.
The notion of a massless Galilean system goes back to Souriau
\cite{Souriau:1970}, with masslessness understood as the vanishing of the
Bargmann mass central charge; see also \cite{Batlle:2017cfa}.
Both massless and massive Carroll fermion models have also been studied
in the literature
\cite{Bagchi:2019clu,Koutrolikos:2023evq,Bergshoeff:2023vfd,
Stakenborg:2023bmw,Bagchi:2025vri,Banerjee:2022ocj,Bagchi:2022eui}.

In this paper, we use a common contraction scheme to
organize Galilean and Carrollian fermion theories and their interactions. Our construction starts from Hermitian and complex Dirac functionals,
obtaining Hermitian limiting actions either directly or through a
two-color completion. A common mass- and $c$-dependent field redefinition
determines the contracted boost representations. Keeping either $m$ or
$\Epsilon=mc^2$ fixed then selects the limiting action densities.
Exponent diagrams and tables organize both conjugation branches and
show which terms survive together. These free representations supply
the symmetry conditions for constructing invariant Hermitian local
potentials intrinsically, while relativistic interaction limits select particular channels within this construction.

We classify local quartic interactions for one pair of Grassmann-odd
Pauli spinors transforming under a common global $U(1)$ phase symmetry.
Requiring invariance under rotations and the non-diagonal boosts,
with no derivatives or additional discrete-symmetry restrictions,
leaves exactly five independent Hermitian quartic scalar operators.
Three are products of invariant daggered--undaggered bilinears,
while two involve \emph{spin-singlet fermion-pairing bilinears}.
We prove completeness of this basis of local potentials in a fixed
field representation and also construct families of higher-order and
color-mixed interactions.

We develop one of the two pairing channels. Its exact
L\'evy--Leblond reduction generates derivative interactions, and an
explicit periodic free solution establishes its first-order independence
from the other quartics, modulo free equations of motion and total
derivatives. In the magnetic Carroll theory,
we also obtain an exact nonlinear solution in which the pairing
interaction sources initially vanishing spinor components.
This pairing channel may enrich the phase structure of interacting
Carrollian systems, motivating further investigation beyond the
classical analysis presented here.

We apply the construction to known free representatives.
The L\'evy--Leblond theory \cite{Levy-Leblond:1967eic} admits nonlinear
Schr\"odinger reductions; earlier interacting Galilean fermion models
provide benchmarks \cite{deMontigny:2007fqc,Saradzhev:2014qta}.
Explicit field dictionaries identify our minimal and non-minimal
magnetic Carroll representatives with those of Bergshoeff
\textit{et al.}~\cite{Bergshoeff:2023vfd}, complementing the constructions of
\cite{Bergshoeff:2022qkx,Koutrolikos:2023evq,Ekiz:2025hdn,
Bagchi:2025vri,Bagchi:2026lgk}.
Our potentials provide interacting extensions of these systems,
including the real-mass non-minimal theory. For representative
potentials we obtain exact classical Grassmann-valued solutions.
The dictionaries and the extended models
in their conventions will be presented in
Appendix~\ref{app:Bergshoeff-dictionary}.

A distinctive Galilean example is a colored spin-$\tfrac12$ model
whose singular contraction produces a local fermionic gauge symmetry.
The free fields are gauge-trivial on each spatial slice.
Two quartic deformations change
this structure in different ways: one leaves algebraic constraints
on the surviving fields, while the other yields nonlinear spatial
differential equations
with nonzero invariant local data. A bosonic local phase symmetry
survives in both. These deformations govern the spatial field content
at each time.

Finally, contractions of Nambu--Jona-Lasinio (NJL) interactions connect
selected invariant channels to relativistic parents. In the colored
Galilean gauge model, a jointly normalized contraction of the Hermitian
parent and its NJL interaction gives a quartic that preserves a
deformed fermionic shift symmetry. This complements the intrinsic
construction by identifying interactions inherited from the same
relativistic limit as the free action.

The paper is organized as follows.
Section~\ref{sec:limits} introduces the relativistic action principles,
the field redefinition, and the contracted boost representations.
Sections~\ref{sec:S1-limits} and \ref{sec:S2-limits} apply this framework
to the $C_1$ and $C_2$ branches, deriving the free actions from their
leading powers of $c$ at fixed $m$ or fixed $\Epsilon$, and constructing
the Hermitian two-color Carroll completions.
Section~\ref{sec:interactions} constructs local interactions and states
the complete single-color quartic basis.
Section~\ref{sec:models} develops the representative
interacting models and the auxiliary-field reductions.
Section~\ref{sec:NJL} gives the NJL contractions as a relativistic
illustration. Section~\ref{sec:conclusions} summarizes the results and
their scope.
Appendix~\ref{app:interaction-catalog} summarizes representative
bilinear-generated interaction families. Appendix~\ref{app:Bergshoeff-dictionary}
collects relations between free Carroll actions and the associated
field dictionaries, together with their interacting extensions.
Appendix~\ref{app:quartic-completeness} proves the quartic
classification, and Appendix~\ref{app:boost-checks} verifies boost
invariance through common building blocks.
Conventions are collected in Appendix~\ref{app:conventions}.

\section{Relativistic action principles and non-Lorentzian limits}

\label{sec:limits}

 The starting point is the Dirac action
\begin{equation}
\label{act-D}
\mathcal{S}^{[r]}=\frac1{\kappa_r}\int d^4x \,
\bar{\psi}^{[r]}\bigl(\gamma^\mu\partial_\mu + m c\bigr)\psi,
\end{equation}
where the label $[r]$ stands for the choice of the Dirac conjugation matrix,
\[
\bar\psi^{[r]}:=\psi^\dagger C_r,\qquad r=1,2,
\]
and $\kappa_r\in\mathbb R$ is the branch-dependent normalization constant. We consider
\be \label{C12}
C_1=i\gamma_0\quad \hbox{(diagonal) and} \quad C_2=C_1 \gamma_5 \quad  \hbox{ (off-diagonal)}.
\ee
We call an action Hermitian if $\mathcal S^\dagger=\mathcal S$, anti-Hermitian if
$\mathcal S^\dagger=-\mathcal S$, and complex otherwise. These relations
are understood modulo boundary terms.

In what follows, we refer to the functional \eqref{act-D} as the
\emph{seed}.\footnote{Here the seed is the uncontracted
Dirac functional; the seed Lagrangian method of
Ref.~\cite{Bergshoeff:2022qkx} starts from a non-Lorentzian Lagrangian
block.} The $C_1$ seed is already hermitian. In the $C_2$ branch,
the kinetic term is hermitian, whereas the mass term is anti-hermitian;
the massive seed is therefore not a physical action in its own right.
For this complex seed, the spinor and its dual are independent
Grassmann variables. Daggered symbols denote the duals in this
complexified variational problem, without imposing ordinary conjugacy,
so the left and right Euler--Lagrange equations are not related by
Hermitian conjugation.

We use two routes to obtain Hermitian actions from the $C_2$ seed.
First, a non-Lorentzian contraction can suppress the rescaled
anti-Hermitian part, leaving a Hermitian action in which ordinary
conjugacy can be imposed. This occurs in the fixed-$m$ Carroll limits
that we will derive in Section~\ref{sec:S2-limits}; we will use the minimal
magnetic model as an example in
Section~\ref{sec:interacting-Bergshoeff}.

Second, we retain the mass mixing by introducing two independent Dirac
fields before contraction and pairing their mixed bilinear with its
Hermitian conjugate. This defines the Hermitian relativistic parent

\begin{equation}
\label{L-D12}
\mathcal S_{12}^{[r]}=\frac1{2\kappa_r}\int d^4x\,
\mathcal L_{12}^{[r]},\qquad
\mathcal L_{12}^{[r]}
=\bar\psi_2^{[r]}(\gamma^\mu\partial_\mu+mc)\psi_1+\mathrm{h.c.},
\qquad
\bar\psi_a^{[r]}=\psi_a^\dagger C_r.
\end{equation}
The internal label $a=1,2$ will be called \emph{color}; it distinguishes
the two fields and implies neither QCD color nor a color gauge symmetry.
Each field now has its ordinary Hermitian conjugate. The complex seed
organizes one off-diagonal term of the Hermitian variational principle
\eqref{L-D12}.\footnote{A different Hermitian parent is obtained by
replacing $mc$ in the single-color $C_2$ seed by a real chiral mass
$mc\gamma_5$, since $C_2\gamma_5=C_1$. This changes the mass shell to
the tachyonic form $(\Box+m^2c^2)\psi=0$.}

\subsection{Field redefinition}
\label{sec:field-redefinition}

With these action principles specified, we now implement the
Carrollian and Galilean contractions through
a $c$- and time-dependent field redefinition,
\begin{equation}
\label{S}
\Phi:=S\,\psi =\begin{pmatrix} \zeta \\[2pt]\chi\end{pmatrix},
\qquad
S:=e^{+i m c^2 t}\begin{pmatrix}c^\beta&0\\[2pt]0&1\end{pmatrix},
\qquad
S^{-1}=\begin{pmatrix}c^{-\beta}&0\\[2pt]0&1\end{pmatrix}e^{-i m c^2 t},
\end{equation}
Here $\beta$ is a constant. The order of the Pauli blocks in $\Phi$
is $(\zeta,\chi)$ throughout all four-component matrices below.
For $m>0$, the phase removes the positive rest energy of the lower
Dirac block: $E_{\rm rel}=\sqrt{m^2c^4+c^2\mathbf p^2}$ becomes
$\omega=E_{\rm rel}-mc^2\to\mathbf p^2/(2m)$ in the Galilean limit.
The mass term of the relativistic parent remains $+mc$.
For the colored parent the same
transformation acts on both fields,
\begin{equation}
\label{S12}
\Phi_a=S\psi_a=\begin{pmatrix}\zeta_a\\\chi_a\end{pmatrix},
\qquad a=1,2.
\end{equation}
Substituting
$\psi=S^{-1}\Phi$ and $x^0=ct$ into the Dirac action, the same functional
becomes
\begin{equation}
\label{act-12b}
\mathcal{S}^{[r]}\equiv\mathcal{S}_c^{(r,\beta)}
=\frac1{\kappa_r}\int dt\,d^3x\,
\Phi^\dagger\widetilde C_r\Upsilon\Phi ,
\end{equation}
where
\be\label{Ups}
 \Upsilon:=S\bigl(c\gamma^\mu\partial_\mu+mc^2\bigr)S^{-1},
 \qquad
 \widetilde C_r:=S^{-1\dagger}C_rS^{-1}.
\ee
Here $\widetilde C_r$ and $\Upsilon$ are the conjugation matrix and Dirac
operator in the new field variables. In agreement with \eqref{C12}, we get
\be \label{Ct}
\widetilde C_1=
\begin{pmatrix}
c^{-2\beta}\I&0\\[2pt]
0&-\I
\end{pmatrix},  \qquad
\widetilde C_2=c^{-\beta}\begin{pmatrix}0& i \I \\[2pt]i \I &0\end{pmatrix} ,
\ee
and
\begin{equation}
\label{Ups-mat}
\Upsilon=
\begin{pmatrix}
i\I\partial_t+2mc^2\I
&
c^{1+\beta}\nablahat
\\[2pt]
c^{1-\beta}\nablahat
&
-i\I\partial_t
\end{pmatrix},\qquad \text{where } \nablahat:= \sigma^i \partial_i.
\end{equation}
The choices $r=1,2$ in \eqref{act-12b} define the two families with normalization constants
$\kappa_1$ and $\kappa_2$. In what follows, we study their Carrollian
($c\to0$) and Galilean ($c\to\infty$) limits.

The procedure is to expand the density in powers of $c$, retain the
largest power for $c\to\infty$ or the smallest for $c\to0$, and absorb
that power into $\kappa_r$. Terms with the same leading power survive
together. For example, in the Galilean limit at fixed $m$ and $\beta=1$, the $C_1$ density
retains the $\chi$ time derivative, the mixed spatial derivatives and
the $\zeta$ mass term, while the $\zeta$ time derivative is suppressed
by $c^{-2}$. This gives the L\'evy--Leblond action. Holding
$\Epsilon=mc^2$ fixed instead lowers the mass-term exponent by two
and changes which terms survive. Sections~\ref{sec:S1-limits} and
\ref{sec:S2-limits} will implement this selection graphically and collect
the resulting actions in tables. We first determine the contracted
generators, which will also guide the construction of interactions.

\subsection{Contractions and mass-shell equations}
\label{sec:contractions-mass-shells}
Before the field redefinition, the Poincar\'e generators act on a Dirac
spinor through the differential operators
\begin{equation}
\label{Poincare-generators}
\begin{aligned}
T_\mu&:=\partial_\mu,
& M_{\mu\nu}&:=x_\mu\partial_\nu-x_\nu\partial_\mu,\\
J_{\mu\nu}&:=M_{\mu\nu}+S_{\mu\nu},
& S_{\mu\nu}&:=\tfrac14[\gamma_\mu,\gamma_\nu].
\end{aligned}
\end{equation}
With $x^0=ct$, their temporal and spatial components include
\[
T_0=c^{-1}\partial_t,\qquad T_i=\partial_i,\qquad
J_{0i}=-(ct\,\partial_i+c^{-1}x_i\partial_t)
+\tfrac12\gamma_0\gamma_i.
\]
Our Dirac-matrix conventions and the Poincar\'e commutation relations
are collected in Appendix~\ref{app:Poincare}.

The rest-energy phase in \eqref{S} is naturally described by adjoining
a central mass generator to the Poincar\'e algebra before contraction
\cite{Aldaya:1982xj,Aldaya:1985plo,Duval:2002cw}. It commutes with all
Poincar\'e generators and is represented by the scalar $m$.
Conjugation by $S$ leaves spatial
translations and rotations unchanged, but modifies time translations
and boosts. Writing $\widetilde T_0=ST_0S^{-1}$ and
$\widetilde J_{0i}=SJ_{0i}S^{-1}$, we obtain
\be\label{tT0-J0i}
\widetilde T_0= c^{-1} \partial_t -imc\,,\qquad \widetilde J_{0i}=
\begin{pmatrix}
-ct \partial_i -c^{-1} x_i(\partial_t-imc^2) & -c^{\beta}\frac i2 \sigma_i\\[2pt]
c^{-\beta}\frac{i}{2} \sigma_i & -ct \partial_i -c^{-1} x_i(\partial_t-imc^2)
\end{pmatrix}.
\ee
For Dirac spinors, the quadratic translation Casimir
$\mathcal C_1=T^\mu T_\mu$ and the Pauli--Lubanski Casimir
$\mathcal C_2=W^\mu W_\mu$ satisfy
$\mathcal C_2=\tfrac34\mathcal C_1$, where
$W^\mu=\tfrac12\epsilon^{\mu\nu\rho\sigma}S_{\nu\rho}T_\sigma$
is the Pauli--Lubanski vector, with $\epsilon_{0123}=1$.
The similarity transformation therefore gives
\begin{equation}
\label{Cas-t}
\widetilde{\mathcal C}_1
=m^2c^2+2im\partial_t+\vec\nabla^2-c^{-2}\partial_t^2,
\qquad
\widetilde{\mathcal C}_2
=\frac34\widetilde{\mathcal C}_1 .
\end{equation}
The coefficient
$\frac34=\frac12(\frac12+1)$ reflects the spin-$\frac12$ Dirac
representation. Consequently, every finite contraction of
$\widetilde{\mathcal C}_1$ induces the corresponding contraction of
$\widetilde{\mathcal C}_2$. This relation is independent of $\beta$, which instead controls the
off-diagonal spin-mixing blocks in \eqref{tT0-J0i}
that survive in the contracted boosts.
The Pauli--Lubanski conventions and the spin-Casimir relation will be
detailed in Appendix~\ref{app:conventions}.

Applied to the Dirac action at fixed $m$, the Galilean limit
recovers the L\'evy--Leblond theory for spin-$\tfrac12$
(see Ref. \cite{Levy-Leblond:1967}). A similar construction applied to the
Plyushchay--Cort\'es relativistic anyon equations \cite{Cortes:1992fa}
yields square roots of the Schr\"odinger equation for higher-spin fields
and anyons in $2+1$ dimensions \cite{Horvathy:2006pw,Horvathy:2010vm}.
Beyond the standard Galilean limit, the same
field redefinition also yields Galilean limits with $m\to0$ and Carrollian
limits with $m\to\infty$, after suitable scalings of the boost generators
and/or the action normalization.
Indeed, both limits admit a variation in which
 \be\label{Epsi}
 \Epsilon=mc^2 \quad \text{is fixed when } (c\to 0\;,\;m\to \infty) \text{ \phantom{a} or \phantom{a} } (c\to\infty\;,\;m\to 0).
 \ee
We shall call this prescription a \emph{fixed-$\Epsilon$ scaling}. This term
refers only to the way the limit is taken.

The relativistic Dirac equation implies the Klein--Gordon equation
$(\mathcal C_1-m^2c^2)\psi=0$. After the field redefinition \eqref{S},
this becomes
\[
\left(\widetilde{\mathcal C}_1-m^2c^2\right)\Phi
=\left(\vec\nabla^{\,2}+2im\partial_t-c^{-2}\partial_t^2\right)\Phi=0.
\]
We define the contracted Casimirs by finite rescaled limits of
$\widetilde{\mathcal C}_1$ itself:
\begin{eqnarray}
\text{Galilei:}&&
\mathcal C_m^G:=
\lim_{\substack{c\to\infty\\ {\rm fixed-}m}}
c^{-2}\widetilde{\mathcal C}_1=m^2,
\qquad
\mathcal C_0^G:=
\lim_{\substack{c\to\infty\\{\rm fixed-}\Epsilon}}
\widetilde{\mathcal C}_1=\vec{\nabla}^{\,2},
\label{Galilei-Cas-limits}\\[5pt]
\text{Carroll:}&&
\mathcal C_0^C:=
\lim_{\substack{c\to0\\{\rm fixed-}m}}
c^2\widetilde{\mathcal C}_1
=-\partial_t^2,
\qquad
\mathcal C_\Epsilon^C:=
\lim_{\substack{c\to0\\{\rm fixed-}\Epsilon}}
c^2\widetilde{\mathcal C}_1
=\left(i\partial_t+\Epsilon\right)^2.
\label{Carroll-Cas-limits}
\end{eqnarray}
Their values on a regular limit of Klein--Gordon solutions are fixed by
the corresponding limits of the mass-shell eigenvalue:
\[
\mathcal C_m^G\Phi=m^2\Phi,\qquad
\mathcal C_0^G\Phi=0,\qquad
\mathcal C_0^C\Phi=0,\qquad
\mathcal C_\Epsilon^C\Phi=\Epsilon^2\Phi.
\]
By \eqref{Cas-t}, the same rescalings of the spin Casimir
$\widetilde{\mathcal C}_2$ give $3/4$ of these operators and eigenvalues.
All these limits are independent of $\beta$.
While $\mathcal C_\Epsilon^C$ reduces to $\mathcal C_0^C$ at
$\Epsilon=0$, the two Galilean prescriptions select different orders
in $c$ and must be distinguished.

At fixed $m$, the leading Casimir $\mathcal C_m^G=m^2$ records only the
squared Bargmann mass and does not impose a differential equation.
The subleading $c^0$ term in \eqref{Cas-t},
$\vec\nabla^{\,2}+2im\partial_t$, is also Galilean invariant and does
define a differential constraint: after the two $m^2c^2$ terms cancel,
the Klein--Gordon equation requires
$(\vec\nabla^{\,2}+2im\partial_t)\Phi=0$ in the fixed-$m$ and
$c\to\infty$ limit. For $m\neq0$ this gives
\be
\label{KG-Galilei-limit}
\text{(Galilean) Schr\"odinger equation:}\qquad
\left(i\partial_t+\frac{\vec{\nabla}^{\,2}}{2m}\right)\Phi=0.
\ee
The analogous equation in the fixed-$\Epsilon$ Carroll limit
follows by multiplying the Klein--Gordon equation by $c^2$ and then taking
$c\to0$:
\begin{equation}
\label{Cas-CE-mass-shell}
\text{Carroll equation:}\qquad
\left(\mathcal C_\Epsilon^C-\Epsilon^2\right)\Phi
=-\partial_t\left(\partial_t-2i\Epsilon\right)\Phi=0.
\end{equation}
We call this the Carroll equation, in analogy with the Schr\"odinger
equation \eqref{KG-Galilei-limit}, as both arise from limits of the
relativistic Klein--Gordon equation.
In the fixed-$\Epsilon$ Galilean limit, the Casimir operator
in \eqref{Galilei-Cas-limits} gives \(\vec\nabla^{\,2}\Phi=0\),
whereas in the fixed-$m$ Carroll limit, the operator in
\eqref{Carroll-Cas-limits} gives \(\partial_t^2\Phi=0\).

For $|\beta|>1$, making the dominant spin-mixing off-diagonal block in
\eqref{tT0-J0i} finite suppresses the orbital part. The resulting
transformations are therefore not Galilean or Carrollian boosts.
We therefore restrict attention to $|\beta|\leq1$, for which the
contracted boosts retain their orbital terms. The continuous scaling
parameter then gives three boost classes,
\begin{equation}
\label{beta-star}
\beta_\star=
\begin{cases}
-1,& \beta=-1,\\
0,& -1<\beta<1,\\
+1,& \beta=1.
\end{cases}
\end{equation}

We call $\beta=\pm1$ the \emph{endpoint} cases and
$-1<\beta<1$ the \emph{central} sector. The same terms apply to the
action principles obtained in these ranges. We retain $\beta$ when the
scaling range matters and use $\beta_\star$ to label the boost class.

\subsection{Carroll boosts and energy}
\label{sec:Carroll-representations}
For $c\to0$, finite generators follow by multiplying
$\widetilde T_0$ and $\widetilde J_{0i}$ by $c$ throughout
$|\beta|\leq1$.
At fixed $m$, time translations satisfy
  \[
  \lim_{c\to0}c\,S T_0 S^{-1}=\partial_t.
  \]
At fixed $\Epsilon=mc^2$, the same rescaling yields the inherited
Carroll time-translation generator
\begin{equation}
\label{eqn:Dt}
D_t^{(\Epsilon)}
:=\lim_{\substack{c\to0\\mc^2=\Epsilon}}c\,S T_0 S^{-1}
=e^{i\Epsilon t}\partial_t e^{-i\Epsilon t}
=\partial_t-i\Epsilon.
\end{equation}

  The fixed-$\Epsilon$ boosts are
  \[
  K_{i,\Epsilon}^{C,(\beta_\star)}
  =
  \lim_{\substack{c\to0\\mc^2=\Epsilon}}
  c\,\widetilde J_{0i},
  \]
 for $\beta_{\star} \in \{-1,0,1\}$. Explicitly,
  \begin{equation}
  \label{boost-CE-beta}
  K_{i,\Epsilon}^{C,(\beta_\star)}
  =
  \begin{cases}
  \begin{pmatrix}
  -x_iD_t^{(\Epsilon)} & 0\\
  0 & -x_iD_t^{(\Epsilon)}
  \end{pmatrix},
  & \beta_\star=0,\\[8pt]
  \begin{pmatrix}
  -x_iD_t^{(\Epsilon)} & -\frac{i}{2}\sigma_i\\
  0 & -x_iD_t^{(\Epsilon)}
  \end{pmatrix},
  & \beta_\star=-1,\\[8pt]
  \begin{pmatrix}
  -x_iD_t^{(\Epsilon)} & 0\\
  \frac{i}{2}\sigma_i & -x_iD_t^{(\Epsilon)}
  \end{pmatrix},
  & \beta_\star=1 .
  \end{cases}
  \end{equation}
 The infinitesimal transformations generated by
  $K_{i,\Epsilon}^{C,(\beta_\star)}$ are
  \begin{equation}
  \label{Cboost-CE-beta-components}
  \begin{array}{ll}
  \beta_\star=-1:&
  \displaystyle
  \delta_b\chi=-b^ix_iD_t^{(\Epsilon)}\chi,\qquad
  \delta_b\zeta=-b^ix_iD_t^{(\Epsilon)}\zeta
  -\frac{i}{2}\,b^i\sigma_i\chi,\\[8pt]
  \beta_\star=0:&
  \displaystyle
  \delta_b\chi=-b^ix_iD_t^{(\Epsilon)}\chi,\qquad
  \delta_b\zeta=-b^ix_iD_t^{(\Epsilon)}\zeta,\\[8pt]
  \beta_\star=1:&
  \displaystyle
  \delta_b\chi=-b^ix_iD_t^{(\Epsilon)}\chi
  +\frac{i}{2}\,b^i\sigma_i\zeta,\qquad
  \delta_b\zeta=-b^ix_iD_t^{(\Epsilon)}\zeta.
  \end{array}
  \end{equation}
  The fixed-$m$ Carroll limit  is obtained from \eqref{boost-CE-beta} at $\Epsilon=0$, 
  \[
  K_{i,0}^{C,(\beta_\star)}
  :=\lim_{c\to0}c\,\widetilde J_{0i}
  =
  K_{i,\Epsilon}^{C,(\beta_\star)}\big|_{\Epsilon=0},
  \]
  since $mc^2\to0$ at fixed $m$. Its component transformations follow
  from \eqref{Cboost-CE-beta-components} by setting $\Epsilon=0$.
  
  Note that the fixed-$\Epsilon$ Carroll boosts can be obtained from 
  the $\Epsilon=0$ Carroll
  boosts by a time-dependent phase similarity transformation,
  \begin{equation}
  \label{phase-sim-C}
  \mathcal P_\Epsilon\,\partial_t\,\mathcal P_\Epsilon^{-1}
  =
  D_t^{(\Epsilon)},\qquad \mathcal P_\Epsilon(t):=
  e^{i\Epsilon t}\,\mathbf 1_4 .
  \end{equation}
  Hence, for
  $-1\leq\beta\leq1$,
  \begin{equation}
  \label{boost-C-phase-sim}
  K_{i,\Epsilon}^{C,(\beta_\star)}
  =
  \mathcal P_\Epsilon\,K_{i,0}^{C,(\beta_\star)}\,
  \mathcal P_\Epsilon^{-1}.
  \end{equation}
  For an action and its generators transported together, this map
  also transports the invariance statement. The direct checks for our
  contracted actions are collected in Appendix~\ref{app:boost-checks}.

  The representative Carroll boosts are also related by a constant similarity
  transformation on the four-component spinor
  $\Phi=(\zeta,\chi)^T$. In this block basis, define
  \begin{equation}
  \label{D-map}
  \mathfrak D:=-i\gamma_5
  =
  \begin{pmatrix}
  0&\I\\[2pt]
  -\I&0
  \end{pmatrix},
  \qquad
  \Phi'=\mathfrak D\Phi
  =
  \begin{pmatrix}
  \chi\\[2pt]
  -\zeta
  \end{pmatrix},
  \qquad
  \mathfrak D^{-1}=-\mathfrak D=\mathfrak D^\dagger .
  \end{equation}
  Conjugation by $\mathfrak D$ exchanges the two off-diagonal
  spin-mixing blocks in \eqref{boost-CE-beta} and leaves the diagonal
  term $-x_iD_t^{(\Epsilon)}$ unchanged. Hence
  \begin{equation}
  \label{boost-C-duality}
  \mathfrak D K_{i,\Epsilon}^{C,(\beta_\star)}\mathfrak D^{-1}
  =K_{i,\Epsilon}^{C,(-\beta_\star)},\qquad
  \beta_\star=-1,0,1 .
  \end{equation}
  They realize the Carroll algebra, with the abstract time-translation
  generator represented by $D_t^{(\Epsilon)}$ of \eqref{eqn:Dt}:
  \begin{equation}
  \label{Carroll-algebra-Dt}
  [K_{i,\Epsilon}^{C,(\beta_\star)},T_j]=\delta_{ij}\,D_t^{(\Epsilon)},\qquad
  [K_{i,\Epsilon}^{C,(\beta_\star)},D_t^{(\Epsilon)}]=0,\qquad
  [K_{i,\Epsilon}^{C,(\beta_\star)},K_{j,\Epsilon}^{C,(\beta_\star)}]=0,
  \end{equation}
  where the spin-mixing blocks drop out of the commutators since they are
  nilpotent and commute with the orbital part. 

For this contracted representation, the transported energy
generator associated with \eqref{eqn:Dt} and
\eqref{Carroll-algebra-Dt} is
\begin{equation}
\label{Carroll-energy-operator}
H_C^{(\Epsilon)}:=iD_t^{(\Epsilon)}
=i\partial_t+\Epsilon.
\end{equation}
Hence $\mathcal C_\Epsilon^C=\bigl(H_C^{(\Epsilon)}\bigr)^2$ and 
the Carroll equation \eqref{Cas-CE-mass-shell} reads
\begin{equation}
\label{Carroll-energy-mass-shell}
\left(\mathcal C_\Epsilon^C-\Epsilon^2\right)\Phi
=\left[\bigl(H_C^{(\Epsilon)}\bigr)^2-\Epsilon^2\right]\Phi=0.
\end{equation}

A different field basis, useful for the comparisons below, is obtained through the time-dependent redefinition
\begin{equation}
\label{Carroll-energy-basis}
\Phi=e^{i\Epsilon t}\Psi,\qquad
e^{-i\Epsilon t}H_C^{(\Epsilon)}e^{i\Epsilon t}
=i\partial_t
\end{equation}
which puts the inherited time-translation generator in its ordinary differential $i\partial_t$ form and brings the Carroll equation \eqref{Carroll-energy-mass-shell} to
\begin{equation}
\label{Carroll-equation-oscillator}
\left[(i\partial_t)^2-\Epsilon^2\right]\Psi
=-\left(\partial_t^2+\Epsilon^2\right)\Psi=0.
\end{equation}
For $\Epsilon\neq0$, the mode decompositions in the two field bases
take the respective forms
\[
\Psi(t,\vec x)
=e^{-i\Epsilon t}\Psi_+(\vec x)
+e^{i\Epsilon t}\Psi_-(\vec x),\qquad
\Phi(t,\vec x)
=\Psi_+(\vec x)+e^{2i\Epsilon t}\Psi_-(\vec x).
\]
Thus $\Psi$ has frequencies $\pm\Epsilon$, whereas $\Phi$ combines a
time-independent profile with a mode oscillating at angular frequency
$2|\Epsilon|$.

Equation~\eqref{Carroll-equation-oscillator} was already obtained by
quantizing the Carroll-particle mass-shell constraint in
\cite{Bergshoeff:2014jla}. Its derivation as a Klein--Gordon limit and
relation to the first-order quantum equation at fixed energy are discussed in
\cite{Marsot:2021tvq};
see also \cite{deBoer:2021jej,Figueroa-OFarrill:2023vbj,Banerjee:2023jpi}. Furthermore, upon quantization, the modes provide the operator expansion basis for the Hilbert space of the scalar Carrollian system.
Scalar dynamics combining electric and magnetic sectors are studied
more generally in \cite{Ciambelli:2023xqk}.

\subsection{Galilean boosts and Bargmann mass}
\label{sec:Galilean-representations}
For $c\to\infty$, the boosts require the inverse rescaling
$c^{-1}$. At fixed $m$, subtracting the central rest-energy
term gives a finite time-translation generator,
  \[
  \lim_{c\to\infty}c\bigl(\widetilde T_0+imc\bigr)=\partial_t,
  \]
  while the boosts are
  \[
  K_{i,m}^{G,(\beta_\star)}
  =\lim_{c\to\infty}c^{-1}\widetilde J_{0i},
  \]
 giving the three boost classes
  \begin{equation}
  \label{boost-G-beta}
  K_{i,m}^{G,(\beta_\star)}
  =
  \begin{cases}
  \begin{pmatrix}
  -t\partial_i+imx_i & 0\\[2pt]
  0 & -t\partial_i+imx_i
  \end{pmatrix},
  & \beta_\star=0,\\[16pt]
  \begin{pmatrix}
  -t\partial_i+imx_i & 0\\[2pt]
  \frac{i}{2}\sigma_i & -t\partial_i+imx_i
  \end{pmatrix},
  & \beta_\star=-1,\\[16pt]
  \begin{pmatrix}
  -t\partial_i+imx_i & -\frac{i}{2}\sigma_i\\[2pt]
  0 & -t\partial_i+imx_i
  \end{pmatrix},
  & \beta_\star=1 .
  \end{cases}
  \end{equation}
  The infinitesimal transformations generated by
  $K_{i,m}^{G,(\beta_\star)}$ are
  \begin{equation}
  \label{Gboost-G-beta-components}
  \begin{array}{ll}
  \beta_\star=-1:&
  \displaystyle
  \delta_b\chi=-b^iD_i^{(m)}\chi+\frac{i}{2}\,b^i\sigma_i\zeta,\qquad
  \delta_b\zeta=-b^iD_i^{(m)}\zeta,\\[8pt]
  \beta_\star=0:&
  \displaystyle
  \delta_b\chi=-b^iD_i^{(m)}\chi,\qquad
  \delta_b\zeta=-b^iD_i^{(m)}\zeta,\\[8pt]
  \beta_\star=+1:&
  \displaystyle
  \delta_b\chi=-b^iD_i^{(m)}\chi,\qquad
  \delta_b\zeta=-b^iD_i^{(m)}\zeta
  -\frac{i}{2}\,b^i\sigma_i\chi,
  \end{array}
  \end{equation}
  where
  \begin{equation}
  \label{eqn:Di}
  D_i^{(m)}:=t\partial_i-imx_i .
  \end{equation}
  The fixed-$\Epsilon$ scaling instead has
  $m=\Epsilon/c^2\to0$ and hence vanishing Bargmann mass.
  In this scaling, time translations are again represented by
  $D_t^{(\Epsilon)}$ of \eqref{eqn:Dt}, whereas $\Epsilon$ drops out of
  the boosts. With the same boost rescaling,
  \[
  K_{i,0}^{G,(\beta_\star)}
  :=
  \lim_{\substack{c\to\infty\\mc^2=\Epsilon}}
  c^{-1}\widetilde J_{0i},
  \]
  one gets the $m=0$ specialization of \eqref{boost-G-beta}.

  The duality transformation \eqref{D-map} also relates the
  Galilei boost sectors
   \begin{equation}
  \label{boost-G-duality}
\mathfrak D K_{i,m}^{G,(\beta_\star)}\mathfrak D^{-1}
=
K_{i,m}^{G,(-\beta_\star)},\qquad
\beta_\star=-1,0,1 ,
\end{equation}
for either finite or vanishing $m$. 

Here $\Epsilon$ enters the time-translation generator as a finite
internal shift, not as a Bargmann mass. The boost--time-translation
commutator has the usual form: in the fixed-$m$ realization,
  \[
  [K_{i,m}^{G,(\beta_\star)},\partial_t]=\partial_i.
  \]
  The same commutator holds in the fixed-$\Epsilon$ realization,
  where $T_t$ acts as $D_t^{(\Epsilon)}$, since the constant
  internal shift commutes with the boosts.

The commutator with spatial translations detects the Bargmann central charge.
Since $T_j$ is not changed by the similarity transformation,
  \[
  [c^{-1}\widetilde J_{0i},T_j]
  =\delta_{ij}c^{-1}\widetilde T_0 .
  \]
  Therefore, for fixed $m$,
  \[
  [K_{i,m}^{G,(\beta_\star)},T_j]
  = -im\,\delta_{ij},
  \]
  while in the fixed-$\Epsilon$ scaling
  \[
  [K_{i,0}^{G,(\beta_\star)},T_j]=0.
  \]
 Thus, the
  mass parameter $m$ fixes the Bargmann central charge, represented by
  $-im$ in these differential-generator conventions. The physical mass
  is $m$, with Schr\"odinger dispersion $\omega=\mathbf p^2/(2m)$ for
  $m>0$. We shall refer to these nonzero-central-charge
  sectors as Bargmann sectors. In the fixed-$\Epsilon$ Galilei scaling this
  central charge is absent; we shall call the resulting sectors
  \textit{massless} Galilei sectors, equivalently the zero-central-charge sector of the
  Bargmann algebra.

With the contracted representations fixed, we now determine the
corresponding leading action densities. The next two sections apply
the same power-counting procedure to the $C_1$ and $C_2$ branches,
respectively. These free actions will provide the setting for the
interaction construction and its applications. A common set of
boost-invariance proofs will be given in Appendix~\ref{app:boost-checks}.

\section{Free-action contractions: the \texorpdfstring{$C_1$}{C1} branch}
\label{sec:S1-limits}
We begin with the Hermitian $C_1$ branch and compare the fixed-$m$
and fixed-$\Epsilon$ scalings through their powers of $c$.
The $r=1$ branch of \eqref{act-12b} is
\begin{eqnarray}
\label{act-1-massive}
\mathcal{S}^{(1,\beta)}_{c}=
\frac1{\kappa_1}\int dt\, d^3x\,
\Big[
\chi^\dagger i\partial_t\chi
+c^{1-\beta}(\zeta^\dagger\hat\nabla\chi
-\chi^\dagger\hat\nabla\zeta)
+i c^{-2\beta}\zeta^\dagger\partial_t\zeta
+2m c^{2-2\beta}\zeta^\dagger\zeta
\Big].
\end{eqnarray}
To identify the leading terms, write
\be\label{act-1-ABDG}
\mathcal{S}^{(1,\beta)}_{c}
=\frac1{\kappa_1}\int dt\, d^3x\,
\Big[c^0 A + c^{1-\beta} B + c^{-2\beta} D + 2m c^{2-2\beta} G\Big],
\ee
where
\be\label{ABDG}
A = \chi^\dagger i\partial_t\chi,\qquad
B=\zeta^\dagger\hat\nabla\chi
-\chi^\dagger\hat\nabla\zeta,\qquad
D=\zeta^\dagger i\partial_t\zeta,\qquad
G=\zeta^\dagger\zeta.
\ee
At fixed $\Epsilon=mc^2$, the mass term instead scales as
$+2\Epsilon c^{-2\beta}\zeta^\dagger\zeta$, giving
\be\label{act-1-ABDGE}
\mathcal{S}^{(1,\beta)}_{c}
=\frac1{\kappa_1}\int dt\, d^3x\,
\Big[c^0 A + c^{1-\beta} B + c^{-2\beta} (D + 2\Epsilon G)\Big],
\ee
so that $D$ and $G$ have the same weight. Figure~\ref{fig:cpowers-1}
plots these powers of $c$ against $\beta$. Rescaling
$\kappa_1=c^p\tilde\kappa_1$ shifts every exponent by $-p$ and
keeps the selected leading density finite. We restrict attention to
$|\beta|\leq1$, as justified in Section~\ref{sec:limits}.
\begin{figure}[ht!]
\centering
\begin{tikzpicture}[x=1.35cm,y=0.48cm,>=stealth]
\fill[gray!18]
  (-2.1,{2-2*(-2.1)})
  -- (1,0)
  -- (2.1,0)
  -- (2.1,{-2*(2.1)})
  -- (0,0)
  -- (-2.1,0)
  -- cycle;
\begin{scope}
\clip
  (-2.1,{-2*(-2.1)})
  -- (-1,2)
  -- (1,0)
  -- (2.1,0)
  -- (2.1,{-2*(2.1)})
  -- (0,0)
  -- (-2.1,0)
  -- cycle;
\foreach \s in {-7,-6.5,...,7}
  \draw[gray!65,thin] (-2.45,\s) -- (2.45,\s+2.45);
\end{scope}
\draw[->] (-2.35,0) -- (3.2,0) node[right] {$\beta$};
\draw[->] (0,-4.45) -- (0,6.25)
  node[above right,align=left,xshift=3pt,yshift=-20pt] {power of $c$\\as function of $\beta$};
\foreach \x/\lab in {-2/-2,-1/-1,1/1,2/2}
  \draw (\x,0.08) -- (\x,-0.08) node[below=1pt,font=\scriptsize] {$\lab$};
\foreach \y/\lab in {-4/-4,-2/-2,2/2,4/4,6/6}
  \draw (0.06,\y) -- (-0.06,\y) node[left=1pt,font=\scriptsize] {$\lab$};
\draw[densely dotted,gray] (-1,-4.35) -- (-1,4.35);
\draw[densely dotted,gray] (1,-4.35) -- (1,0.35);
\draw[thick,orange!85!black,domain=-2.15:2.15,samples=2]
  plot (\x,{1-\x}) node[right] {$1-\beta$};
\draw[thick,purple!80!black,domain=-2.15:2.15,samples=2]
  plot (\x,{0}) node[above] {$0$};
\draw[thick,cyan!70!black,domain=-2.15:2.15,samples=2]
  plot (\x,{-2*\x}) node[right] {$-2\beta$};
\draw[thick,green!45!black,domain=-2.15:2.15,samples=2]
  plot (\x,{2-2*\x}) node[right] {$2-2\beta$};
\node[orange!85!black,anchor=south east] at (-2.2,2.1) {$c^{1-\beta}$};
\node[purple!80!black,anchor=south west] at (-2.6,0) {$c^0$};
\node[cyan!70!black,anchor=south west] at (-4.2,4) {$c^{-2\beta}\,,\,\Epsilon=mc^2$};
\node[green!45!black,anchor=south west] at (-3,6) {$c^{2-2\beta}$};
\fill[black] (-1,2) circle (1.7pt);
\fill[black] (1,0) circle (2.2pt);
\fill[black] (1,0) circle (1.7pt);
\draw[ultra thick,green!45!black,domain=-2.15:1,samples=2]
  plot (\x,{2-2*\x});
\draw[ultra thick,purple!80!black,domain=-2.15:0,samples=2]
  plot (\x,{0});
\draw[ultra thick,purple!80!black,domain=1:2.15,samples=2]
  plot (\x,{0});
\draw[ultra thick,cyan!70!black,domain=0:2.15,samples=2]
  plot (\x,{-2*\x});
\draw[thick,black,dashed,domain=-2.15:2.15,samples=2]
  plot (\x,{-2*\x}) ;
\end{tikzpicture}
\caption{Powers of $c$ in the $C_1$ actions
\eqref{act-1-ABDG} and \eqref{act-1-ABDGE}. The gray and hatched
regions correspond to fixed $m$ and fixed $\Epsilon$, respectively;
thick lines mark their leading envelopes.}
\label{fig:cpowers-1}
\end{figure}

The upper envelope selects the Galilean limit $c\to\infty$;
the lower envelope selects the Carrollian limit $c\to0$.
Intersections retain several terms with the same leading weight,
whereas each open segment selects a fixed combination. Holding
$\Epsilon$ fixed lowers the mass exponent by two, placing it on the
$D$ line. These observations give the tables below without a separate
calculation for every value of $\beta$. The tables record all surviving
densities and their boosts; the discussion then groups related rows
and develops the representatives used in the interacting models.

\subsection{Galilean limits}
The various Galilean limits of $\mathcal{S}^{(1,\beta)}_{c,\mu}$, represented
by \eqref{act-1-ABDG} and \eqref{act-1-ABDGE}, can be written compactly as
\begin{eqnarray}
\label{GralAct1}
\mathcal{S}^{(1,\beta)}_{\infty,\mu}
&=&
\frac1{\tilde\kappa_1}\int dt\, d^3x\,
\mathcal{L}^{(1,\beta)}_{\infty,\mu},\nonumber\\
\mathcal{L}^{(1,\beta)}_{\infty,\mu}
&:=&
\epsilon_A A+\epsilon_B B+\epsilon_D D+\epsilon_G G,
\qquad \mu=m,\Epsilon,
\end{eqnarray}
Here $\mu$ records which parameter is held fixed. The coefficients
in Table~\ref{tab:S1-Galilei-coefficients} select the corresponding
leading combinations of \eqref{ABDG}.
\begin{table}[ht]
\centering
\begingroup
\small
\setlength{\tabcolsep}{2.5pt}
\renewcommand{\arraystretch}{1.15}
\begin{adjustbox}{max width=\textwidth}
\begin{tabular}{|c|c|>{\scriptsize}c|cccc|c||c|}
\hline
Fixed & Case & $\beta$ & $\epsilon_A$ & $\epsilon_B$ & $\epsilon_D$ & $\epsilon_G$ & $\mathcal{L}^{(1,\beta)}_{\infty,\mu}$, $\mu=m,\Epsilon$ & Boost\\
\hline\hline
 & i) & $-1\leq\beta<1$ & $0$ & $0$ & $0$ & $2m$
 & $+2m\,\zeta^\dagger\zeta$
 & $\begin{array}{@{}c@{}}
 K_{i,m}^{G,(-1)}\;,
 K_{i,m}^{G,(0)}
 \end{array}$\\[4pt]
\cline{2-9}
$m$ & ii) & $\beta=1$ & $1$ & $1$ & $0$ & $2m$
 & $\begin{array}{c}
 i\chi^\dagger \partial_t\chi+\zeta^\dagger\hat\nabla\chi-\chi^\dagger\hat\nabla\zeta\\
 +2m\,\zeta^\dagger\zeta
 \end{array}$
 & $K_{i,m}^{G,(1)}$\\[8pt]
\hline
 & iii) & $\beta=-1$ & $0$ & $1$ & $1$ & $2\Epsilon$
 & $\begin{array}{c}
i\zeta^\dagger\partial_t\zeta+\zeta^\dagger\hat\nabla\chi-\chi^\dagger\hat\nabla\zeta\\
 +2\Epsilon\,\zeta^\dagger\zeta
 \end{array}$
 & $K_{i,0}^{G,(-1)}$\\[8pt]
\cline{2-9}
$\Epsilon$ & iv) & $-1<\beta<1$ & $0$ & $1$ & $0$ & $0$
 & $\zeta^\dagger\hat\nabla\chi-\chi^\dagger\hat\nabla\zeta$
 & $K_{i,0}^{G,(0)}$\\[4pt]
\cline{2-9}
  & v) & $\beta=1$ & $1$ & $1$ & $0$ & $0$
 & $i\chi^\dagger \partial_t\chi+\zeta^\dagger\hat\nabla\chi-\chi^\dagger\hat\nabla\zeta$
 & $K_{i,0}^{G,(1)}$\\
\hline
\end{tabular}
\end{adjustbox}
\endgroup
\caption{Galilean $C_1$ densities and coefficients in
\eqref{GralAct1}. Boosts are those of \eqref{boost-G-beta}, with
the class label defined in \eqref{beta-star}.}
\label{tab:S1-Galilei-coefficients}
\end{table}

Row i) contains only an algebraic mass term and is invariant under
arbitrary unitary transformations preserving $\zeta^\dagger\zeta$;
the listed boost records its inherited Lorentz contraction.
The remaining rows contain the L\'evy--Leblond system, its massless
representatives, and a purely spatial central sector. Their covariance
will be verified in Appendix~\ref{app:boost-checks}.

\subsubsection{L\'evy--Leblond and its massless limits}
Rows ii), iii), and v) of
Table~\ref{tab:S1-Galilei-coefficients} are related to the
L\'evy--Leblond system \cite{Levy-Leblond:1967}.
Row ii), at fixed $m$ and $\beta=1$, is
\begin{equation}
\label{act-1-b1-NR}
\mathcal{L}^{(1,1)}_{\infty,m}
=
\chi^\dagger i\partial_t\chi
+\zeta^\dagger\hat\nabla\chi
-\chi^\dagger\hat\nabla\zeta
+2m\zeta^\dagger\zeta ,
\end{equation}
Its equations are
\begin{equation}
\label{eom-1-b1-NR}
i\partial_t\chi-\hat\nabla\zeta=0,\qquad
2m\zeta+\hat\nabla\chi=0.
\end{equation}
For $m\ne0$, eliminating the auxiliary field
$\zeta=-\hat\nabla\chi/(2m)$ gives the free Schr\"odinger equation
$$
i\partial_t\chi+\frac{1}{2m}\hat{\nabla}^2\chi=0,
$$
for $\chi$. Its nonlinear extension will be developed in
Section~\ref{sec:interacting-LL}.
Together with the auxiliary relation in \eqref{eom-1-b1-NR}, this
reproduces the fixed-$m$ Klein--Gordon limit \eqref{KG-Galilei-limit}
for both components; the leading Casimir has
$\mathcal C_m^G\Phi=m^2\Phi$.
The boost transformations for $\beta_\star=1$, generated by
$K_{i,m}^{G,(1)}$, are given in \eqref{Gboost-G-beta-components}.
The representative endpoint cancellation will be given in
Appendix~\ref{app:boost-G-endpoints}.
For row iii) of Table~\ref{tab:S1-Galilei-coefficients}, the combined
phase redefinition and field exchange
\be
\chi\longrightarrow-e^{2i\Epsilon t}\zeta,\qquad
\zeta\longrightarrow e^{2i\Epsilon t}\chi
\ee
maps the density and equations directly to row v), the massless
L\'evy--Leblond system \cite{Levy-Leblond:1967} obtained from row ii)
at $m=0$. The corresponding $m=0$ boosts in \eqref{boost-G-beta} transform
under the same field exchange.
The field equations of these massless representatives imply the
zero-Casimir condition $\mathcal C_0^G\Phi=0$.

\subsubsection{The spatial central sector}
Row iv) of Table~\ref{tab:S1-Galilei-coefficients} follows by
taking $\kappa_1=c^{1-\beta}\tilde\kappa_1$ at fixed $\Epsilon$:
\begin{equation}
\label{act-1-between-NRE}
\mathcal{L}^{(1,-1<\beta<1)}_{\infty,\Epsilon}
=
\zeta^\dagger\hat\nabla\chi
-\chi^\dagger\hat\nabla\zeta .
\end{equation}
This is the ``magnetic Galilei'' density of, for example,
Sec.~IV of \cite{Koutrolikos:2023evq}. Its spatial Dirac equations imply
$\mathcal C_0^G\Phi=0$.
Its active fields carry the diagonal boost representation; the
corresponding invariance argument will be given in
Appendix~\ref{app:boost-diagonal}.
\subsection{Carrollian limits}
The lower envelope of Figure~\ref{fig:cpowers-1} retains one or
both temporal kinetic terms and, at fixed $\Epsilon$, the mass term.
The fixed-$m$ results are obtained by setting $\Epsilon=0$ in the
following compact expression:
\begin{eqnarray}
\label{CarrAct1}
\mathcal{S}^{(1,\beta)}_{0,\Epsilon}
&=&
\frac1{\tilde\kappa_1}\int dt\, d^3x\,
\mathcal{L}^{(1,\beta)}_{0,\Epsilon},\nonumber\\
\mathcal{L}^{(1,\beta)}_{0,\Epsilon}
&:=&
\epsilon_A A+\epsilon_B B+\epsilon_D D+\epsilon_G G .
\end{eqnarray}
The coefficients, densities, and inherited boosts are listed in
Table~\ref{tab:S1-Carroll-coefficients}.
\begin{table}[ht]
\centering
\begingroup
\small
\setlength{\tabcolsep}{2.5pt}
\renewcommand{\arraystretch}{1.15}
\begin{adjustbox}{max width=\textwidth}
\begin{tabular}{|c|c|>{\scriptsize}c|cccc|c||c|}
\hline
Fixed & Case & $\beta$ & $\epsilon_A$ & $\epsilon_B$ & $\epsilon_D$ & $\epsilon_G$ & $\mathcal{L}^{(1,\beta)}_{0,\mu}$, $\mu=m,\Epsilon$ & Boost\\
\hline\hline
 & i) & $-1\leq\beta<0$ & $1$ & $0$ & $0$ & $0$
 & $\chi^\dagger i\partial_t\chi$
 & $\begin{array}{@{}c@{}}
 K_{i,0}^{C,(-1)}\\
 K_{i,0}^{C,(0)}
 \end{array}$\\[4pt]
\cline{2-9}
$m$ & ii) & $\beta=0$ & $1$ & $0$ & $1$ & $0$
 & $\chi^\dagger i\partial_t\chi+\zeta^\dagger i\partial_t\zeta$
 & $K_{i,0}^{C,(0)}$\\[4pt]
\cline{2-9}
 & iii) & $0<\beta\leq1$ & $0$ & $0$ & $1$ & $0$
 & $\zeta^\dagger i\partial_t\zeta$
 & $\begin{array}{@{}c@{}}
 K_{i,0}^{C,(0)}\\
 K_{i,0}^{C,(1)}
 \end{array}$\\[4pt]
\hline
 & iv) & $-1\leq\beta<0$ & $1$ & $0$ & $0$ & $0$
 & $\chi^\dagger i\partial_t\chi$
 & $\begin{array}{@{}c@{}}
 K_{i,\Epsilon}^{C,(-1)}\\
 K_{i,\Epsilon}^{C,(0)}
 \end{array}$\\[4pt]
\cline{2-9}
$\Epsilon$ & v) & $\beta=0$ & $1$ & $0$ & $1$ & $2\Epsilon$
 & $
 \chi^\dagger i\partial_t\chi+\zeta^\dagger i \partial_t\zeta
 +2\Epsilon\,\zeta^\dagger\zeta
 $
 & $K_{i,\Epsilon}^{C,(0)}$\\[8pt]
\cline{2-9}
 & vi) & $0<\beta\leq1$ & $0$ & $0$ & $1$ & $2\Epsilon$
 & $\zeta^\dagger i\partial_t\zeta+2\Epsilon\,\zeta^\dagger\zeta$
 & $\begin{array}{@{}c@{}}
 K_{i,\Epsilon}^{C,(0)}\\
 K_{i,\Epsilon}^{C,(1)}
 \end{array}$\\
\hline
\end{tabular}
\end{adjustbox}
\endgroup
\caption{Carrollian $C_1$ densities and coefficients in
\eqref{CarrAct1}, with boosts defined in \eqref{boost-CE-beta}.}
\label{tab:S1-Carroll-coefficients}
\end{table}
All rows realize the Carroll algebra
together with the internal phase symmetry generated by $i\mathbf 1$, namely
$\mathfrak{carroll}\oplus\mathfrak u(1)$: rotations and translations act
diagonally, the boosts are the entries of \eqref{boost-CE-beta}, and the only
non-vanishing boost commutator is \eqref{Carroll-algebra-Dt}.
The abstract time-translation generator $T_t$ is represented by
$D_t^{(\Epsilon)}$ of \eqref{eqn:Dt} in the fixed-$\Epsilon$
rows and by $\partial_t$ in the fixed-$m$ rows.
Rows i)--iii) are the familiar \emph{electric} Carroll fermions
\cite{Stakenborg:2023bmw, Bagchi:2022eui, Bergshoeff:2023vfd,
Bagchi:2025vri, Ekiz:2025hdn, Bagchi:2026lgk}. Row ii) contains
two Pauli kinetic terms; rows i), iii), and iv) share the same
single-field density up to relabeling. Their contracted
boost representations are listed separately in the table.
Rows v) and vi) also retain the $\Epsilon$-dependent term
\cite{Bagchi:2025vri}. A common representative of these densities is
row v),
\begin{equation}
\label{CarrAct10}
\mathcal{L}^{(1,0)}_{0,\Epsilon}
=
\chi^\dagger i\partial_t\chi
+i\zeta^\dagger\partial_t\zeta
+2\Epsilon\,\zeta^\dagger\zeta.
\end{equation}
The other densities follow by setting one field or $\Epsilon$ to zero.
At $\Epsilon=0$, this agrees with \cite{Bagchi:2026lgk}.
The equations are
\begin{equation}
\label{eom-1-CarrAct10}
\partial_t\chi=0,\qquad
(i\partial_t+2\Epsilon)\zeta=0 .
\end{equation}
These first-order equations imply
$\mathcal C_\Epsilon^C\Phi=\Epsilon^2\Phi$, equivalently the
Carroll equation \eqref{Cas-CE-mass-shell}; at fixed
$m$, where $\Epsilon=0$, this reduces to $\mathcal C_0^C\Phi=0$.
In the original variables, $\chi$ is time independent
and $\zeta$ has frequency $-2\Epsilon$. The component rephasing
\[
\zeta=e^{2i\Epsilon t}\zeta',
\qquad
i\zeta^\dagger\partial_t\zeta+2\Epsilon\zeta^\dagger\zeta
=i\zeta'^\dagger\partial_t\zeta'
\]
puts the density in the electric Carroll kinetic form: both redefined
fields are time independent on shell and its canonical Hamiltonian
vanishes. The effect of this map on mixed potentials will be illustrated
in Appendix~\ref{app:interaction-catalog}.
The Carroll boost symmetries are generated by the corresponding entries
$K_{i,\Epsilon}^{C,(\beta_\star)}$ in \eqref{boost-CE-beta}, with
$\Epsilon=0$ for the fixed-$m$ rows and with the appropriate component
projection when one of the two fields is absent. Their common covariance
argument is provided in Appendix~\ref{app:boost-diagonal}.

\section{Free-action contractions: the \texorpdfstring{$C_2$}{C2} branch}
\label{sec:S2-limits}

We apply the same leading-power analysis to the off-diagonal
conjugation $C_2$. Here the surviving mass-mixing term distinguishes
Hermitian limiting actions from complex seeds, for which
\eqref{L-D12} supplies the two-color completion.
The transformed seed \eqref{act-12b} is
\begin{eqnarray}
\label{act-2-massive}
\mathcal{S}^{(2,\beta)}_{c}
=
\frac1{\kappa_2}\int dt\, d^3x\,
\Big[
c^{-\beta}
\big(
\zeta^\dagger\partial_t\chi
-\chi^\dagger\partial_t\zeta
\big)
+i c\,\chi^\dagger\hat\nabla\chi
+i c^{1-2\beta}\zeta^\dagger\hat\nabla\zeta
+2im c^{2-\beta}\chi^\dagger\zeta
\Big].
\end{eqnarray}

As in Section~\ref{sec:S1-limits}, isolate the four blocks
\be\label{ABDG2}
A_2=\zeta^\dagger\partial_t\chi-\chi^\dagger\partial_t\zeta,\qquad
B_2=i\chi^\dagger\hat\nabla\chi,\qquad
D_2=i\zeta^\dagger\hat\nabla\zeta,\qquad
G_2=i\chi^\dagger\zeta .
\ee
The integrated $A_2$, $B_2$, and $D_2$ terms are Hermitian in
the convention of Section~\ref{sec:limits}; $G_2$ is not. Only limits
retaining this one-sided mixing need the completion \eqref{L-D12}.
With these blocks, the seed becomes
\be\label{act-2-ABDG}
\mathcal{S}^{(2,\beta)}_{c}
=\frac1{\kappa_2}\int dt\, d^3x\,
\Big[
c^{-\beta}A_2+cB_2+c^{1-2\beta}D_2
+2m c^{2-\beta}G_2
\Big].
\ee
At fixed $\Epsilon$, the $G_2$ exponent drops by two and coincides
with that of $A_2$:
\be\label{act-2-ABDGE}
\mathcal{S}^{(2,\beta)}_c
=\frac1{\kappa_2}\int dt\, d^3x\,
\Big[
c^{-\beta}(A_2+2\Epsilon G_2)+cB_2+c^{1-2\beta}D_2
\Big].
\ee

\begin{figure}[ht]
\centering
\begin{tikzpicture}[x=1.35cm,y=0.48cm,>=stealth]
\fill[gray!18]
  (-2.1,1)
  -- (-1,1)
  -- (1,-1)
  -- (2.1,{1-2*(2.1)})
  -- (2.1,1)
  -- (1,1)
  -- (-1,3)
  -- (-2.1,{1-2*(-2.1)})
  -- cycle;

\begin{scope}
\clip
  (-2.1,1)
  -- (-1,1)
  -- (1,-1)
  -- (2.1,{1-2*(2.1)})
  -- (2.1,1)
  -- (0,1)
  -- (-2.1,{1-2*(-2.1)})
  -- cycle;
\foreach \s in {-7,-6.5,...,7}
  \draw[gray!65,thin] (-2.45,\s) -- (2.45,\s+2.45);
\end{scope}

\draw[->] (-2.35,0) -- (3.5,0) node[right] {$\beta$};
\draw[->] (0,-4.45) -- (0,6.25)
  node[above right,align=left,xshift=3pt,yshift=-20pt] {power of $c$\\as function of $\beta$};

\foreach \x/\lab in {-2/-2,-1/-1,1/1,2/2}
  \draw (\x,0.08) -- (\x,-0.08) node[below=1pt,font=\scriptsize] {$\lab$};
\foreach \y/\lab in {-4/-4,-2/-2,1/1,2/2,4/4,6/6}
  \draw (0.06,\y) -- (-0.06,\y) node[left=1pt,font=\scriptsize] {$\lab$};

\draw[densely dotted,gray] (0,-4.35) -- (0,2.35);
\draw[densely dotted,gray] (1,-4.35) -- (1,1.35);

\draw[thin,orange!85!black,domain=-2.15:2.15,samples=2]
  plot (\x,{-\x}) node[right] {$-\beta$};
\draw[thick,purple!80!black,domain=-2.15:2.15,samples=2]
  plot (\x,{1}) node[right] {$1$};
\draw[thick,cyan!70!black,domain=-2.15:2.15,samples=2]
  plot (\x,{1-2*\x}) node[right] {$1-2\beta$};
\draw[thick,green!45!black,domain=-2.15:2.15,samples=2]
  plot (\x,{2-\x}) node[right,yshift=-5pt] {$2-\beta$};
\node[green!45!black,anchor=south east] at (-2.2,3.7) {$c^{2-\beta}$};
\node[purple!80!black,anchor=south west] at (-2.6,0.6) {$c$};
\node[cyan!70!black,anchor=south west] at (-3,5) {$c^{1-2\beta}$};
\node[orange!85!black,anchor=north west] at (-4,3) {$c^{-\beta}\,,\Epsilon=mc^2$};

\draw[line width=1.8pt,cyan!70!black,domain=-2.15:-1,samples=2]
  plot (\x,{1-2*\x});
\draw[line width=1.8pt,orange!85!black,domain=-1:1,samples=2]
  plot (\x,{-\x});
\draw[line width=1.8pt,green!45!black,domain=-1:1,samples=2]
  plot (\x,{2-\x});
\draw[line width=1.8pt,purple!80!black,domain=-2.15:-1,samples=2]
  plot (\x,{1});
\draw[line width=1.8pt,purple!80!black,domain=1:2.15,samples=2]
  plot (\x,{1});
\draw[line width=1.8pt,cyan!70!black,domain=1:2.15,samples=2]
  plot (\x,{1-2*\x});
\draw[thick,black,dashed,domain=-2.15:2.15,samples=2]
  plot (\x,{-\x});
\end{tikzpicture}
\caption{Powers of $c$ in the $C_2$ seeds
\eqref{act-2-ABDG} and \eqref{act-2-ABDGE}. As in
Figure~\ref{fig:cpowers-1}, the gray and hatched regions correspond
to fixed $m$ and fixed $\Epsilon$, and thick lines mark the leading
envelopes.}
\label{fig:cpowers-2}
\end{figure}

The upper and lower envelopes of Figure~\ref{fig:cpowers-2}
select the Galilean and Carrollian limits, respectively. Choosing
$\kappa_2=c^p\tilde\kappa_2$ then removes the common leading power.

\subsection{Galilean limits}

The resulting Galilean densities have the form
\begin{eqnarray}
\label{GralAct2}
\mathcal{S}^{(2,\beta)}_{\infty,\mu}
&=&
\frac1{\tilde\kappa_2}\int dt\, d^3x\,
\mathcal{L}^{(2,\beta)}_{\infty,\mu},\nonumber\\
\mathcal{L}^{(2,\beta)}_{\infty,\mu}
&:=&
\epsilon_A A_2+\epsilon_B B_2+\epsilon_D D_2+\epsilon_G G_2,
\qquad \mu=m,\Epsilon .
\end{eqnarray}
As in the $C_1$ catalog, $\mu$ indicates the fixed parameter.
Table~\ref{tab:S2-Galilei-coefficients} gives the coefficients in
the basis \eqref{ABDG2}. A common proof of covariance will be given in
Appendix~\ref{app:boost-checks}.

\begin{table}[ht]
\centering
\begingroup
\small
\setlength{\tabcolsep}{2.5pt}
\renewcommand{\arraystretch}{1.15}
\begin{adjustbox}{max width=\textwidth}
\begin{tabular}{|c|c|>{\scriptsize}c|cccc|c||c|}
\hline
Fixed & Case & $\beta$ & $\epsilon_A$ & $\epsilon_B$ & $\epsilon_D$ & $\epsilon_G$ & $\mathcal{L}^{(2,\beta)}_{\infty,\mu}$, $\mu=m,\Epsilon$ & Boost\\
\hline\hline
 & i) & $\beta=-1$ & $0$ & $0$ & $1$ & $2m$
 & $i\zeta^\dagger\hat\nabla\zeta+2im\,\chi^\dagger\zeta$
 & $K_{i,m}^{G,(-1)}$\\[4pt]
\cline{2-9}
$m$ & ii) & $-1<\beta<1$ & $0$ & $0$ & $0$ & $2m$
 & $2im\,\chi^\dagger\zeta$
 & $K_{i,m}^{G,(0)}$\\[4pt]
\cline{2-9}
 & iii) & $\beta=1$ & $0$ & $1$ & $0$ & $2m$
 & $i\chi^\dagger\hat\nabla\chi+2im\,\chi^\dagger\zeta$
 & $K_{i,m}^{G,(1)}$\\[4pt]
\hline
 & iv) & $-1\leq\beta<0$ & $0$ & $0$ & $1$ & $0$
 & $i\zeta^\dagger\hat\nabla\zeta$
 & $\begin{array}{@{}c@{}}
 K_{i,0}^{G,(-1)}\\
 K_{i,0}^{G,(0)}
 \end{array}$\\[4pt]
\cline{2-9}
$\Epsilon$ & v) & $\beta=0$ & $0$ & $1$ & $1$ & $0$
 & $i\chi^\dagger\hat\nabla\chi+i\zeta^\dagger\hat\nabla\zeta$
 & $K_{i,0}^{G,(0)}$\\[4pt]
\cline{2-9}
 & vi) & $0<\beta\leq1$ & $0$ & $1$ & $0$ & $0$
 & $i\chi^\dagger\hat\nabla\chi$
 & $\begin{array}{@{}c@{}}
 K_{i,0}^{G,(0)}\\
 K_{i,0}^{G,(1)}
 \end{array}$\\
\hline
\end{tabular}
\end{adjustbox}
\endgroup
\caption{Galilean $C_2$ densities and coefficients in
\eqref{GralAct2}, with boosts defined in \eqref{boost-G-beta}.}
\label{tab:S2-Galilei-coefficients}
\end{table}

Row ii) contains only an algebraic mixing term. We therefore
concentrate on the endpoints at fixed $m$ and the spatial sectors at
fixed $\Epsilon$.

\subsubsection{Endpoint Bargmann seeds}
Rows i) and iii) of Table~\ref{tab:S2-Galilei-coefficients}
are the fixed-$m$ endpoints $\beta=\pm1$, where the mass mixing and
one spatial kinetic term have the same leading weight. At
$\beta=-1$, the fixed-$m$ upper envelope in Fig.~\ref{fig:cpowers-2} is the
double intersection of the lines $c^{1-2\beta}$ and $c^{2-\beta}$ at order
$c^3$. Taking
$\kappa_2=c^3\tilde\kappa_2$ gives
\begin{equation}
\label{act-2-bm1-NR}
\mathcal{L}^{(2,-1)}_{\infty,m}
=
i\zeta^\dagger\hat\nabla\zeta
+2im\chi^\dagger\zeta .
\end{equation}
At $\beta=1$, the fixed-$m$ upper envelope in Fig.~\ref{fig:cpowers-2}
contains the $\chi$ spatial term and the mass mixing. Taking
$\kappa_2=c\tilde\kappa_2$ gives
\begin{equation}
\label{act-2-NRb1}
\mathcal{L}^{(2,1)}_{\infty,m}
=
i\chi^\dagger\hat\nabla\chi
+2im\chi^\dagger\zeta .
\end{equation}
These are the $\beta_\star=-1$ and $+1$ boost classes in
\eqref{Gboost-G-beta-components}; their common invariance check is
deferred to Appendix~\ref{app:boost-G-endpoints}.
In either case the $G_2$ term makes the limiting action
non-hermitian, as anticipated above. As left--right variational systems,
their nontrivial equations obtained by varying the independent dual and
ordinary fields are, respectively,
\begin{equation}
\label{eom-2-endpoint-NR}
\begin{array}{ll}
\beta=-1: &
2m\,\zeta=0,\qquad
\hat\nabla\zeta=0;\qquad
\zeta^\dagger\overleftarrow{\hat\nabla}-2m\chi^\dagger=0,\\[6pt]
\beta=+1: &
\hat\nabla\chi+2m\zeta=0;\qquad
\chi^\dagger\overleftarrow{\hat\nabla}=0,\qquad
2m\chi^\dagger=0 .
\end{array}
\end{equation}
The left and right equations follow from the independent-dual
variational principle. We will use these seeds to construct Hermitian
completions and analyze their fermionic gauge symmetry and quartic
deformations in Section~\ref{sec:extended-real-actions}.

\subsubsection{Massless spatial sectors}

Rows iv) and vi) of Table~\ref{tab:S2-Galilei-coefficients}
are related by field exchange. Taking
$\kappa_2=c^{1-2\beta}\tilde\kappa_2$ and
$\kappa_2=c\tilde\kappa_2$, respectively, gives
\begin{equation}
\label{act-2-neg-NRE}
\mathcal{L}^{(2,-1\leq\beta<0)}_{\infty,\Epsilon}
=
i\zeta^\dagger\hat\nabla\zeta \,,\qquad \mathcal{L}^{(2,0<\beta\leq1)}_{\infty,\Epsilon}
=
i\chi^\dagger\hat\nabla\chi .
\end{equation}
Row v), obtained with $\kappa_2=c\tilde\kappa_2$, is their
direct sum. None retains $G_2$, so all three actions are Hermitian.
The corresponding field equations are
\begin{equation}
\label{eom-2-GE-spatial}
 \hat\nabla\zeta=0\,,\qquad
 \hat\nabla\chi=0.
\end{equation}
Squaring the spatial Dirac operator gives
$\mathcal C_0^G\Phi=0$.
The active fields in each action transform only by the diagonal massless
Galilei transport, $\delta=-t\,b^i\partial_i$; see
Appendix~\ref{app:boost-diagonal}. Hence all fixed-$\Epsilon$ densities
transform as spatial scalar densities.

\subsection{Carrollian limits}
For the Carroll limit $c\to0$, the lower envelope in Fig.~\ref{fig:cpowers-2}
for $-1\leq\beta\leq1$ gives the compact form
\begin{eqnarray}
\label{CarrAct2}
\mathcal{S}^{(2,\beta)}_{0,\mu}
&=&
\frac1{\tilde\kappa_2}\int dt\, d^3x\,
\mathcal{L}^{(2,\beta)}_{0,\mu},\nonumber\\
\mathcal{L}^{(2,\beta)}_{0,\mu}
&:=&
\epsilon_A A_2+\epsilon_B B_2+\epsilon_D D_2+\epsilon_G G_2,
\qquad \mu=m,\Epsilon .
\end{eqnarray}
Table~\ref{tab:S2-Carroll-coefficients} lists the coefficients.
The fixed-$m$ rows are Hermitian; at fixed nonzero $\Epsilon$, the
surviving $2\Epsilon G_2$ makes each row a complex seed.
\begin{table}[ht]
\centering
\begingroup
\small
\setlength{\tabcolsep}{2.5pt}
\renewcommand{\arraystretch}{1.15}
\begin{adjustbox}{max width=\textwidth}
\begin{tabular}{|c|c|>{\scriptsize}c|cccc|c||c|}
\hline
Fixed & Case & $\beta$ & $\epsilon_A$ & $\epsilon_B$ & $\epsilon_D$ & $\epsilon_G$ & $\mathcal{L}^{(2,\beta)}_{0,\mu}$, $\mu=m,\Epsilon$ & Boost\\
\hline\hline
 & i) & $\beta=-1$ & $1$ & $1$ & $0$ & $0$
 & $\begin{array}{c}
 \zeta^\dagger\partial_t\chi-\chi^\dagger\partial_t\zeta\\
 +i\chi^\dagger\hat\nabla\chi
 \end{array}$
 & $K_{i,0}^{C,(-1)}$\\[8pt]
\cline{2-9}
$m$ & ii) & $-1<\beta<1$ & $1$ & $0$ & $0$ & $0$
 & $\zeta^\dagger\partial_t\chi-\chi^\dagger\partial_t\zeta$
 & $K_{i,0}^{C,(0)}$\\[4pt]
\cline{2-9}
 & iii) & $\beta=1$ & $1$ & $0$ & $1$ & $0$
 & $\begin{array}{c}
 \zeta^\dagger\partial_t\chi-\chi^\dagger\partial_t\zeta\\
 +i\zeta^\dagger\hat\nabla\zeta
 \end{array}$
 & $K_{i,0}^{C,(1)}$\\[8pt]
\hline
 & iv) & $\beta=-1$ & $1$ & $1$ & $0$ & $2\Epsilon$
 & $\begin{array}{c}
 \zeta^\dagger\partial_t\chi-\chi^\dagger\partial_t\zeta
 +i\chi^\dagger\hat\nabla\chi\\
 +2i\Epsilon\,\chi^\dagger\zeta
 \end{array}$
 & $K_{i,\Epsilon}^{C,(-1)}$\\[8pt]
\cline{2-9}
$\Epsilon$ & v) & $-1<\beta<1$ & $1$ & $0$ & $0$ & $2\Epsilon$
 & $\zeta^\dagger\partial_t\chi-\chi^\dagger\partial_t\zeta
 +2i\Epsilon\,\chi^\dagger\zeta$
 & $K_{i,\Epsilon}^{C,(0)}$\\[4pt]
\cline{2-9}
 & vi) & $\beta=1$ & $1$ & $0$ & $1$ & $2\Epsilon$
 & $\begin{array}{c}
 \zeta^\dagger\partial_t\chi-\chi^\dagger\partial_t\zeta
 +i\zeta^\dagger\hat\nabla\zeta\\
 +2i\Epsilon\,\chi^\dagger\zeta
 \end{array}$
 & $K_{i,\Epsilon}^{C,(1)}$\\
\hline
\end{tabular}
\end{adjustbox}
\endgroup
\caption{Carrollian $C_2$ densities and coefficients in
\eqref{CarrAct2}, with boosts defined in \eqref{boost-CE-beta}.}
\label{tab:S2-Carroll-coefficients}
\end{table}

\subsubsection{Central sectors}

Rows ii) and v) of Table~\ref{tab:S2-Carroll-coefficients}
share the central boost representation. At fixed $m$, taking
$\kappa_2=c^{-\beta}\tilde\kappa_2$ gives
\begin{equation}
\label{act-2-C}
\mathcal{L}^{(2,-1<\beta<1)}_{0,m}
=
\zeta^\dagger\partial_t\chi
-\chi^\dagger\partial_t\zeta .
\end{equation}
This fixed-$m$ action is hermitian.
Varying the daggered fields gives
\begin{equation}
\label{eom-2-C}
\partial_t\zeta=0,\qquad
\partial_t\chi=0.
\end{equation}
The action is invariant under the diagonal boosts generated by
$K_{i,0}^{C,(0)}$; the direct check is provided in
Appendix~\ref{app:boost-diagonal}.
The same normalization at fixed $\Epsilon$ gives row v),
\begin{equation}
\label{act-2-CE}
\mathcal{L}^{(2,-1<\beta<1)}_{0,\Epsilon}
=
\zeta^\dagger\partial_t\chi
-\chi^\dagger\partial_t\zeta
+2i\Epsilon\chi^\dagger\zeta .
\end{equation}
This belongs to the central boost class $\beta_\star=0$ and, because it
retains $2\Epsilon G_2$, is a non-hermitian complex seed. Varying the
independent dual fields gives
\begin{equation}
\label{eom-2-CE}
(\partial_t-2i\Epsilon)\zeta=0,\qquad
\partial_t\chi=0,
\end{equation}
while varying the undaggered fields gives the complementary equations
\begin{equation}
\label{eom-2-CE-dagger}
(\partial_t+2i\Epsilon)\chi^\dagger=0,\qquad
\partial_t\zeta^\dagger=0.
\end{equation}
The two sets are the left and right equations of the complex seed. For
$\Epsilon\neq0$, if the dual fields are instead identified with the ordinary Hermitian
conjugates of $\chi$ and $\zeta$, they overconstrain the system to
$\chi=\zeta=0$.
In the independent-dual formulation, however,
$\zeta=e^{2i\Epsilon t}\zeta'$ and
$\chi^\dagger=e^{-2i\Epsilon t}\chi'{}^\dagger$, with
$\chi$ and $\zeta^\dagger$ unchanged, map the seed and its left and
right equations to the parameter-free form. This relates
the complexified free seeds. We will construct the corresponding
Hermitian two-color completion and its kinetic form in
Section~\ref{sec:colored-Carroll-central}.
\subsubsection{Endpoints at fixed mass}
Rows i) and iii) of Table~\ref{tab:S2-Carroll-coefficients}
are related by the chiral map \eqref{D-map} and the boost duality
\eqref{boost-C-duality}. For the first, choosing
$\kappa_2=c\tilde\kappa_2$ in \eqref{act-2-massive} gives the Hermitian
density
\begin{equation}
\label{act-2-bm1-C}
\mathcal L^{(2,-1)}_{0,m}
=\zeta^\dagger\partial_t\chi-\chi^\dagger\partial_t\zeta
+i\chi^\dagger\hat\nabla\chi.
\end{equation}
The $\zeta$ spatial term and the non-Hermitian mass mixing are
suppressed by $c^2$ after normalization, so the limit is Hermitian
without color doubling. Varying the daggered fields gives
\begin{equation}
\label{eom-2-bm1-C}
\partial_t\zeta-i\hat\nabla\chi=0,\qquad
\partial_t\chi=0 .
\end{equation}
The boost is the $\beta_\star=-1$ entry of
\eqref{boost-CE-beta}. The invariance of the action will be proved in
Appendix~\ref{app:boost-C-endpoints}.
Its dual, row iii), follows with
$\kappa_2=c^{-1}\tilde\kappa_2$:
\begin{equation}
\label{act-2-b1-C}
\mathcal{L}^{(2,1)}_{0,m}
=
\zeta^\dagger\partial_t\chi
-\chi^\dagger\partial_t\zeta
+i\zeta^\dagger\hat\nabla\zeta .
\end{equation}
Varying the daggered fields gives
\begin{equation}
\label{eom-2-b1-C}
\partial_t\zeta=0,\qquad
\partial_t\chi+i\hat\nabla\zeta=0 .
\end{equation}
Together with \eqref{eom-2-C}, the two endpoint systems
\eqref{eom-2-bm1-C} and \eqref{eom-2-b1-C} show that every fixed-$m$
Carroll sector satisfies $\mathcal C_0^C\Phi=0$.
This condition does not require every solution to be time
independent: for example, \eqref{eom-2-bm1-C} admits
$\chi=\chi_0(\vec x)$ and
$\zeta=\zeta_0(\vec x)+it\hat\nabla\chi_0(\vec x)$.
Such solutions obey $(i\partial_t)^2\Phi=0$, but need not
obey $i\partial_t\Phi=0$.
The duality preserves both Hermiticity and covariance, so the
opposite endpoint needs no separate invariance calculation.
\subsubsection{Endpoints at fixed energy parameter}
Rows iv) and vi) of Table~\ref{tab:S2-Carroll-coefficients}
retain the one-sided mixing. With $\kappa_2=c\tilde\kappa_2$, the
$\beta=-1$ density is
\begin{equation}
\label{act-2-bm1-CE}
\mathcal{L}^{(2,-1)}_{0,\Epsilon}
=
\zeta^\dagger\partial_t\chi
-\chi^\dagger\partial_t\zeta
+i\chi^\dagger\hat\nabla\chi
+2i\Epsilon\chi^\dagger\zeta.
\end{equation}
The $\beta=1$ density, with
$\kappa_2=c^{-1}\tilde\kappa_2$, is
\begin{equation}
\label{act-2-b1-CE}
\mathcal{L}^{(2,1)}_{0,\Epsilon}
=
\zeta^\dagger\partial_t\chi
-\chi^\dagger\partial_t\zeta
+i\zeta^\dagger\hat\nabla\zeta
+2i\Epsilon\chi^\dagger\zeta .
\end{equation}
Their boosts are the $\beta_\star=\mp1$ entries of
\eqref{boost-CE-beta}, with component form
\eqref{Cboost-CE-beta-components}. The phase similarity
\eqref{boost-C-phase-sim} relates these representations to the
connection-free ones; the common covariance check is in
Appendix~\ref{app:boost-C-endpoints}. For nonzero $\Epsilon$, both
densities remain complex seeds. We next construct their Hermitian
two-color completions.

\subsection{Hermitian two-color Carroll completions}
\label{sec:colored-Carroll-free}

For Carroll limits in which the massive $C_2$ seed \eqref{act-D}
retains a non-Hermitian mass-mixing term, we first form the
Hermitian two-color completion \eqref{L-D12} and then take the
contraction. Equivalently, we may complete the contracted seed:
we assign color $1$ to the undaggered fields and color $2$ to
their independent duals, then add the Hermitian conjugate.
Both procedures give the same limiting action. We now apply this
equivalence to the central and endpoint seeds and determine their
free solutions. The endpoint solution will provide the starting point
for the color-mixed deformation in
Section~\ref{sec:colored-Carroll}.

\subsubsection{Central free action}
\label{sec:colored-Carroll-central}

At fixed $\Epsilon$ and $\beta=0$, the single-color $C_2$ density
\eqref{act-2-CE} retains
the non-Hermitian mixing term $2i\Epsilon\chi^\dagger\zeta$.
Following the completion procedure described above, we introduce
two independent sets of fields $(\chi_a,\zeta_a)$, $a=1,2$,
and add the Hermitian conjugate to obtain the Hermitian central action
\begin{equation}
\label{act-2-C-real}
\mathcal S_{12;\,0,\Epsilon}^{(2,0)}
=
\frac1{2\tilde\kappa_2}\int dt\,d^3x\,
\Big[
-\chi_2^\dagger\partial_t\zeta_1
+\zeta_2^\dagger\partial_t\chi_1
+2i\Epsilon\chi_2^\dagger\zeta_1
+\zeta_1^\dagger\partial_t\chi_2
-\chi_1^\dagger\partial_t\zeta_2
-2i\Epsilon\zeta_1^\dagger\chi_2
\Big].
\end{equation}
The action is invariant under the central Carroll boosts; the diagonal
transport check will be given in Appendix~\ref{app:boost-diagonal}.

The field equations are
\begin{equation}
\label{eom-2-C-real}
(\partial_t-2i\Epsilon)\zeta_1=0,\qquad
\partial_t\chi_1=0,\qquad
(\partial_t-2i\Epsilon)\chi_2=0,\qquad
\partial_t\zeta_2=0.
\end{equation}
With the block ordering
$\Phi_a=(\zeta_a,\chi_a)^T$ of \eqref{S12},
%
the field equations imply that both colors satisfy
$\mathcal C_\Epsilon^C\Phi_a=\Epsilon^2\Phi_a$ in this representation.

Free Carroll actions with nonzero and vanishing mass parameters can
be related by time-dependent field redefinitions. In the present
central model, the component rephasing
\[
\zeta_1=e^{2i\Epsilon t}\zeta'_1,\qquad
\chi_2=e^{2i\Epsilon t}\chi'_2
\]
with $\chi_1$ and $\zeta_2$ unchanged, maps
\eqref{act-2-C-real} to its $\Epsilon=0$, purely kinetic form,
\begin{equation}
\label{act-2-C-real-kinetic}
\mathcal S_{12;\,0,\Epsilon}^{(2,0)}
=
\frac1{2\tilde\kappa_2}\int dt\,d^3x\,
\Big[
-\chi_2^{\prime\dagger}\partial_t\zeta'_1
+\zeta_2^\dagger\partial_t\chi_1
+\zeta_1^{\prime\dagger}\partial_t\chi'_2
-\chi_1^\dagger\partial_t\zeta_2
\Big].
\end{equation}

The free equations make all four redefined Pauli fields
time independent, and the canonical Hamiltonian of
\eqref{act-2-C-real-kinetic} vanishes.
Its relation to the ultralocal fermion kinetic term
$i\bar\Psi\partial_t\Psi$ of \cite{Stakenborg:2023bmw}, Section~4.1,
is explicit: set
\[
\Psi_A=\begin{pmatrix}\zeta'_1\\-i\chi'_2\end{pmatrix},\qquad
\Psi_B=\begin{pmatrix}\chi_1\\i\zeta_2\end{pmatrix},\qquad
\bar\Psi_{A,B}=\Psi_{A,B}^\dagger\Lambda,\qquad
\Lambda=\begin{pmatrix}0&\mathbf1_2\\\mathbf1_2&0\end{pmatrix}.
\]
Then the same action reads
\[
\mathcal S_{12;\,0,\Epsilon}^{(2,0)}
=\frac{i}{2\tilde\kappa_2}\int dt\,d^3x\,
\left(\bar\Psi_A\partial_t\Psi_A+
\bar\Psi_B\partial_t\Psi_B\right).
\]
This identifies the free kinetic structure with a standard Carroll
form.

\subsubsection{Endpoint free actions}
\label{sec:Carroll with finite E}

At fixed $\Epsilon$ and $\beta=-1$, the single-color $C_2$ density
\eqref{act-2-bm1-CE} retains
both the non-Hermitian mixing $2i\Epsilon\chi^\dagger\zeta$ and
the spatial term $i\chi^\dagger\hat\nabla\chi$.
Introducing the two colors as above and adding the Hermitian
conjugate gives the Hermitian endpoint action
\begin{equation}
\label{act-2-bm1-CE-real}
\begin{aligned}
\mathcal S_{12;\,0,\Epsilon}^{(2,-1)}
=
\frac1{2\tilde\kappa_2}\int dt\,d^3x\,
\Big[{}&
-\chi_2^\dagger\partial_t\zeta_1
+\zeta_2^\dagger\partial_t\chi_1
+i\chi_2^\dagger\hat\nabla\chi_1
+2i\Epsilon\chi_2^\dagger\zeta_1\\
&+\zeta_1^\dagger\partial_t\chi_2
-\chi_1^\dagger\partial_t\zeta_2
+i\chi_1^\dagger\hat\nabla\chi_2
-2i\Epsilon\zeta_1^\dagger\chi_2
\Big].
\end{aligned}
\end{equation}

The field equations are
\begin{equation}
\label{eom-2-bm1-CE-real}
\begin{gathered}
(\partial_t-2i\Epsilon)\zeta_1-i\hat\nabla\chi_1=0,\qquad
\partial_t\chi_1=0,\\
(\partial_t-2i\Epsilon)\chi_2=0,\qquad
\partial_t\zeta_2-i\hat\nabla\chi_2=0.
\end{gathered}
\end{equation}

The two frequency branches follow directly from
\eqref{eom-2-bm1-CE-real}. Applying
$\partial_t$ to the first equation of \eqref{eom-2-bm1-CE-real} and
$\partial_t-2i\Epsilon$ to the last, then using the equations for
$\chi_a$, gives
\be \label{finiteE}
\partial_t(\partial_t-2i\Epsilon)\zeta_a=0,\qquad a=1,2 .
\ee
Together with \eqref{eom-2-bm1-CE-real}, this gives
$\mathcal C_\Epsilon^C\Phi_a=\Epsilon^2\Phi_a$ for both colors: the first-order system
constitutes a square root of the Carroll equation
\eqref{Cas-CE-mass-shell}.
For $\Epsilon\neq0$, each $\zeta_a$ therefore has the form
\be\label{solutionEC}
\zeta_a(t,\vec x)=\zeta_a^{(0)}(\vec x)
+e^{2i\Epsilon t}\zeta_a^{(-2\Epsilon)}(\vec x).
\ee
The two profiles have frequencies $0$ and $-2\Epsilon$. The common phase \eqref{Carroll-energy-basis} redefinition centers the latter frequencies at
$+\Epsilon$ and $-\Epsilon$, respectively.
Writing the corresponding mode decomposition for $\chi_a$,
\[
\chi_a(t,\vec x)=\chi_a^{(0)}(\vec x)
+e^{2i\Epsilon t}\chi_a^{(-2\Epsilon)}(\vec x),
\]
the first-order equations \eqref{eom-2-bm1-CE-real} imply
$\chi_1^{(-2\Epsilon)}=0$ and $\chi_2^{(0)}=0$; the remaining components satisfy
\[
2\Epsilon\,\zeta_1^{(0)}+\hat\nabla\chi_1^{(0)}=0,\qquad
2\Epsilon\,\zeta_2^{(-2\Epsilon)}-\hat\nabla\chi_2^{(-2\Epsilon)}=0.
\]
For $\Epsilon\neq0$, these relations determine
$\zeta_1^{(0)}$ and
$\zeta_2^{(-2\Epsilon)}$,
and the complete solution
for the $\chi$ fields is
$\chi_1(t,\vec x)=\chi_1^{(0)}(\vec x)$ and
$\chi_2(t,\vec x)=e^{2i\Epsilon t}\chi_2^{(-2\Epsilon)}(\vec x)$,
with arbitrary spatial functions $\chi_1^{(0)}$ and
$\chi_2^{(-2\Epsilon)}$.

In the degenerate case $\Epsilon=0$ the same
equations allow linear Carroll-time dependence,
$\zeta_1=\zeta_1^{(0)}+it\,\hat\nabla\chi_1^{(0)}$ and
$\zeta_2=\zeta_2^{(0)}+it\,\hat\nabla\chi_2^{(0)}$, unless additional boundary
or regularity conditions are imposed.

The discrete color-and-chiral map
\begin{equation}
\label{dual-U}
\mathcal U:\quad
\chi_1\mapsto\zeta_2,\qquad
\zeta_1\mapsto-\chi_2,\qquad
\chi_2\mapsto\zeta_1,\qquad
\zeta_2\mapsto-\chi_1
\end{equation}
relates \eqref{act-2-bm1-CE-real} to the opposite endpoint,
$\beta_\star=+1$. Appendix~\ref{app:Bergshoeff-dictionary} will give
the explicit dual action and the field dictionaries relating both
endpoints to the non-minimal magnetic model of
Ref.~\cite{Bergshoeff:2023vfd}.
Appendix~\ref{app:spatial-field-redefinition} will also relate the free
endpoint and central actions by a spatial field redefinition for
$\Epsilon\neq0$, and explain its effect on interactions.

The two branches thus organize the free actions by their leading
densities and contracted boost representations. In the Carroll limit,
the $C_1$ branch selects electric kinetic terms, while the $C_2$
endpoints retain the spatial terms of magnetic representatives.
We now use these representations to construct invariant interactions
for the resulting theories.

\section{Boost-invariant local interactions}
\label{sec:interactions}

The free actions obtained in Sections~\ref{sec:S1-limits} and
\ref{sec:S2-limits} carry the boost representations derived in
Section~\ref{sec:limits}. We construct their local interactions by
requiring the added potentials to respect the same representations.
Here symmetry determines the allowed interactions intrinsically; the NJL
contractions studied later will select the channels inherited from that
relativistic model.
The construction gives sufficient families at arbitrary polynomial order
and a complete single-color quartic basis under the assumptions stated
below. The starting point is a rotational-scalar bilinear
\be\label{coloredbilinear}
\mathcal B_{ab}[\Gamma]:=\Phi_a^\dagger\Gamma\Phi_b ,
\ee
where $a,b$ are fermion ``color'' labels and $\Gamma$ acts on spinor indices.
For one color the labels can be suppressed. The diagonal part of a
Galilean or Carrollian boost gives
\begin{align}
\delta_b^G\mathcal B_{ab}
&=-\partial_i(t b^i\mathcal B_{ab})
+i(m_b-m_a)b^ix_i\mathcal B_{ab},
\nonumber\\
\delta_b^C\mathcal B_{ab}
&=-\partial_t(b^ix_i\mathcal B_{ab})
+i(\Epsilon_b-\Epsilon_a)b^ix_i\mathcal B_{ab}.
\label{bilinear-diagonal-boost}
\end{align}
Thus, an individual bilinear has no additional phase when $m_a=m_b$ in the
Galilei case or $\Epsilon_a=\Epsilon_b$ in the Carroll case. More generally,
a product of bilinears is admissible whenever its $m$- or
$\Epsilon$-dependent terms cancel in the product, so that its variation in
the action is a boundary term.

\subsection{Single-color self-interactions}
In the central class, $\beta_\star=0$, both Pauli components transform
diagonally, so the preceding criterion applies to all rotational scalars.
At an endpoint, one component acquires an additional spin-mixing term;
see \eqref{Cboost-CE-beta-components} and
\eqref{Gboost-G-beta-components}. The interaction must also be invariant
under this mixing.

Let $u$ denote the component without spin mixing and $v$ the
component whose boost contains the term
$\delta_b^{\rm mix}v=\pm\mathbf S(b)u$, where
$\mathbf S(b):=\frac{i}{2}b^i\sigma_i$ and
$\mathbf S(b)^\dagger=-\mathbf S(b)$. The plus sign applies to the Galilei
$\beta_\star=-1$ and Carroll $\beta_\star=+1$ endpoints, and the minus sign
to the other two endpoints. The two Hermitian
channels
\begin{equation}
\label{endpoint-mixed-bilinear}
N_u:=u^\dagger u,\qquad
M_{uv}:=u^\dagger v+v^\dagger u
\end{equation}
transform as scalars under the boost: their diagonal variations are pure transport
terms, while their spin-mixing variations vanish. In particular,
$\delta_b^{\rm mix}M_{uv}
=\pm u^\dagger\bigl(\mathbf S(b)+\mathbf S(b)^\dagger\bigr)u=0$.
By contrast, $v^\dagger v$ and a generic mixed bilinear do not transform as
boost scalars.
The component assignments are
\[
\begin{array}{c|cc}
 & \beta_\star=-1 & \beta_\star=+1\\ \hline
\text{Galilei}\; (u,v) & (\zeta,\chi) & (\chi,\zeta)\\
\text{Carroll} \; (u,v) & (\chi,\zeta)& (\zeta,\chi)
\end{array}.
\]
Any real polynomial $V(N_u,M_{uv})$, including $V_u(N_u)$, therefore
defines an admissible endpoint interaction. This sufficient family is
not exhaustive: at quartic order, two additional channels complete the
basis.

\subsection{Complete single-color quartic basis}
\label{sec:quartic-basis}

Consider one pair of classical Grassmann-odd Pauli spinors $u,v$, transforming under the endpoint boost above.
We seek Hermitian quartic polynomials with no derivatives, invariant
under rotations and the common global $U(1)$ phase
$u\mapsto e^{i\alpha}u$, $v\mapsto e^{i\alpha}v$.
We use ordinary Hermitian conjugation and impose no  discrete symmetries. Under these assumptions, the invariant space
has five independent real operators.

In addition to $N_u$ and $M_{uv}$, define the spin-singlet pairing bilinears
\begin{equation}
\label{endpoint-pair-bilinears}
P_u:=u^T i\sigma_2u,\qquad Q_{uv}:=u^T i\sigma_2v .
\end{equation}
They are rotational scalars and their spin-mixing variations vanish:
$\delta_b^{\rm mix}P_u=0$ and
$\delta_b^{\rm mix}Q_{uv}
=\pm u^T i\sigma_2\mathbf S(b)u=0$.
The latter identity uses the symmetry of $i\sigma_2\sigma_i$ and
Grassmann antisymmetry. Under this common $U(1)$, both pairing bilinears transform
as $P_u\mapsto e^{2i\alpha}P_u$ and
$Q_{uv}\mapsto e^{2i\alpha}Q_{uv}$. Their neutral products also cancel
the common $m$- or $\Epsilon$-dependent boost phases.
The general quartic potential in this class is
\begin{equation}
\label{endpoint-quartic-complete}
\begin{split}
V_4={}&g_1N_u^2+g_2N_uM_{uv}+g_3M_{uv}^2
+g_4Q_{uv}^\dagger Q_{uv}\\
&+i g_5\left(P_u^\dagger Q_{uv}-Q_{uv}^\dagger P_u\right),
\qquad g_j\in\mathbb R .
\end{split}
\end{equation}
The complementary Hermitian quadrature supplies no additional operator,
since
\[
P_u^\dagger Q_{uv}+Q_{uv}^\dagger P_u=2N_uM_{uv}.
\]
The completeness and independence proof is given in
Appendix~\ref{app:quartic-completeness}: rotations and $U(1)$ leave ten real quartic operators,
and the boost condition reduces this number to five.
The bilinear-generated potential $V(N_u,M_{uv})$ supplies
only the first three channels; neutral products of the pairing bilinears
supply the remaining two pairing channels. Thus the completeness result
identifies deformations that would be
missed by restricting the construction to $N_u$ and $M_{uv}$.
The classification concerns local potentials in a fixed field
representation, before imposing equations of motion or identifying
actions under field redefinitions.

The same result applies to both endpoint signs and to both kinematical
limits, using their respective assignments of $(u,v)$.
The applications below develop both bilinear-generated
potentials and the additional pairing channel $Q_{uv}^\dagger Q_{uv}$. Its
L\'evy--Leblond reduction and first-order nonredundancy will be established
in Section~\ref{sec:interacting-LL}; an exact magnetic Carroll solution
will be constructed in Section~\ref{sec:interacting-Bergshoeff}.
Appendix~\ref{app:interaction-catalog} will summarize interaction
families for the free actions derived in the preceding sections.

\subsection{Color-mixed interactions}
\label{sec:extended-interactions}

For the two-color completion \eqref{L-D12}, the same argument allows
mixed interactions between colors. We continue to consider Hermitian,
derivative-free rotational scalars and take the two colors to carry the
same $m$ or $\Epsilon$. In the central class, the Hermitian real and
imaginary combinations of each off-diagonal bilinear
\eqref{coloredbilinear} are then boost scalars by
\eqref{bilinear-diagonal-boost}.

At the endpoints, retain the notation $u_a$ for the unmixed component
and $v_a$ for its spin-mixed partner. The colored counterparts of
\eqref{endpoint-mixed-bilinear} are
\be\label{Colorendpoint-mixed-bilinear}
N_{u,ab}:=u_a^\dagger u_b,\qquad
M_{uv,ab}:=u_a^\dagger v_b+v_a^\dagger u_b,
\qquad a,b=1,2.
\ee
These bilinears satisfy
$N_{u,ab}^\dagger=N_{u,ba}$ and
$M_{uv,ab}^\dagger=M_{uv,ba}$. Their diagonal boost variations are pure
transport terms, while their spin-mixing variations vanish:
\be\label{Colorendpoint-mixed-bilinear2}
\delta_b^{\rm mix}N_{u,ab}=0,\qquad
\delta_b^{\rm mix}M_{uv,ab}
=\pm u_a^\dagger
\bigl(\mathbf S(b)+\mathbf S(b)^\dagger\bigr)u_b=0,
\ee
where the common endpoint sign is immaterial. Thus a sufficient,
non-exhaustive class of interactions is given by any Hermitian local function
\begin{equation}
\label{colored-endpoint-potential}
V_{\rm col}\!\left(
N_{u,ab},M_{uv,ab}\right),
\qquad
V_{\rm col}^\dagger=V_{\rm col}.
\end{equation}
The unmixed and spin-mixed components in the colored theories are
\[
\begin{array}{c|cc}
 & \beta_\star=-1 & \beta_\star=+1\\ \hline
\text{Galilei} \; (u_a,v_a)
&(\zeta_a,\chi_a)
&(\chi_a,\zeta_a)\\
\text{Carroll} \;(u_a,v_a)
&(\chi_a,\zeta_a)
&(\zeta_a,\chi_a)
\end{array}.
\]
For example, in the $\beta_\star=-1$ Galilei sector,
$N_{\zeta,12}=\zeta_1^\dagger\zeta_2$ is not Hermitian by itself. At
quadratic order in the fermion fields one may use the real color-mixing
density
$V_{\rm col} \sim N_{\zeta,12}+N_{\zeta,12}^\dagger$, whereas a quartic example is
$V_{\rm col} \sim  N_{\zeta,12}^\dagger N_{\zeta,12}$.

\section{Representative actions and interacting deformations}
\label{sec:extensions}
\label{sec:models}
We now combine the free actions derived in Sections~\ref{sec:S1-limits}
and \ref{sec:S2-limits} with the invariant potentials of
Section~\ref{sec:interactions}. The examples develop L\'evy--Leblond
fermions, magnetic Carroll fermions, and a Galilean system with local
fermionic shifts. We follow how each interaction changes the field
equations, the auxiliary-field reduction, and the available exact solutions.
\subsection{Interacting L\'evy--Leblond models}
\label{sec:interacting-LL}

For the Galilean assignment $(u,v)=(\chi,\zeta)$, we first
use a density-channel deformation as a benchmark before developing an
additional pairing channel. The benchmark family is
\begin{equation}
\label{act-LL-interacting}
\mathcal L_{\rm LL}[U]
=i\chi^\dagger\partial_t\chi
+\zeta^\dagger\hat\nabla\chi
-\chi^\dagger\hat\nabla\zeta
+2m\zeta^\dagger\zeta
+U(N_\chi),
\end{equation}
where $N_\chi$ is defined by \eqref{endpoint-mixed-bilinear} with
$u=\chi$. Any real polynomial $U$ is boost compatible; for a single
classical Grassmann Pauli spinor it terminates at $N_\chi^2$, since
$N_\chi^3=0$.
The quartic example is $U(N_\chi)=\frac{g}{2}N_\chi^2$.  The two first-order
field equations are
\begin{equation}
\label{eom-LL-interacting}
i\partial_t\chi-\hat\nabla\zeta+U'(N_\chi)\chi=0,
\qquad
2m\zeta+\hat\nabla\chi=0.
\end{equation}
Consequently, the auxiliary Pauli spinor can still be eliminated exactly,
and the active component obeys
\begin{equation}
\label{nonlinear-Schrodinger-LL}
\left(i\partial_t+\frac{\nabla^2}{2m}\right)\chi
+U'(N_\chi)\chi=0,
\qquad
\zeta=-\frac{1}{2m}\hat\nabla\chi .
\end{equation}
Thus the interacting L\'evy--Leblond equations remain a first-order
factorization of a nonlinear Schr\"odinger system.
The larger class $V(N_\chi,M_{\chi\zeta})$ introduced in
Section~\ref{sec:interactions} generally gives implicit nonlinear
auxiliary-field equations.
For the quartic choice, an exact plane-wave family is
\[
\chi=e^{i\vec k\cdot\vec x
-i\left(\vec k^{\,2}/(2m)-gN_0\right)t}\eta,\qquad
\zeta=-\frac{i\,\vecsigma\cdot\vec k}{2m}\chi,\qquad
N_0=\eta^\dagger\eta .
\]
For generic two-component Grassmann data $\eta$, the cubic quantity
$N_0\eta$ is nonzero, so the interaction contributes genuinely to this
solution.

Quartic self-interactions of Galilei-covariant Fermi fields have been
considered previously, including the channel
$\lambda(\bar\Psi\Psi)^2$ and its Green functions
\cite{deMontigny:2007fqc}; solitary waves in a generalized nonlinear
L\'evy--Leblond system were studied in \cite{Saradzhev:2014qta}.
The density-channel family provides a benchmark for the
additional pairing channel that we now develop.

\paragraph{Exact reduction of a pairing channel.}
Consider the fourth operator in \eqref{endpoint-quartic-complete}, with
its real coupling denoted by $g_Q:=g_4$ and $m\neq0$:
\begin{equation}
\label{ll-pairing-action}
\mathcal L_{\rm LL}^{Q}
=\mathcal L_{\rm LL}[0]+g_Q Q_{\chi\zeta}^\dagger Q_{\chi\zeta},
\qquad Q_{\chi\zeta}=\chi^T i\sigma_2\zeta .
\end{equation}
This interaction is Hermitian and invariant under the same Galilean
boosts. It remains quadratic in the auxiliary spinor
$\zeta$. Writing $\chi=(\chi_\uparrow,\chi_\downarrow)^T$ for its two
spin components, define the even Hermitian matrix
\begin{equation}
\label{ll-pairing-matrix}
\mathsf A_\chi=
\begin{pmatrix}
\chi_\downarrow^\dagger\chi_\downarrow&
-\chi_\downarrow^\dagger\chi_\uparrow\\
-\chi_\uparrow^\dagger\chi_\downarrow&
\chi_\uparrow^\dagger\chi_\uparrow
\end{pmatrix},
\qquad
Q_{\chi\zeta}^\dagger Q_{\chi\zeta}
=\zeta^\dagger\mathsf A_\chi\zeta .
\end{equation}
Ordinary Grassmann multiplication gives
\begin{equation}
\label{ll-pairing-nilpotence}
\mathsf A_\chi^2=-N_\chi\mathsf A_\chi
=-\frac12N_\chi^2\mathbf1_2,
\qquad \mathsf A_\chi^3=0 .
\end{equation}
Varying the action defined by \eqref{ll-pairing-action} with respect to
$\zeta^\dagger$ gives the auxiliary equation below. For $m\neq0$,
\eqref{ll-pairing-nilpotence} makes its coefficient matrix explicitly
invertible, yielding the unique algebraic solution
\begin{equation}
\label{ll-pairing-auxiliary}
\begin{split}
(2m\mathbf1_2+g_Q\mathsf A_\chi)\zeta&=-\hat\nabla\chi,\\
\zeta&=-\frac1{2m}\left(
\mathbf1_2-\frac{g_Q}{2m}\mathsf A_\chi
+\frac{g_Q^2}{4m^2}\mathsf A_\chi^2
\right)\hat\nabla\chi .
\end{split}
\end{equation}
The finite truncation in this inverse, like those used below, follows
from nilpotency in the classical Grassmann algebra and is exact.
In quantum field theory, the corresponding products must instead be
evaluated using the fermionic operator algebra, whose anticommutation
relations can modify these truncations.
Integrating the spatial mixed
term by parts and substituting \eqref{ll-pairing-auxiliary} gives,
up to a spatial divergence,
\begin{equation}
\label{ll-pairing-effective}
\mathcal L_{\rm LL,eff}^{Q}
=i\chi^\dagger\partial_t\chi
-\frac1{2m}(\hat\nabla\chi)^\dagger\hat\nabla\chi
+\frac{g_Q}{4m^2}\mathcal R_\chi^\dagger\mathcal R_\chi+\frac{g_Q^2}{8m^3}N_\chi
\mathcal R_\chi^\dagger\mathcal R_\chi,
\qquad
\mathcal R_\chi:=\chi^T i\sigma_2\hat\nabla\chi .
\end{equation}
Thus a derivative-free quartic in the first-order formulation becomes a
derivative pairing channel with a fixed sextic correction in the
reduced theory. Since $i\sigma_2\sigma_i$ is symmetric,
$\mathcal R_\chi$ is a derivative pairing bilinear whose spin-triplet
structure is contracted with the spatial derivative to form a rotational
scalar. No new time derivatives are introduced by the
auxiliary elimination.

\paragraph{An independent quartic deformation.}
The pairing channel defines an independent first-order deformation under
regular local field redefinitions perturbative about the identity, with
field variations and boundary currents compatible with periodic boundary
conditions. We establish this using a periodic free solution on which the
other four quartic densities vanish while the pairing contribution has a
nonzero integral. At first order, such a redefinition changes the free
action by terms proportional to its Euler--Lagrange equations and a total
derivative. Take $m>0$, $k>0$, and the free solution
\begin{equation}
\label{ll-pairing-test-field}
\begin{split}
\chi&=\begin{pmatrix}f\\0\end{pmatrix},
\qquad
f=e^{-ik^2t/(2m)}
\left(\theta_1e^{ikx}+\theta_2e^{-ikx}\right),\\
\zeta_0&=-\frac{\sigma_1\partial_x\chi}{2m},
\end{split}
\end{equation}
Here $\zeta_0$ denotes the auxiliary field at zero coupling, and
$x=x^1$ is a spatial coordinate. The constants $\theta_1,\theta_2$
and their Hermitian conjugates are independent Grassmann generators.
The fields are periodic with
spatial period $2\pi/k$ and time period $4\pi m/k^2$; the transverse
directions may also be taken periodic with finite volume. On this
solution,
\begin{equation}
\label{ll-pairing-test-invariants}
\begin{split}
N_\chi^2&=M_{\chi\zeta_0}=P_\chi=0,\\
\mathcal R_\chi&=-2ik\theta_1\theta_2e^{-ik^2t/m},\\
Q_{\chi\zeta_0}^\dagger Q_{\chi\zeta_0}
&=\frac{k^2}{m^2}\theta_2^\dagger\theta_1^\dagger
\theta_1\theta_2\neq0 .
\end{split}
\end{equation}
All four other quartic densities in \eqref{endpoint-quartic-complete}
vanish pointwise, whereas the last line of
\eqref{ll-pairing-test-invariants} has a nonzero spacetime integral over
one period. The free equations vanish and the integrated total
derivatives cancel. The pairing channel is therefore linearly independent
of the other four modulo the free equations and total derivatives.

\paragraph{Parity in the L\'evy--Leblond model.}
The free action is invariant under
\begin{equation}
\label{ll-pairing-parity}
\chi(t,\vec x)\longmapsto\chi(t,-\vec x),
\qquad
\zeta(t,\vec x)\longmapsto-\zeta(t,-\vec x).
\end{equation}
Suppressing the reflected arguments, this gives
$N_\chi\mapsto N_\chi$, $M_{\chi\zeta}\mapsto-M_{\chi\zeta}$,
$P_\chi\mapsto P_\chi$, and
$Q_{\chi\zeta}\mapsto-Q_{\chi\zeta}$.
Imposing this additional symmetry sets $g_2=g_5=0$ in
\eqref{endpoint-quartic-complete}, leaving the three-dimensional space
spanned by $N_\chi^2$, $M_{\chi\zeta}^2$, and
$Q_{\chi\zeta}^\dagger Q_{\chi\zeta}$.
The pairing channel therefore survives parity invariance. This
restriction refers to the L\'evy--Leblond parity action
\eqref{ll-pairing-parity}; no discrete symmetry was assumed in the
five-operator classification.

\subsection{Self-interactions of minimal magnetic Carroll fermions}
\label{sec:interacting-Bergshoeff}

We now extend the minimal magnetic Carroll density
\eqref{act-2-bm1-C}, obtained in Section~\ref{sec:S2-limits} from
the fixed-$m$, $\beta=-1$ contraction of the $C_2$ seed.
The chiral map \eqref{D-map} and boost duality
\eqref{boost-C-duality} carry the construction to the opposite endpoint
\eqref{act-2-b1-C}.

Using the invariant bilinears $N_\chi$ and
$M_{\chi\zeta}$ of \eqref{endpoint-mixed-bilinear}, with
$(u,v)=(\chi,\zeta)$, we construct the flat-space interacting family
\begin{equation}
\label{act-Bergshoeff-interacting}
\mathcal L_{\rm MC}[V]
=\mathcal L^{(2,-1)}_{0,m}
+\mu M_{\chi\zeta}
+V_{\rm int}(N_\chi,M_{\chi\zeta}),
\qquad V_{\rm int}=O\!\left((N_\chi,M_{\chi\zeta})^2\right),
\end{equation}
where $\mu\in\mathbb R$ is a mass-deformation
coefficient and $V_{\rm int}$ is a real potential beginning at fourth
order in the fermion fields. The action is invariant under the
$\beta_\star=-1$ Carroll boost in
\eqref{Cboost-CE-beta-components}, with $\Epsilon=0$.
For $V_{\rm int}=0$, this reproduces, up to an overall normalization,
the first minimal magnetic Carroll action of Bergshoeff
\textit{et al.}~\cite{Bergshoeff:2023vfd}, Eq.~(2.24).
The mass and interaction terms are independent boost-invariant
deformations of this free representative.

For example, set $\mu=0$ and take
$V_{\rm int}=U_C(N_\chi)$.  The exact equations and their general local
solution are
\begin{gather}
\partial_t\chi=0,\qquad
\partial_t\zeta-i\hat\nabla\chi-U_C'(N_\chi)\chi=0,
\label{eom-Bergshoeff-interacting}\\
\chi(t,\vec x)=\chi_0(\vec x),\qquad 
\zeta(t,\vec x)=\zeta_0(\vec x)
+t\left[i\hat\nabla\chi_0
+U_C'(N_{\chi_0})\chi_0\right].
\label{solution-Bergshoeff-interacting}
\end{gather}
The choice $U_C=\frac{g_C}{2}N_\chi^2$ therefore gives an explicit
quartic, nonlinear deformation for arbitrary spatial Grassmann-valued data
$\chi_0$ and $\zeta_0$.

A further extension uses the mixed invariant to include sixth- and
eighth-order self-interactions alongside the quartic term. For one pair
of two-component spinors, we consider the polynomial
\begin{equation}
\label{higher-Bergshoeff-potential}
V_{\rm B}(M_{\chi\zeta})
=\mu M_{\chi\zeta}
+\frac{\lambda_4}{2}M_{\chi\zeta}^2
+\frac{\lambda_6}{3}M_{\chi\zeta}^3
+\frac{\lambda_8}{4}M_{\chi\zeta}^4,
\qquad \lambda_4,\lambda_6,\lambda_8\in\mathbb R.
\end{equation}
Higher powers of $M_{\chi\zeta}$ vanish by Grassmann nilpotency.
For spatially homogeneous fields, the equations reduce to
\begin{equation}
\label{homogeneous-Bergshoeff-eom}
\partial_t\chi=-V_{\rm B}'(M_{\chi\zeta})\chi,
\qquad
\partial_t\zeta=V_{\rm B}'(M_{\chi\zeta})\zeta,
\qquad
\partial_tM_{\chi\zeta}=0,
\end{equation}
and hence admit the exact family of solutions
\begin{equation}
\label{homogeneous-Bergshoeff-solution}
\chi(t)=e^{-tV_{\rm B}'(M_0)}\chi_0,
\qquad
\zeta(t)=e^{tV_{\rm B}'(M_0)}\zeta_0,
\qquad
M_0=\chi_0^\dagger\zeta_0+\zeta_0^\dagger\chi_0.
\end{equation}
The nilpotent part of each exponential terminates after finitely many terms,
so \eqref{homogeneous-Bergshoeff-solution} can activate the sixth- and
eighth-order couplings without a perturbative expansion.

To study the dynamics of the pairing channels in
\eqref{endpoint-quartic-complete}, we write the bilinears
\eqref{endpoint-pair-bilinears} in components. With
\[
\chi=\begin{pmatrix}\chi_\uparrow\\\chi_\downarrow\end{pmatrix},
\qquad
\zeta=\begin{pmatrix}\zeta_\uparrow\\\zeta_\downarrow\end{pmatrix},
\]
where the arrows label the Grassmann-odd components with spin
projections along a fixed axis, these bilinears read
\begin{equation}
\label{magnetic-pairing-components}
P_\chi=\chi^T i\sigma_2\chi=2\chi_\uparrow\chi_\downarrow,\qquad
Q_{\chi\zeta}=\chi^T i\sigma_2\zeta
=\chi_\uparrow\zeta_\downarrow-\chi_\downarrow\zeta_\uparrow.
\end{equation}
We set $\mu=0$ and develop the $Q_{\chi\zeta}^\dagger Q_{\chi\zeta}$
channel:
\begin{equation}
\label{magnetic-pairing-action}
\mathcal L_{\rm MC,pair}
=\mathcal L^{(2,-1)}_{0,m}+g_4Q_{\chi\zeta}^\dagger Q_{\chi\zeta},
\qquad g_4\in\mathbb R .
\end{equation}
Varying the daggered Grassmann fields independently gives
\begin{equation}
\label{magnetic-pairing-eom}
\partial_t\chi
=g_4Q_{\chi\zeta}i\sigma_2(\chi^\dagger)^T,
\qquad
\partial_t\zeta-i\hat\nabla\chi
=-g_4Q_{\chi\zeta}i\sigma_2(\zeta^\dagger)^T.
\end{equation}
The conjugate equations follow by ordinary Hermitian conjugation.
For spatially homogeneous initial data
$\chi_0=(a,0)^T$ and $\zeta_0=(0,b)^T$, where
$a,b,a^\dagger,b^\dagger$ generate a four-generator Grassmann algebra,
an exact solution is
\begin{equation}
\label{magnetic-pairing-solution}
\chi(t)=\begin{pmatrix}a\\-g_4t\,a^\dagger a b\end{pmatrix},
\qquad
\zeta(t)=\begin{pmatrix}-g_4t\,b^\dagger a b\\b\end{pmatrix}.
\end{equation}
Indeed, $\chi^\dagger\zeta=\zeta^\dagger\chi=0$ and
$Q_{\chi\zeta}=ab$ throughout this evolution, while
$Q_{\chi\zeta}^\dagger Q_{\chi\zeta}=b^\dagger a^\dagger ab\ne0$.
Substitution into \eqref{magnetic-pairing-eom} verifies the solution
without a perturbative expansion; higher corrections vanish by
Grassmann nilpotency.

This pairing channel thus sources an initially vanishing second
\emph{spinor component} of $\chi$. By contrast, for spatially homogeneous
fields with any potential $V(N_\chi,M_{\chi\zeta})$, the equation is
$\partial_t\chi=-(\partial V/\partial M_{\chi\zeta})\chi$ and preserves a vanishing
second component. This gives a concrete dynamical distinction between
the pairing channel and the previously illustrated potential family
in the fixed field representation. 

Let us observe that the $N_\chi^2$ operator already occurs in the torsion-induced
term of Eq.~(3.26) of Ref.~\cite{Bergshoeff:2023vfd}.
In the present four-dimensional dictionary,
$\Gamma^{123}\Gamma_\star=\operatorname{diag}(\mathbf1_2,-\mathbf1_2)$,
so its contracted quartic bilinear is $6N_\chi^2$.
Here that operator has an independent flat-space coupling within
the larger potential families constructed above.
These interactions provide extensions
of the minimal magnetic Carroll models of Bergshoeff
\textit{et al.}~\cite{Bergshoeff:2023vfd}.
Appendix~\ref{app:Bergshoeff-dictionary} will give the field dictionary,
while Section~\ref{app:Bergshoeff-interacting} of that appendix
will express the resulting interactions directly in the variables of
that reference. This includes the full five-channel quartic potential
and its two pairing channels.

\subsection{A solvable color-mixed Carroll interaction}
\label{sec:colored-Carroll}

For the free action \eqref{act-2-bm1-CE-real}, the invariant bilinears
\eqref{Colorendpoint-mixed-bilinear} with $u_a=\chi_a$ give the
solvable color-mixed potential
\begin{equation}
\label{colored-Bergshoeff-potential}
\mathcal I_\chi:=N_{\chi,12}N_{\chi,21},\qquad
\mathcal U_{\rm col}(\mathcal I_\chi)
=g_4\mathcal I_\chi+\frac{g_8}{2}\mathcal I_\chi^2,
\qquad g_4,g_8\in\mathbb R,
\end{equation}
which is Hermitian and Carroll invariant.
Adding this potential to the
density in \eqref{act-2-bm1-CE-real} gives
\begin{equation}
\label{colored-Bergshoeff-interacting-eom}
\begin{gathered}
\partial_t\chi_1=0,\qquad
(\partial_t-2i\Epsilon)\chi_2=0,\\
(\partial_t-2i\Epsilon)\zeta_1-i\hat\nabla\chi_1
-h(\mathcal I_\chi)N_{\chi,12}\chi_1=0,\\
\partial_t\zeta_2-i\hat\nabla\chi_2
-h(\mathcal I_\chi)\chi_2N_{\chi,21}=0,\qquad
h(s):=g_4+g_8s.
\end{gathered}
\end{equation}
together with their Hermitian conjugates.
These equations can again be integrated exactly. For arbitrary spatial
Grassmann-valued spinors $\alpha_1(\vec x)$ and $\alpha_2(\vec x)$, define
\[
A=\alpha_1^\dagger\alpha_2,\qquad
B=\alpha_2^\dagger\alpha_1,\qquad
h_0=h(AB).
\]
Then a solution with initial data
$\zeta_a(0,\vec x)=\zeta_a^0(\vec x)$ is
\begin{eqnarray}
\chi_1&=&\alpha_1,\qquad
\chi_2=e^{2i\Epsilon t}\alpha_2,\nonumber\\
\zeta_1&=&e^{2i\Epsilon t}\zeta_1^0
-\frac{1-e^{2i\Epsilon t}}{2\Epsilon}\hat\nabla\alpha_1
+t e^{2i\Epsilon t}h_0A\alpha_1,\nonumber\\
\zeta_2&=&\zeta_2^0
-\frac{1-e^{2i\Epsilon t}}{2\Epsilon}\hat\nabla\alpha_2
+t h_0\alpha_2 B.\label{colored-Bergshoeff-interacting-solution}
\end{eqnarray}
The $\Epsilon\to0$ limit is regular.  For generic two-component data,
$(AB)^2$ is nonzero, so the eighth-order coupling $g_8$ contributes both to
the action and to the exact solution. This provides a solvable
color-mixed deformation of the identified magnetic Carroll action.
The solution is a classical Grassmann-valued configuration.

The interacting model constructed above also provides an extension
of the free non-minimal magnetic Carroll theory of
Ref.~\cite{Bergshoeff:2023vfd}.
Appendix~\ref{app:Bergshoeff-interacting} will express this interacting
extension in their field conventions.

\subsection{Galilean fermionic gauge symmetry and quartic deformations}
\label{sec:extended-real-actions}

We now consider a two-color Galilean model with a local fermionic
gauge symmetry. At fixed $m$ and $\beta=-1$, the Hermitian completion gives
the action below
\begin{equation}
\label{act-2-bm1-NR-real}
\mathcal S_{12;\,\infty,m}^{(2,-1)}
=
\frac1{2\tilde\kappa_2}\int dt\,d^3x\,
\Big[
i\zeta_2^\dagger\hat\nabla\zeta_1
+2im\chi_2^\dagger\zeta_1
+i\zeta_1^\dagger\hat\nabla\zeta_2
-2im\zeta_1^\dagger\chi_2
\Big].
\end{equation}
Its corresponding non-hermitian seed is given in equation \eqref{act-2-bm1-NR}.

The $\beta_\star=-1$ representation in
\eqref{Gboost-G-beta-components} acts identically on both colors. The
direct check of Galilei covariance is given in
Appendix~\ref{app:boost-checks}.

The action \eqref{act-2-bm1-NR-real} possesses the fermionic gauge symmetry
\be\label{gaugetransf} 
\delta_\epsilon\zeta_2=2m\,\epsilon,\qquad
\delta_\epsilon\chi_2=\hat\nabla\epsilon ,
\ee
with $\epsilon(t,\vec x)$ an arbitrary Grassmann-odd Pauli spinor parameter, together with the
Hermitian-conjugate transformations. Then
$\delta_\epsilon \mathcal S_{12;\,\infty,m}^{(2,-1)}=0$.
The variation of the first two terms is a total derivative,
\[
\delta_\epsilon\!\left(
i\zeta_2^\dagger\hat\nabla\zeta_1
+2im\chi_2^\dagger\zeta_1
\right)
=
2im\,\epsilon^\dagger\hat\nabla\zeta_1
+2im\,(\partial_i\epsilon^\dagger)\sigma^i\zeta_1
=
2im\,\partial_i\!\left(\epsilon^\dagger\sigma^i\zeta_1\right),
\]
while the variation of the last two terms vanishes:
\[
\delta_\epsilon\!\left(
i\zeta_1^\dagger\hat\nabla\zeta_2
-2im\zeta_1^\dagger\chi_2
\right)
=
i\zeta_1^\dagger\hat\nabla(2m\,\epsilon)
-2im\,\zeta_1^\dagger\hat\nabla\epsilon
=0.
\]
The field equations are
\begin{equation}
\label{eom-2-bm1-NR-real}
\begin{gathered}
2im\,\zeta_1=0,\qquad
\hat\nabla\zeta_1=0,\qquad
\hat\nabla\zeta_2-2m\chi_2=0.
\end{gathered}
\end{equation}
For nonzero $m$ the first equation sets $\zeta_1=0$, so the second
is redundant. Since $\chi_1$ is absent from
\eqref{act-2-bm1-NR-real}, it admits an arbitrary local shift,
\begin{equation}
\label{chi1-null-shift}
\delta_\rho\chi_1=\rho(t,\vec x),
\end{equation}
together with its Hermitian conjugate, for an arbitrary Grassmann-odd
Pauli spinor $\rho$.
For $m\neq0$, choosing $\epsilon=-\zeta_2/(2m)$ gauges away
$\zeta_2$, and the field equations \eqref{eom-2-bm1-NR-real}
set $\chi_2=0$.
Thus, on every fixed-time slice, the active solution space is a single point
modulo \eqref{gaugetransf}; after also quotienting the local shift
\eqref{chi1-null-shift}, all four Pauli fields can be set to zero.
This gauge symmetry emerges because a singular contraction discards
terms present at finite $c$ and can reduce the rank of the field operator, allowing new local
identities. The off-shell variation above establishes the resulting
symmetry directly.

For completeness, the same prescription at $\beta=+1$ gives the
Hermitian action
\begin{equation}
\label{act-2-NRb1-real}
\mathcal S_{12;\,\infty,m}^{(2,1)}
=
\frac1{2\tilde\kappa_2}\int dt\,d^3x\,
\Big[
i\chi_2^\dagger\hat\nabla\chi_1
+2im\chi_2^\dagger\zeta_1
+i\chi_1^\dagger\hat\nabla\chi_2
-2im\zeta_1^\dagger\chi_2
\Big],
\end{equation}
with associated seed listed in equation \eqref{act-2-NRb1}.
It is the image of \eqref{act-2-bm1-NR-real} under the discrete
color-and-chiral map $\mathcal U$ defined in \eqref{dual-U}.
The field equations and local fermionic gauge symmetry of
\eqref{act-2-NRb1-real} follow from those of
\eqref{act-2-bm1-NR-real} through the same map; the two systems are
therefore equivalent.

We next add invariant quartic potentials and study their effect on
the fermionic gauge symmetry, the reduced equations and their solutions.

\subsubsection{The mixed quartic and its algebraic reduction} \label{sec:symbreak}

For the Galilean assignment $(u_a,v_a)=(\zeta_a,\chi_a)$, choose the
following member of \eqref{colored-endpoint-potential}:
\begin{equation}
\label{NJL-12-P}
\frac{\lambda_M}{4}\,
M_{\zeta\chi,21}M_{\zeta\chi,21}^\dagger,
\qquad \lambda_M\in\mathbb R ,
\end{equation}
and add it to the Galilean- and gauge-invariant free action
\eqref{act-2-bm1-NR-real}.  The resulting action is
\begin{equation}
\label{act-2-bm1-NR-real-NJL}
\mathcal S_{12;\,\infty,m}^{(2,-1),\lambda_M}
=
\frac1{2\tilde\kappa_2}\int dt\,d^3x\,
\Big[
i\zeta_2^\dagger\hat\nabla\zeta_1
+2im\chi_2^\dagger\zeta_1
+i\zeta_1^\dagger\hat\nabla\zeta_2
-2im\zeta_1^\dagger\chi_2
+\frac{\lambda_M}{4}\,
M_{\zeta\chi,21}M_{\zeta\chi,21}^\dagger
\Big].
\end{equation}
Its Bargmann covariance follows from
\eqref{Colorendpoint-mixed-bilinear}--\eqref{Colorendpoint-mixed-bilinear2}.

Under the free gauge transformation \eqref{gaugetransf}, the mixed
bilinears vary as
\begin{align}
\delta_\epsilon M_{\zeta\chi,21}
=(\partial_i\epsilon^\dagger)\sigma^i\zeta_1
+2m\,\epsilon^\dagger\chi_1,\qquad
\delta_\epsilon M_{\zeta\chi,12}
=\zeta_1^\dagger\hat\nabla\epsilon
+2m\,\chi_1^\dagger\epsilon .
\label{M-gauge-variation}
\end{align}
While the free sector is gauge invariant, the interaction term varies as
\(
\delta_\epsilon
\bigl(M_{\zeta\chi,21}M_{\zeta\chi,12}\bigr)
=\bigl(\delta_\epsilon M_{\zeta\chi,21}\bigr)M_{\zeta\chi,12}
+M_{\zeta\chi,21}\bigl(\delta_\epsilon M_{\zeta\chi,12}\bigr)
\)
which is not a total derivative. Thus the original linear gauge
symmetry of the free action is explicitly broken for $\lambda_M\neq0$.
The interaction also breaks the spectator shift \eqref{chi1-null-shift},
since
$\delta_\rho M_{\zeta\chi,21}=\zeta_2^\dagger\rho$ and
$\delta_\rho M_{\zeta\chi,12}=\rho^\dagger\zeta_2$.
Consequently, neither of the local transformations that emptied the free
spatial field content can be used to identify configurations of the
interacting theory.

This breaks the former fermionic shifts, but not all local symmetries.
For $m\neq0$, every potential made from $N_{\zeta,ab}$ and
$M_{\zeta\chi,ab}$ preserves the following bosonic local $U(1)$:
\begin{equation}
\label{Bargmann-residual-phase}
\zeta_a\longmapsto e^{i\alpha(t,\vec x)}\zeta_a,\qquad
\chi_a\longmapsto e^{i\alpha(t,\vec x)}
\left[\chi_a+\frac{i}{2m}\sigma^i(\partial_i\alpha)\zeta_a\right],
\qquad \alpha\in\mathbb R .
\end{equation}
Indeed, the combination
\begin{equation}
\label{Bargmann-free-gauge-invariant}
\mathcal F_2:=\hat\nabla\zeta_2-2m\chi_2
\end{equation}
transforms by the same phase, and the free density is
$i\zeta_1^\dagger\mathcal F_2-i\mathcal F_2^\dagger\zeta_1$
up to a spatial divergence. All $N_{\zeta,ab}$ are invariant.
In each $M_{\zeta\chi,ab}$, the two derivative-of-$\alpha$ terms cancel.
Thus \eqref{Bargmann-residual-phase} is an off-shell symmetry of the
complete deformed action.

Varying the daggered fields gives
\begin{equation}
\label{eom-2-bm1-NR-real-NJL}
\begin{aligned}
\left(2im+\frac{\lambda_M}{4}M_{\zeta\chi,12}\right)\zeta_1&=0,
&\qquad
i\hat\nabla\zeta_1
+\frac{\lambda_M}{4}M_{\zeta\chi,12}\chi_1&=0,\\
i\hat\nabla\zeta_2
-\left(2im-\frac{\lambda_M}{4}M_{\zeta\chi,21}\right)\chi_2&=0,
&
\frac{\lambda_M}{4}M_{\zeta\chi,21}\zeta_2&=0 .
\end{aligned}
\end{equation}
For $m\neq0$, the first coefficient is invertible in the classical Grassmann
algebra, so the first equation implies $\zeta_1=0$. On this
constraint surface $M_{\zeta\chi,12}=\chi_1^\dagger\zeta_2$.
Substituting into
\eqref{eom-2-bm1-NR-real-NJL} gives
\begin{equation}
\label{eom-2-bm1-NR-real-NJL-zeta1-zero}
\frac{\lambda_M}{4}
(\chi_1^\dagger\zeta_2)\chi_1=0,\qquad
i\hat\nabla\zeta_2
-\left(2im-\frac{\lambda_M}{4}\zeta_2^\dagger\chi_1\right)\chi_2=0,
\qquad
\frac{\lambda_M}{4}
(\zeta_2^\dagger\chi_1)\zeta_2=0.
\end{equation}
Together with their Hermitian conjugates, these are the residual
equations. They still
contain nilpotent zero divisors.
The auxiliary pair can be eliminated exactly:
$\zeta_1=0$ and
$\chi_2=(2im-\frac{\lambda_M}{4}\zeta_2^\dagger\chi_1)^{-1}
i\hat\nabla\zeta_2$, together with their conjugates.

The residual equations admit a simple branch with $\chi_1=\zeta_1=0$,
an arbitrary Grassmann-odd profile $\zeta_2(t,\vec x)$, and
$\chi_2=\hat\nabla\zeta_2/(2m)$. In the free theory, this profile
could be removed by the local fermionic shift \eqref{gaugetransf}.
The mixed quartic breaks that symmetry, changing the gauge structure
that made the free model empty.

\subsubsection{A quartic with spatial differential equations}
\label{sec:spatial-quartic}

The mixed interaction of Section~\ref{sec:symbreak} leaves
algebraic constraints on the remaining fields after auxiliary-field
elimination. To obtain a reduced equation that governs the spatial
profile of $\zeta_2$, we now consider the diagonal potential
\begin{equation}
\label{Bargmann-spatial-potential}
V_s=\frac{g_s}{2}M_{\zeta\chi,22}^2,\qquad g_s\in\mathbb R,
\end{equation}
added to the same free density \eqref{act-2-bm1-NR-real}.
Varying $\zeta_1^\dagger$, $\chi_2^\dagger$ and
$\zeta_2^\dagger$ gives
\begin{equation}
\label{Bargmann-spatial-full-eom}
\begin{aligned}
i\hat\nabla\zeta_2-2im\chi_2&=0,\\
2im\zeta_1+g_sM_{\zeta\chi,22}\zeta_2&=0,\\
i\hat\nabla\zeta_1+g_sM_{\zeta\chi,22}\chi_2&=0,
\end{aligned}
\end{equation}
together with the Hermitian-conjugate equations. For $m\neq0$, the
first two determine the auxiliary fields and give
\begin{equation}
\label{Bargmann-spatial-auxiliary}
\chi_2=\frac{\hat\nabla\zeta_2}{2m},\qquad
\zeta_1=\frac{i g_s}{2m}M_{\zeta\chi,22}\zeta_2,\qquad
M_{\zeta\chi,22}
=\frac1{2m}\partial_i(\zeta_2^\dagger\sigma^i\zeta_2).
\end{equation}
Substitution into the third equation of
\eqref{Bargmann-spatial-full-eom} yields
\[
\begin{aligned}
0&=-\frac{g_s}{2m}\hat\nabla
\bigl(M_{\zeta\chi,22}\zeta_2\bigr)
+\frac{g_s}{2m}M_{\zeta\chi,22}\hat\nabla\zeta_2\\
&=-\frac{g_s}{2m}
\bigl(\partial_iM_{\zeta\chi,22}\bigr)\sigma^i\zeta_2.
\end{aligned}
\]
For $g_s\neq0$, inserting the expression for $M_{\zeta\chi,22}$ in
\eqref{Bargmann-spatial-auxiliary} therefore gives directly
\begin{equation}
\label{Bargmann-spatial-eom}
\partial_i\!\left[\partial_j
\bigl(\zeta_2^\dagger\sigma^j\zeta_2\bigr)\right]\,
\sigma^i\zeta_2=0,
\end{equation}
together with its Hermitian conjugate. Both spatial derivatives act
on the bilinear inside the brackets.
Substituting the auxiliary-field solutions into the action and
discarding a spatial divergence, we obtain the effective Lagrangian
density, with the overall normalization of \eqref{act-2-bm1-NR-real} omitted:
\begin{equation}
\label{Bargmann-spatial-reduced}
\mathcal L_{s,\mathrm{red}}
=\frac{g_s}{8m^2}
\left[\partial_i(\zeta_2^\dagger\sigma^i\zeta_2)\right]^2.
\end{equation}
The local phase symmetry
\eqref{Bargmann-residual-phase} remains, and acts on the reduced field as
$\zeta_2\mapsto e^{i\alpha(t,\vec x)}\zeta_2$. The bilinear
$\zeta_2^\dagger\sigma^i\zeta_2$ and its divergence are invariant.

A local solution with nonzero quartic density illustrates this
spatially differential theory. Choose
constant orthonormal Pauli spinors $u_\pm$ with
$\sigma^1u_\pm=\pm u_\pm$, and independent Grassmann generators
$\theta_\pm$ and their conjugates. On an interval where both radicands
are positive, let
\begin{equation}
\label{Bargmann-spatial-solution}
\zeta_2=\theta_+\sqrt{a_+x^1+b_+}\,u_+
+\theta_-\sqrt{a_-x^1+b_-}\,u_-,
\qquad a_\pm,b_\pm\in\mathbb R .
\end{equation}
Its invariant divergence is constant,
\[
\partial_i(\zeta_2^\dagger\sigma^i\zeta_2)
=a_+\theta_+^\dagger\theta_+
-a_-\theta_-^\dagger\theta_- .
\]
The reduced equation therefore holds, and \eqref{Bargmann-spatial-auxiliary}
reconstructs the other fields; one may set the spectator $\chi_1=0$.
For $a_+a_-\neq0$, the square of this divergence is
$-2a_+a_-\,\theta_+^\dagger\theta_+\theta_-^\dagger\theta_-\neq0$.

For comparison, in the mixed channel \eqref{NJL-12-P},
the first and third equations in
\eqref{eom-2-bm1-NR-real-NJL-zeta1-zero} impose algebraic constraints
on $\chi_1$ and $\zeta_2$, while the second reconstructs the auxiliary
field $\chi_2$ from these fields and $\hat\nabla\zeta_2$. In particular,
the branch $\chi_1=0$ admits an arbitrary profile $\zeta_2(t,\vec x)$.
For the diagonal channel \eqref{Bargmann-spatial-potential}, the reduced equation
\eqref{Bargmann-spatial-eom} instead imposes a spatial differential
condition on $\zeta_2$, satisfied by the exact local solution
\eqref{Bargmann-spatial-solution} with nonzero quartic density.
Both deformations preserve the bosonic local phase symmetry
\eqref{Bargmann-residual-phase}.

\section{Relativistic NJL channels and their contractions}
\label{sec:NJL}
We now use relativistic NJL interactions to identify
which channels in the intrinsic families of
Section~\ref{sec:interactions} arise through contraction. The $c$-power
expansions select the leading operators in each limit. We then
establish the joint normalization of the free
and interacting actions for the colored Bargmann model in
Section~\ref{sec:NJL-joint-Bargmann}.
\subsection{Single-color NJL interaction}
For either Dirac-conjugation branch, consider the relativistic
NJL interaction
\begin{equation}
\label{NJL-rel}
\mathcal{L}_{\rm NJL}^{[r]}
=
-(-1)^r\frac{\lambda}{4}
\left[
\left(\bar\psi^{[r]}\psi\right)^2
-\left(\bar\psi^{[r]}\gamma^5\psi\right)^2
\right].
\end{equation}

The field redefinition \eqref{S} expresses the scalar
and pseudoscalar bilinears in \eqref{NJL-rel} as
\begin{equation}
\label{NJL-scalar-bilinear}
\bar\psi^{[1]}\psi=c^{-2\beta}N_\zeta-N_\chi \,,\qquad 
\bar\psi^{[1]}\gamma^5\psi
=ic^{-\beta}\bigl(\chi^\dagger\zeta+\zeta^\dagger\chi\bigr).
\end{equation}

The interaction therefore becomes
\begin{equation}
\label{NJL-expanded}
\mathcal{L}_{\rm NJL}^{[1]}
=
\frac{\lambda}{4}
\left[
c^0N_\chi^2
+c^{-2\beta}\left\{
\bigl(\chi^\dagger\zeta+\zeta^\dagger\chi\bigr)^2
-2N_\chi N_\zeta
\right\}
+c^{-4\beta}N_\zeta^2
\right].
\end{equation}
The three channels carry powers $c^0$, $c^{-2\beta}$ and
$c^{-4\beta}$. The first two also multiply the time-derivative blocks
of the free action; its fixed-$m$ and fixed-$\Epsilon$ decompositions
were obtained in Section~\ref{sec:S1-limits},
Eqs.~\eqref{act-1-ABDG} and \eqref{act-1-ABDGE}, respectively.
Which NJL channel leads depends on the sign of $\beta$ and the limit.

Since $C_2=C_1\gamma^5$ and $(\gamma^5)^2=1$, the second-branch
bilinears satisfy $\bar\psi^{[2]}\psi=\bar\psi^{[1]}\gamma^5\psi$
and $\bar\psi^{[2]}\gamma^5\psi=\bar\psi^{[1]}\psi$.
The compensating sign in \eqref{NJL-rel} therefore gives
\begin{align}
\mathcal L_{\rm NJL}^{[2]}
=\mathcal L_{\rm NJL}^{[1]} .
\label{NJL-expanded-C2}
\end{align}
For real $\lambda$, the common interaction is Hermitian, although
adding it does not complete the non-Hermitian free $C_2$ seed.
Table~\ref{tab:NJL-limits} lists its leading channels after factoring
out their overall powers of $c$. Each belongs to the invariant classes
of Section~\ref{sec:interactions}.

\begin{table}[ht]
\centering
\begingroup
\small
\setlength{\tabcolsep}{2.5pt}
\renewcommand{\arraystretch}{1.15}
\begin{adjustbox}{max width=\textwidth}
\begin{tabular}{|c|>{\scriptsize}c|c||c|}
\hline
Limit & $\beta$ & Leading NJL channel & Boost\\
\hline\hline
 & $-1\leq\beta<0$ & $\dfrac{\lambda}{4}N_\zeta^2$
 & $K_{i,m/0}^{G,(-1)},\ K_{i,m/0}^{G,(0)}$\\[6pt]
\cline{2-4}
$c\to\infty$ & $\beta=0$ &
$\dfrac{\lambda}{4}
\left[
N_\chi^2+\bigl(\chi^\dagger\zeta+\zeta^\dagger\chi\bigr)^2
-2N_\chi N_\zeta+N_\zeta^2
\right]$
 & $K_{i,m/0}^{G,(0)}$\\[6pt]
\cline{2-4}
 & $0<\beta\leq1$ & $\dfrac{\lambda}{4}N_\chi^2$
 & $K_{i,m/0}^{G,(0)},\ K_{i,m/0}^{G,(1)}$\\[6pt]
\hline
 & $-1\leq\beta<0$ & $\dfrac{\lambda}{4}N_\chi^2$
 & $K_{i,\Epsilon/0}^{C,(-1)},\ K_{i,\Epsilon/0}^{C,(0)}$\\[6pt]
\cline{2-4}
$c\to0$ & $\beta=0$ &
$\dfrac{\lambda}{4}
\left[
N_\chi^2+\bigl(\chi^\dagger\zeta+\zeta^\dagger\chi\bigr)^2
-2N_\chi N_\zeta+N_\zeta^2
\right]$
 & $K_{i,\Epsilon/0}^{C,(0)}$\\[6pt]
\cline{2-4}
 & $0<\beta\leq1$ & $\dfrac{\lambda}{4}N_\zeta^2$
 & $K_{i,\Epsilon/0}^{C,(0)},\ K_{i,\Epsilon/0}^{C,(1)}$\\[6pt]
\hline
\end{tabular}
\end{adjustbox}
\endgroup
\caption{Leading NJL channels, with overall powers of $c$ factored
out, and their boost generators. Both conjugation branches give the same
result.  In the Galilei entries, the subscript $m/0$ denotes the fixed-$m$ boost or its $m=0$ specialization for the fixed-$\Epsilon$ scaling.
In the Carroll entries, $\Epsilon/0$ denotes the fixed-$\Epsilon$ boost or its
$\Epsilon=0$ fixed-$m$ specialization.  The boost families follow
\eqref{beta-star}, \eqref{boost-CE-beta} and  \eqref{boost-G-beta}.}
\label{tab:NJL-limits}
\end{table}
At the endpoints, these results take the universal form
\begin{equation}
\label{NJL-endpoint-single}
\mathcal L_{{\rm NJL},{\rm end}}
=\frac{\lambda}{4}N_u^2,
\end{equation}
where $u$ is the component without spin mixing and the leading power
of $c$ has been factored out. For the minimal magnetic Carroll model
of Section~\ref{sec:interacting-Bergshoeff}, this selects the channel
$\frac{\lambda}{4}N_\chi^2$ for either conjugation branch.

Although $\bigl(\chi^\dagger\zeta+\zeta^\dagger\chi\bigr)^2$ is itself
compatible with the endpoint spin mixing, the full middle combination
$\bigl(\chi^\dagger\zeta+\zeta^\dagger\chi\bigr)^2
-2N_\chi N_\zeta$ is not endpoint invariant in general because the
$N_\chi N_\zeta$ term contains the component that transforms with spin
mixing. The leading endpoint limits thus select the density-square
interactions of Section~\ref{sec:interactions}, rather than the full
middle channel.

\subsection{Colored NJL channels}
\label{sec:NJL-Bargmann}

The colored Dirac action \eqref{L-D12} can be complemented by the Hermitian
color-mixed NJL interaction
\begin{align}
\label{NJL-12-full}
\mathcal L_{{\rm NJL},12}^{[r]}
&=-(-1)^r\frac{\lambda}{4}
\Big[
\bigl(\bar\psi_2^{[r]}\psi_1\bigr)
\bigl(\bar\psi_1^{[r]}\psi_2\bigr)
-\bigl(\bar\psi_2^{[r]}\gamma^5\psi_1\bigr)
\bigl(\bar\psi_1^{[r]}\gamma^5\psi_2\bigr)
\Big]\nonumber\\
&=\frac{\lambda}{4}\Big[
\bigl(\chi_2^\dagger\chi_1
-c^{-2\beta}\zeta_2^\dagger\zeta_1\bigr)
\bigl(\chi_1^\dagger\chi_2
-c^{-2\beta}\zeta_1^\dagger\zeta_2\bigr)
\nonumber\\
&\hspace{33mm}
+c^{-2\beta}
\bigl(\chi_2^\dagger\zeta_1+\zeta_2^\dagger\chi_1\bigr)
\bigl(\chi_1^\dagger\zeta_2+\zeta_1^\dagger\chi_2\bigr)
\Big].
\end{align}
Both conjugation branches give the same component expression.
Its three powers of $c$ are most easily read using
\[
X:=\chi_2^\dagger\chi_1,\qquad
Z:=\zeta_2^\dagger\zeta_1,\qquad
\mathcal M:=\chi_2^\dagger\zeta_1+\zeta_2^\dagger\chi_1.
\]
The expansion is then
\begin{equation}
\label{NJL-12-decomposition}
\mathcal L_{{\rm NJL},12}^{[r]}
=\frac{\lambda}{4}\Big[
XX^\dagger
+c^{-2\beta}\bigl(\mathcal M\mathcal M^\dagger
-XZ^\dagger-ZX^\dagger\bigr)
+c^{-4\beta}ZZ^\dagger\Big].
\end{equation}
As in the single-color case, the relative powers of $c$ identify the
leading channels. Below we factor out the common leading power and
retain $\lambda$ as the coefficient of the resulting operator.
At $\beta=0$ the common leading interaction is
\begin{equation}
\label{NJL-colored-central}
\mathcal L_{{\rm NJL},\beta=0}
=\frac{\lambda}{4}\Big[
(X-Z)(X-Z)^\dagger+\mathcal M\mathcal M^\dagger\Big].
\end{equation}
This is common to the Galilean and Carrollian limits: at $\beta=0$,
both colors transform diagonally and all three bilinears are invariant
under the internal part of the boost.
At the four endpoints, the leading channels are
\begin{align}
\mathcal L_{{\rm NJL},{\rm G},\beta=-1}
=\mathcal L_{{\rm NJL},{\rm C},\beta=+1}
&=\frac{\lambda}{4}\,
N_{\zeta,12}^\dagger N_{\zeta,12}
=\frac{\lambda}{4}
\bigl(\zeta_2^\dagger\zeta_1\bigr)
\bigl(\zeta_1^\dagger\zeta_2\bigr),
\label{NJL-G-leading}\\
\mathcal L_{{\rm NJL},{\rm G},\beta=+1}
=\mathcal L_{{\rm NJL},{\rm C},\beta=-1}
&=\frac{\lambda}{4}\,
N_{\chi,12}^\dagger N_{\chi,12}
=\frac{\lambda}{4}
\bigl(\chi_2^\dagger\chi_1\bigr)
\bigl(\chi_1^\dagger\chi_2\bigr).
\end{align}
Thus every endpoint selects the Hermitian product
$N_{u,12}^\dagger N_{u,12}$ from \eqref{colored-endpoint-potential}.
We now fix the normalization of the Bargmann channel jointly with
the free action by contracting the full interacting Hermitian parent.

\subsubsection{A jointly scaled Bargmann action}
\label{sec:NJL-joint-Bargmann}

To obtain a finite interacting Bargmann action, the inherited quartic
must survive at the same order as the free density. This fixes the
coupling scaling relative to the action normalization. Start from
\begin{equation}
\label{NJL-joint-parent}
\mathcal S_{12,\mathrm{NJL}}^{[2]}(c)
=\frac1{2\kappa_2(c)}\int d^4x\,
\left[\mathcal L_{12}^{[2]}
+\mathcal L_{\mathrm{NJL},12}^{[2]}\bigl(\lambda(c)\bigr)\right].
\end{equation}
The free density is that of \eqref{L-D12}, including its Hermitian
conjugate; the NJL density \eqref{NJL-12-full} is already Hermitian
for real $\lambda(c)$ and is added only once. Apply \eqref{S12} at
$\beta=-1$ and fixed nonzero $m$, and choose
\begin{equation}
\label{NJL-joint-scaling}
\kappa_2(c)=c^3\tilde\kappa_2,
\qquad \lambda(c)=c^{-2}\lambda_{\rm G},
\qquad \lambda_{\rm G}\in\mathbb R .
\end{equation}
The first scaling is the one used for the free Bargmann action
\eqref{act-2-bm1-NR-real}. Let $\mathcal L_0$ denote its bracketed
free density,
and set
\[
\mathcal L_{\rm sub}:=
\zeta_2^\dagger\partial_t\chi_1
-\chi_2^\dagger\partial_t\zeta_1
+i\chi_2^\dagger\hat\nabla\chi_1+\mathrm{h.c.}
\]
Using $d^4x=c\,dt\,d^3x$, the transformed free density contributes
$c\mathcal L_{12}^{[2]}=c^3\mathcal L_0+c\mathcal L_{\rm sub}$,
up to boundary terms. Equations~\eqref{NJL-12-decomposition} and
\eqref{NJL-joint-scaling} then give the exact finite-$c$ expression
\begin{equation}
\label{NJL-joint-expansion}
\begin{split}
\mathcal S_{12,\mathrm{NJL}}^{[2]}(c)
=\frac1{2\tilde\kappa_2}\int dt\,d^3x\,\Bigl[&
\mathcal L_0+c^{-2}\mathcal L_{\rm sub}
+\frac{\lambda_{\rm G}}4\bigl\{
ZZ^\dagger+c^{-4}XX^\dagger\\
&+c^{-2}\bigl(\mathcal M\mathcal M^\dagger
-XZ^\dagger-ZX^\dagger\bigr)\bigr\}\Bigr].
\end{split}
\end{equation}
Thus the free and interacting terms survive together as $c\to\infty$:
\begin{equation}
\label{NJL-joint-limit}
\mathcal S_{\rm G,NJL}
=\frac1{2\tilde\kappa_2}\int dt\,d^3x\,
\left[\mathcal L_0+\frac{\lambda_{\rm G}}4
(\zeta_2^\dagger\zeta_1)(\zeta_1^\dagger\zeta_2)\right].
\end{equation}
The extra factor $c$ in the integration measure is essential for
comparing the coupling with the free normalization. The full middle
combination, including both crossed terms, is suppressed by $c^{-2}$;
the strict limit selects $ZZ^\dagger$. The intrinsic construction of
Section~\ref{sec:symbreak} instead allows
$\mathcal M\mathcal M^\dagger$ with its own independent coupling.
For comparison with the mixed quartic of Section~\ref{sec:symbreak},
the NJL-induced interaction in \eqref{NJL-joint-limit} preserves a
deformed fermionic shift. Its variation is cancelled by the
field-dependent correction
\[
\delta_\epsilon\chi_2=\hat\nabla\epsilon
-\frac{i\lambda_{\rm G}}{4}(\zeta_2^\dagger\zeta_1)\epsilon,\qquad
\delta_\epsilon\chi_2^\dagger
=(\hat\nabla\epsilon)^\dagger
+\frac{i\lambda_{\rm G}}{4}\epsilon^\dagger(\zeta_1^\dagger\zeta_2),
\]
while $\delta_\epsilon\zeta_2=2m\epsilon$ is unchanged, as is the
spectator shift \eqref{chi1-null-shift}. The NJL channel thus
deforms the fermionic gauge symmetry, whereas the intrinsic mixed
quartic breaks its original linear realization.

\section{Conclusions and outlook}
\label{sec:conclusions}

We have organized Galilean and Carrollian fermion theories within a
common contraction scheme and constructed their local interactions.
The limits of the starting relativistic models recover the L\'evy--Leblond theory
\cite{Levy-Leblond:1967eic,Hagen:1972pd,Hagen:1970wn,Hagen:1971dk},
its massless descendants \cite{Banerjee:2022uqj,Sharma:2023chs},
the magnetic Galilei representative \cite{Koutrolikos:2023evq},
and known Carrollian models
\cite{Stakenborg:2023bmw,Bagchi:2025vri,Bagchi:2022eui,Ekiz:2025hdn,
Koutrolikos:2023evq,Bergshoeff:2023vfd}.

We classified the
five-parameter space of derivative-free Hermitian fermion quartic interactions for
one pair of Grassmann-odd Pauli spinors, subject to rotations, a common
$U(1)$ and an endpoint boost. The basis is given in
\eqref{endpoint-quartic-complete}, and its completeness proof is collected in
Appendix~\ref{app:quartic-completeness}. Invariant bilinears also generate
families of higher-order and color-mixed interactions.

An important outcome is the appearance of spin-singlet fermion-pairing
channels in the complete single-color quartic basis. These enlarge the
interaction space beyond products of invariant daggered--undaggered
bilinears and lead to distinct nonlinear dynamics. The channel developed
here remains independent at first order under perturbative local field
redefinitions. Its exact reductions and classical solutions illustrate
how pairing interactions can generate derivative couplings and activate
spinor components that remain absent for other potential families.
More generally, the examples show how the choice of interaction shapes
the reduced equations and the symmetries that survive the contraction.

The nonlinear L\'evy--Leblond reduction connects with earlier
Galilean self-interactions and solitary-wave studies
\cite{deMontigny:2007fqc,Saradzhev:2014qta}.
For the additional pairing channel
$Q_{\chi\zeta}^\dagger Q_{\chi\zeta}$, the derivative quartic is accompanied
by a fixed sixth-order correction, and the channel survives the parity
restriction of the L\'evy--Leblond model. Its first-order independence
is established using an explicit periodic free solution.
Magnetic Carroll potentials
admit exact classical Grassmann-valued solutions with quartic,
sixth-order and eighth-order vertices, and the colored theory admits
a solvable mixed deformation. The familiar $N_\chi^2$ channel has a
torsion-induced antecedent in Ref.~\cite{Bergshoeff:2023vfd};
here it is included with an independent flat-space coupling in a
broader operator construction. Explicit field dictionaries connect our
variables to the minimal and non-minimal magnetic Carroll theories of
Ref.~\cite{Bergshoeff:2023vfd} and give their interacting extensions.

We also found a two-color Galilean model \eqref{act-2-bm1-NR-real}
 with a gauge symmetry parametrized by a fermionic spinor. This symmetry makes the free model gauge-trivial. Only suitable interaction channels break this redundancy and allow nontrivial spinor configurations to survive. The NJL-derived interaction channel, for example, preserves the gauge freedom through a deformation of its realization, and the system remains gauge-trivial.

At the interacting level, extending the free-field dictionaries requires
tracking any time dependence or spatial derivatives induced in the
potentials.
Further work should determine independent deformations after auxiliary
elimination, extend the operator basis to derivatives and multiple
colors, and investigate quantum consistency, correlation functions and
vacuum structure. These questions may be relevant to Carroll holography
\cite{Duval:2014uva,Alday:2024yyj,Donnay:2022aba,
Ruzziconi:2026bix,Mason:2023mti}.

The pairing channels also suggest a connection with collective phenomena
in fermionic systems. Fermion pairing lies at the heart of conventional
superconductivity \cite{Bardeen:1957mv} and superfluidity in ultracold
Fermi gases \cite{Zwierlein:2005jfw}, where interactions organize the
constituents into coherent many-body phases. Their non-Lorentzian
counterparts raise a natural question: how does this mechanism change
when the dynamics is constrained by Galilean or Carrollian symmetry?
The Carrollian case is particularly intriguing because its free
frequencies are independent of spatial momentum, changing the usual
setting in which pairing is studied. The classical theories constructed
here provide concrete starting points for investigating pair
correlations, possible paired phases, and the role of interactions in
this unusual kinematic regime.

The derivative spinor shift symmetry is structurally reminiscent
of the gravitino transformation \cite{Freedman:1976xh,Deser:1976eh},
motivating the search for a supersymmetric completion. The construction may also
be extended to higher-spin fields
\cite{Hagen:1970wn,Hagen:1971dk,Hagen:1972pd,Campoleoni:2026wja}
or to the distinct Hermitian chiral-mass parent noted in
Section~\ref{sec:limits}.

\section*{Acknowledgements}

This work is supported by ANID FONDECYT grants N$^{\circ}$1252053,  N$^{\circ}$1250672, N$^{\circ}$1262197 and  N$^{\circ}$3240060.

\appendix

\section{Conventions and representations}
\label{app:conventions}

\subsection{Spinors and Poincar\'e generators}\label{app:Poincare}

We set $\hbar=1$, retain $c$ through $x^0=ct$ and
$\partial_0=c^{-1}\partial_t$, and use the mostly-plus metric
$\eta_{\mu\nu}=\mathrm{diag}(-1,+1,+1,+1)$. Our gamma matrices are
\begin{equation}\label{gamma}
\gamma_0=\begin{pmatrix}\,-i\,\mathbf{1}_2&0\\[2pt]0&i\,\mathbf{1}_2\end{pmatrix},
\quad
\gamma_i=\begin{pmatrix}0&\sigma_i\\[2pt]\sigma_i&0\end{pmatrix},
\quad
\gamma_5=\begin{pmatrix}0&i\mathbf{1}_2\\[2pt]-i\mathbf{1}_2&0\end{pmatrix},
\qquad
\{\gamma^\mu,\gamma^\nu\}=2\eta^{\mu\nu}\mathbf{1}_4.
\end{equation}
Here $\sigma^i$ are Pauli matrices and
$\gamma_5=i\gamma_0\gamma_1\gamma_2\gamma_3$, with $\gamma_5^2=1$.

The Poincar\'e generators were defined in \eqref{Poincare-generators}.
We use differential generators without the quantum-mechanical
factors of $i$. Translations and the orbital generators are formally
anti-Hermitian with the appropriate integration measure, but this does
not make the finite-dimensional spinor boost matrices anti-Hermitian
under ordinary dagger.

The translation and orbital generators satisfy
\[
[T_\mu,T_\nu]=0,\quad 
[M_{\mu\nu},T_\rho]=\eta_{\nu\rho}T_\mu-\eta_{\mu\rho}T_\nu,\quad
[M_{\mu\nu},M_{\rho\sigma}]
=\eta_{\nu\rho}M_{\mu\sigma}-\eta_{\mu\rho}M_{\nu\sigma}
 -\eta_{\nu\sigma}M_{\mu\rho}+\eta_{\mu\sigma}M_{\nu\rho}.
\]

Writing $K_i:=J_{0i}$ for the boosts, the full $3{+}1$ commutators are
\[
[J_{ij},J_{kl}]=\delta_{jk}J_{il}-\delta_{ik}J_{jl}-\delta_{jl}J_{ik}+\delta_{il}J_{jk},\quad
[J_{ij},K_k]=\delta_{jk}K_i-\delta_{ik}K_j,\quad [K_i,K_j]=J_{ij},
\]
\[
[J_{ij},T_k]=\delta_{jk}T_i-\delta_{ik}T_j,\quad [K_i,T_0]=T_i,\quad [K_i,T_j]=\delta_{ij}T_0.
\]

\subsection{Galilei, Bargmann, and Carroll algebras}

We retain the differential-generator convention for the contracted
algebras. Let $T_t$ and $T_i$ denote the abstract temporal and spatial
translations, and $B_i$ a generic boost. Their common spatial sector is
\[
[J_{ij},J_{kl}]
=\delta_{jk}J_{il}-\delta_{ik}J_{jl}
 -\delta_{jl}J_{ik}+\delta_{il}J_{jk},
\qquad
[J_{ij},T_k]=\delta_{jk}T_i-\delta_{ik}T_j,
\]
\[
[J_{ij},B_k]=\delta_{jk}B_i-\delta_{ik}B_j,\qquad
[T_i,T_j]=[T_t,T_i]=0.
\]
For the Galilei algebra one has, in addition,
\[
[B_i,T_t]=T_i,\qquad [B_i,T_j]=0,\qquad [B_i,B_j]=0.
\]
The Bargmann algebra is the central extension of the Galilei algebra by a
central generator $M$,
\[
[B_i,T_t]=T_i,\qquad [B_i,T_j]=\delta_{ij}M,\qquad [B_i,B_j]=0,
\qquad [M,\cdot]=0.
\]
In our differential-generator conventions, the massive Galilei limits
represent the central generator by $M=-im\,\mathbf1$.
The corresponding Hermitian mass operator is therefore
$iM=m\,\mathbf1$ and has eigenvalue $m$.

For the Carroll algebra the boost commutators are instead
\[
[B_i,T_t]=0,\qquad [B_i,T_j]=\delta_{ij}T_t,\qquad [B_i,B_j]=0.
\]
The contractions exchange the translation on the right-hand side
of the boost commutator: $T_i$ for Galilei, $T_t$ for Carroll.
The degenerate boost scalings excluded in Section~\ref{sec:limits}
lose one or both of these defining relations.

For the Carroll models considered here, these abstract generators have
the differential realizations
\[
T_i\ \mapsto\ \partial_i,\qquad
T_t\ \mapsto
\begin{cases}
\partial_t,&\text{fixed }m,\\[2pt]
D_t^{(\Epsilon)},&\text{fixed }\Epsilon.
\end{cases}
\]
Here $D_t^{(\Epsilon)}$ is defined in \eqref{eqn:Dt}.
Both realize the same Carroll algebra. The translation generators carry
no quantum-mechanical factors of $i$: the associated energy and momentum
operators are obtained by multiplying their differential realizations by
$i$ and $-i$, respectively. For this chosen realization,
the transported energy generator is \eqref{Carroll-energy-operator}.
The field maps relating these free actions to standard Carroll forms
are discussed in Section~\ref{sec:colored-Carroll-free} and
Appendix~\ref{app:Bergshoeff-dictionary}.

\subsection{Poincar\'e Casimirs}

For Dirac fields, the two Poincar\'e Casimirs are
\be\label{Cas-P}
\mathcal C_1:=T^\mu T_\mu\,, \qquad \mathcal C_2:=W^\mu W_\mu,
\ee
where $W^\mu=\frac12\epsilon^{\mu\nu\rho\sigma}
S_{\nu\rho}T_\sigma$ is the Pauli--Lubanski vector and
$\epsilon_{0123}=1$. Writing
$\gamma^{\mu\nu}:=\frac12[\gamma^\mu,\gamma^\nu]$, this is equivalently
\[
W^\mu=-\frac{i}{2}\,\gamma_5\,\gamma^{\mu\nu}\,T_\nu.
\]
The sign follows from $\epsilon^{0123}=-1$ and our definition
of $\gamma_5$. With $T_\mu=\partial_\mu$,
\[
W^2\psi=\frac34\,T^\mu T_\mu\,\psi.
\]
Thus, using the Dirac equation,
\[
T^\mu T_\mu\,\psi=m^2c^2\psi,\qquad
W^2\psi=\frac34\,m^2c^2\,\psi .
\]
If instead one uses the Hermitian momentum convention
$P_\mu=-iT_\mu$, the same statement reads
$P^2\psi=-m^2c^2\psi$ and $W(P)^2\psi=-\frac34m^2c^2\psi$.
The spin-rotation generators are
$S_i=\frac14\epsilon_{ijk}\gamma^{jk}$. The similarity-transformed
Casimirs and their contractions are given in \eqref{Cas-t},
\eqref{Galilei-Cas-limits}, and \eqref{Carroll-Cas-limits}.

\section{Representative interaction densities}
\label{app:interaction-catalog}
Table~\ref{tab:admissible-interactions} applies the bilinear
construction of Section~\ref{sec:interactions} to the free catalogs.
It collects bilinear-generated interaction families across the different
boost representations. The complete single-color quartic basis,
including the pairing channels, is given in Section~\ref{sec:quartic-basis}.
At an endpoint, $V(N_u,M_{uv})$ uses the unmixed component and the
bilinears defined in \eqref{endpoint-mixed-bilinear}. In the central
rows, $V_{\zeta\chi}$ denotes a real local function of the allowed
rotational scalars involving both components, such as $N_\chi$,
$N_\zeta$, and $M_{\chi\zeta}$. Interactions start at quadratic
order in these bilinears.
\begingroup
\small
\setlength{\tabcolsep}{2.5pt}
\renewcommand{\arraystretch}{1.15}
\setlength{\LTcapwidth}{\textwidth}
\begin{longtable}{|
>{\raggedright\arraybackslash}p{0.3\textwidth}|
>{\centering\arraybackslash}p{0.10\textwidth}|
c|
>{\raggedright\arraybackslash}p{0.45\textwidth}|}
\hline
Representative density plus interaction & Boost & Source & Notes\\
\hline\hline
\endfirsthead
\multicolumn{4}{c}{\tablename~\thetable\ (continued)}\\[2pt]
\hline
Representative density plus interaction & Boost & Source & Notes\\
\hline\hline
\endhead
\hline
\multicolumn{4}{r}{Continued on the next page}\\
\endfoot
\hline
\caption{Representative free densities from
Tables~\ref{tab:S1-Galilei-coefficients}--\ref{tab:S2-Carroll-coefficients}
and boost-invariant, bilinear-generated interaction families. We group free densities
related by endpoint duality, field relabeling or phase redefinitions;
the corresponding field maps also act on the generators
and interaction potentials.
Rows distinguished by the form of the displayed interaction are retained
and identified in the Notes column.
The purely algebraic rows
T.\ref{tab:S1-Galilei-coefficients}.i) and
T.\ref{tab:S2-Galilei-coefficients}.ii) are omitted.}
\label{tab:admissible-interactions}\\
\endlastfoot
$\begin{aligned}
{}&i\chi^\dagger\partial_t\chi
+\zeta^\dagger\hat\nabla\chi-\chi^\dagger\hat\nabla\zeta\\[-2pt]
&+2mN_\zeta+V(N_\chi,M_{uv})
\end{aligned}$
& $K_{i,m}^{G,(1)}$
& T.\ref{tab:S1-Galilei-coefficients}.ii)
& Massive L\'evy--Leblond representative; its massless descendants are not
listed separately.\\[8pt]
\cline{1-4}
$\zeta^\dagger\hat\nabla\chi
-\chi^\dagger\hat\nabla\zeta+V_{\zeta\chi}$
& $K_{i,0}^{G,(0)}$
& T.\ref{tab:S1-Galilei-coefficients}.iv)
& Magnetic Galilei sector.\\[6pt]
\hline\hline
$i\chi^\dagger\partial_t\chi+V(N_\chi,M_{uv})$
& $\begin{array}{c}
K_{i,0}^{C,(-1)}\\ K_{i,0}^{C,(0)}
\end{array}$
& T.\ref{tab:S1-Carroll-coefficients}.i)
& Groups the free densities of
T.\ref{tab:S1-Carroll-coefficients}.iii), T.\ref{tab:S1-Carroll-coefficients}.iv) and T.\ref{tab:S1-Carroll-coefficients}.vi), up to field label and phase
redefinition; these maps also transform the generators and potentials.\\[8pt]
\cline{1-4}
$i\chi^\dagger\partial_t\chi
+i\zeta^\dagger\partial_t\zeta+V_{\zeta\chi}$
& $K_{i,0}^{C,(0)}$
& T.\ref{tab:S1-Carroll-coefficients}.ii)
& Central two-component Carroll sector.\\[6pt]
\cline{1-4}
$\begin{aligned}
{}&i\chi^\dagger\partial_t\chi
+i\zeta^\dagger\partial_t\zeta\\[-2pt]
&+2\Epsilon \zeta^\dagger\zeta+V_{\zeta\chi}
\end{aligned}$
& $K_{i,\Epsilon}^{C,(0)}$
& T.\ref{tab:S1-Carroll-coefficients}.v)
& A mixed potential can obstruct removal of explicit $\Epsilon$
while keeping the same time-independent potential.\\[8pt]
\hline\hline
$i\zeta^\dagger\hat\nabla\zeta
+2im\chi^\dagger\zeta+V(N_\zeta,M_{uv})$
& $K_{i,m}^{G,(-1)}$
& T.\ref{tab:S2-Galilei-coefficients}.i)
& Complex seed: its hermitian realization is the colored Hermitian
completion. The dual endpoint is omitted.\\[8pt]
\cline{1-4}
$i\zeta^\dagger\hat\nabla\zeta+V(N_\zeta,M_{uv})$
& $\begin{array}{c}
K_{i,0}^{G,(-1)}\\ K_{i,0}^{G,(0)}
\end{array}$
& T.\ref{tab:S2-Galilei-coefficients}.iv)
& The T.\ref{tab:S2-Galilei-coefficients}.vi) row is related by field exchange.\\[8pt]
\cline{1-4}
$i\chi^\dagger\hat\nabla\chi
+i\zeta^\dagger\hat\nabla\zeta+V_{\zeta\chi}$
& $K_{i,0}^{G,(0)}$
& T.\ref{tab:S2-Galilei-coefficients}.v)
& The free density is a direct sum, but $V_{\zeta\chi}$ can couple its two
sectors.\\[8pt]
\hline\hline
$\begin{aligned}
{}&\zeta^\dagger\partial_t\chi
-\chi^\dagger\partial_t\zeta\\[-2pt]
&+i\chi^\dagger\hat\nabla\chi+V(N_\chi,M_{uv})
\end{aligned}$
& $K_{i,0}^{C,(-1)}$
& T.\ref{tab:S2-Carroll-coefficients}.i)
& Hermitian endpoint action; the opposite endpoint is dual.\\[8pt]
\cline{1-4}
$\zeta^\dagger\partial_t\chi
-\chi^\dagger\partial_t\zeta+V_{\zeta\chi}$
& $K_{i,0}^{C,(0)}$
& T.\ref{tab:S2-Carroll-coefficients}.ii)
& Hermitian central Carroll action; it may also be extended in
color.\\[6pt]
\cline{1-4}
$\begin{aligned}
{}&\zeta^\dagger\partial_t\chi
-\chi^\dagger\partial_t\zeta+i\chi^\dagger\hat\nabla\chi\\[-2pt]
&+2i\Epsilon\chi^\dagger\zeta+V(N_\chi,M_{uv})
\end{aligned}$
& $K_{i,\Epsilon}^{C,(-1)}$
& T.\ref{tab:S2-Carroll-coefficients}.iv)
& Complex seed: its free colored Hermitian completion is
identified with the non-minimal magnetic Carroll model; the opposite
endpoint is dual.\\[8pt]
\cline{1-4}
$\begin{aligned}
{}&\zeta^\dagger\partial_t\chi
-\chi^\dagger\partial_t\zeta\\[-2pt]
&+2i\Epsilon\chi^\dagger\zeta+V_{\zeta\chi}
\end{aligned}$
& $K_{i,\Epsilon}^{C,(0)}$
& T.\ref{tab:S2-Carroll-coefficients}.v)
& After colored Hermitian completion, a mixed potential can obstruct
phase removal of explicit $\Epsilon$ at fixed potential.\\[8pt]
\end{longtable}
\endgroup

A phase map between free densities need not preserve a chosen
interaction. For example, $\zeta=e^{2i\Epsilon t}\zeta'$ removes
the explicit $\Epsilon$ term from
T.\ref{tab:S1-Carroll-coefficients}.v), but sends
\[
\chi^\dagger\zeta+\zeta^\dagger\chi
\longmapsto
e^{2i\Epsilon t}\chi^\dagger\zeta'
+e^{-2i\Epsilon t}\zeta'{}^\dagger\chi .
\]
A relative-phase-sensitive $V_{\zeta\chi}$ therefore acquires
explicit time dependence. This explains why the table retains some
phase-related free representatives. The same observation
applies to mixed potentials in the central colored Carroll model.
Similarly, $V_{\zeta\chi}$ can couple the two otherwise independent
spatial sectors in T.\ref{tab:S2-Galilei-coefficients}.v).

\section{Relations between Carroll actions and their interactions}
\label{app:Bergshoeff-dictionary}
This appendix presents field dictionaries expressing our models in the
conventions of Bergshoeff et al.~\cite{Bergshoeff:2023vfd}, together with
the color-and-chiral duality connecting the two endpoint realizations.
The non-minimal dictionary checks are collected in
Section~\ref{app:magnetic-dictionary-checks}, and the interacting
extensions in Section~\ref{app:Bergshoeff-interacting}. We close with
a spatial field redefinition connecting the free endpoint and central
actions and its effect on interaction potentials.

\subsection{Comparison with the minimal magnetic Carroll actions}
At the single-color level, the fixed-$m$ endpoint
\eqref{act-2-bm1-C} agrees with the first minimal magnetic Carroll action
of Ref.~\cite{Bergshoeff:2023vfd}
after its mass channel is set to zero.  Choose
$\Gamma^0=\gamma_0$, $\Gamma^i=\gamma_i$ and
$\Gamma_\star=\gamma_5$, so that
$P_\pm=\frac12(1\pm i\Gamma^0)$.  The projected fields in their
Eq.~(2.24) are then
\begin{equation}
\label{Bergshoeff-endpoint-map}
\psi_+=\begin{pmatrix}\chi\\0\end{pmatrix},\qquad
\psi_-=\begin{pmatrix}0\\\zeta\end{pmatrix},
\end{equation}
and direct substitution gives
\begin{equation}
\label{Bergshoeff-endpoint-equivalence}
-\frac12\mathcal L_{\mathrm{magnetic\ Carroll},1}
=\mathcal L^{(2,-1)}_{0,m}
+\mu M_{\chi\zeta},\qquad
\mu=\operatorname{Im}(m_{\rm B}).
\end{equation}
Here $m_{\rm B}$ denotes the complex mass parameter
of Ref.~\cite{Bergshoeff:2023vfd}.
The quantities $N_\chi$ and $M_{\chi\zeta}$ are the bilinears of
\eqref{endpoint-mixed-bilinear} with $(u,v)=(\chi,\zeta)$.
Thus, at $\mu=0$, the two free densities agree up to the irrelevant
overall factor $-2$.  Their second minimal magnetic action, Eq.~(2.27),
belongs to the same endpoint class:
$-\frac12\mathcal L_{\mathrm{magnetic\ Carroll},2}
=\mathcal L^{(2,-1)}_{0,m}+\operatorname{Re}(m_{\rm B})N_\chi$.
The two minimal mass deformations are therefore
the linear $M_{\chi\zeta}$ and $N_\chi$ channels used in the interaction
analysis.

\subsection{Color-and-chiral endpoint duality}
At $\beta=+1$, completing the seed Lagrangian of
Section~\ref{sec:S2-limits}, Eq.~\eqref{act-2-b1-CE}, gives
\begin{equation}
\label{act-2-b1-CE-real}
\begin{aligned}
\mathcal S_{12;\,0,\Epsilon}^{(2,1)}
=
\frac1{2\tilde\kappa_2}\int dt\,d^3x\,
\Big[{}&
-\chi_2^\dagger\partial_t\zeta_1
+\zeta_2^\dagger\partial_t\chi_1
+i\zeta_2^\dagger\hat\nabla\zeta_1
+2i\Epsilon\chi_2^\dagger\zeta_1\\
&+\zeta_1^\dagger\partial_t\chi_2
-\chi_1^\dagger\partial_t\zeta_2
+i\zeta_1^\dagger\hat\nabla\zeta_2
-2i\Epsilon\zeta_1^\dagger\chi_2
\Big].
\end{aligned}
\end{equation}
The field equations are
\begin{equation}
\label{eom-2-b1-CE-real}
\begin{gathered}
(\partial_t-2i\Epsilon)\zeta_1=0,\qquad
\partial_t\chi_1+i\hat\nabla\zeta_1=0,\\
(\partial_t-2i\Epsilon)\chi_2+i\hat\nabla\zeta_2=0,\qquad
\partial_t\zeta_2=0.
\end{gathered}
\end{equation}

The action \eqref{act-2-b1-CE-real}, and the corresponding field equations,
are related to those of the $\beta=-1$ case
\eqref{act-2-bm1-CE-real} by the discrete map
$\mathcal U$ of \eqref{dual-U}.
In terms of the actions,
\begin{equation}
\label{act-2-Carroll-endpoint-duality}
\mathcal U:\quad
\mathcal S_{12;\,0,\Epsilon}^{(2,-1)}
\longmapsto
\mathcal S_{12;\,0,\Epsilon}^{(2,1)} ,\qquad
\mathcal U:=-\mathcal P_{12}\circ
(\mathbf 1_{\rm col}\otimes\mathfrak D)
\end{equation}
defined in terms of the permutation
$\mathcal P_{12}:\Phi_1\leftrightarrow\Phi_2$ and the duality map
\eqref{D-map}, acting diagonally on both colors. The common minus sign
matches the component convention in \eqref{dual-U} for
$\Phi_a=(\zeta_a,\chi_a)^T$ and cancels in the boost conjugation.
Since the boost itself acts
identically on the two colors, define its doubled representation by
\[
\mathbb K_{i,\Epsilon}^{C,(\beta_\star)}
:=\mathbf 1_{\rm col}\otimes K_{i,\Epsilon}^{C,(\beta_\star)} .
\]
The two endpoint representations are then related by
\be
\mathcal U\mathbb K_{i,\Epsilon}^{C,(\beta_\star)}\mathcal U^{-1}
=\mathbb K_{i,\Epsilon}^{C,(-\beta_\star)},
\ee 
The color permutation commutes with $\mathbb K$, so this relation reduces to
the boost duality \eqref{boost-C-duality}. The common
covariance argument for the two hermitian endpoint actions is given in
Appendix~\ref{app:boost-checks}.

\subsection{Comparison with the non-minimal magnetic Carroll action}
The four independent Pauli blocks of the two-color actions correspond
to the untruncated magnetic action, Eq.~(2.20) of
Ref.~\cite{Bergshoeff:2023vfd}. We establish the relation directly
between the contracted actions and their time translations.

For the colored fixed-$\Epsilon$ comparison, denote the nonzero Pauli blocks
of their projected fields by
$\psi_\pm^{\rm B}$ and $\xi_\pm^{\rm B}$, renaming their second parent
spinor to avoid confusion with our $\chi_a$.  We choose
$\Gamma^0=\gamma_0$ and $\Gamma^i=\gamma_i$ only for this comparison of
contracted actions.  Then $P_+$ selects the upper block, $P_-$ the lower
block, and their Dirac conjugates are
$\bar\xi_+=\xi_+^\dagger$ and $\bar\xi_-=-\xi_-^\dagger$.
Their Eq.~(2.20) becomes
\begin{equation}
\label{Bergshoeff-magnetic-components}
\begin{split}
\mathcal L_{\rm B}={}&
-i\xi_+^{{\rm B}\dagger}\partial_t\psi_+^{\rm B}
-i\xi_-^{{\rm B}\dagger}\partial_t\psi_-^{\rm B}
-\xi_-^{{\rm B}\dagger}\hat\nabla\psi_+^{\rm B}\\
&-m_{\rm B}\xi_+^{{\rm B}\dagger}\psi_+^{\rm B}
+m_{\rm B}\xi_-^{{\rm B}\dagger}\psi_-^{\rm B}
+\mathrm{h.c.},
\end{split}
\end{equation}
where $m_{\rm B}$ is allowed to be complex in Ref.~\cite{Bergshoeff:2023vfd}.
With the positive-frequency convention \eqref{S12}, our endpoint
corresponds to the real subfamily $m_{\rm B}=-\Epsilon$.
For $\beta_\star=-1$, use the invertible dictionary
\begin{equation}
\label{Bergshoeff-colored-field-map}
\begin{aligned}
\psi_+^{\rm B}&=e^{-i\Epsilon t}\chi_1,
&\psi_-^{\rm B}&=e^{-i\Epsilon t}\zeta_1,\\
\xi_+^{\rm B}&=-i e^{-i\Epsilon t}\zeta_2,
&\xi_-^{\rm B}&=i e^{-i\Epsilon t}\chi_2,
&m_{\rm B}&=-\Epsilon.
\end{aligned}
\end{equation}
At $m_{\rm B}=-\Epsilon$, this map turns
\eqref{Bergshoeff-magnetic-components} into the density of
\eqref{act-2-bm1-CE-real}, including its conjugate terms, up to a
boundary term and overall normalization. This identifies our endpoint
with the real-mass subfamily of Ref.~\cite{Bergshoeff:2023vfd}.

For the $\beta_\star=+1$ endpoint, compose
\eqref{Bergshoeff-colored-field-map} with the color-and-chiral duality
$\mathcal U$ of \eqref{dual-U}:
\begin{equation}
\label{Bergshoeff-colored-field-map-plus}
\begin{aligned}
\psi_+^{\rm B}&=e^{-i\Epsilon t}\zeta_2,
&\psi_-^{\rm B}&=-e^{-i\Epsilon t}\chi_2,\\
\xi_+^{\rm B}&=i e^{-i\Epsilon t}\chi_1,
&\xi_-^{\rm B}&=i e^{-i\Epsilon t}\zeta_1,
&m_{\rm B}&=-\Epsilon.
\end{aligned}
\end{equation}
The action identification at this endpoint follows from
\eqref{act-2-Carroll-endpoint-duality}. Both dictionaries retain all
four independent Pauli blocks.

The frequencies $(-m_{\rm B},+m_{\rm B})$ in the variables of
\eqref{Bergshoeff-magnetic-components} become $(0,-2\Epsilon)$ in our
rephased basis, implementing
the change between the centered and uncentered forms in
\eqref{Carroll-energy-basis}.
The constant block maps commute with $\partial_t$, so the common phase
transports the time-translation generators as
\[
e^{i\Epsilon t}\partial_t e^{-i\Epsilon t}=D_t^{(\Epsilon)},\qquad
e^{i\Epsilon t}(i\partial_t)e^{-i\Epsilon t}=H_C^{(\Epsilon)}.
\]
Together with the action dictionaries and the remaining transported
symmetries, these identities establish free classical equivalence on
compatible field domains. Quantum equivalence additionally requires
an inner product and a quantization prescription.

In our $C_2$ parametrization, the common scaling
\eqref{S12} places both $(\chi_1,\zeta_1)$ and $(\chi_2,\zeta_2)$ in the
same contracted boost representation; the dictionary
\eqref{Bergshoeff-colored-field-map}, explicitly
$(\psi_+^{\rm B},\psi_-^{\rm B})=e^{-i\Epsilon t}(\chi_1,\zeta_1)$ and
$(\xi_+^{\rm B},\xi_-^{\rm B})=i e^{-i\Epsilon t}(-\zeta_2,\chi_2)$,
reverses the $P_\pm$ assignment for the second color and includes a common
time-dependent phase, intertwining these boosts with the oppositely triangular Carroll
boosts of Eq.~(2.21) in Ref.~\cite{Bergshoeff:2023vfd}.
These are equivalent realizations of the same Carroll symmetry, while
the uniform color transformation in our variables makes the invariant
bilinears \eqref{Colorendpoint-mixed-bilinear} directly available for
constructing the color-mixed potential
\eqref{colored-Bergshoeff-potential}.

\subsubsection{Checks of the field dictionaries}
\label{app:magnetic-dictionary-checks}
Direct substitution in the contracted models fixes the signs and
phases and verifies the correspondence of their field equations.

\paragraph{The $\beta_\star=-1$ dictionary.}
Under \eqref{Bergshoeff-colored-field-map}, the three non-conjugate
channels of \eqref{Bergshoeff-magnetic-components} become
\begin{align*}
-i\xi_+^{{\rm B}\dagger}\partial_t\psi_+^{\rm B}
-m_{\rm B}\xi_+^{{\rm B}\dagger}\psi_+^{\rm B}
&=\zeta_2^\dagger\partial_t\chi_1
-i\left(m_{\rm B}+\Epsilon\right)\zeta_2^\dagger\chi_1,\\
-i\xi_-^{{\rm B}\dagger}\partial_t\psi_-^{\rm B}
+m_{\rm B}\xi_-^{{\rm B}\dagger}\psi_-^{\rm B}
&=-\chi_2^\dagger\partial_t\zeta_1
+i\left(\Epsilon-m_{\rm B}\right)\chi_2^\dagger\zeta_1,\\
-\xi_-^{{\rm B}\dagger}\hat\nabla\psi_+^{\rm B}
&=i\chi_2^\dagger\hat\nabla\chi_1.
\end{align*}
At $m_{\rm B}=-\Epsilon$ the first energy coefficient cancels and the
second becomes $2i\Epsilon$.  These are exactly the first four terms of
\eqref{act-2-bm1-CE-real}.  Conjugating them and integrating the derivatives
by parts gives the remaining four terms,
\[
\zeta_1^\dagger\partial_t\chi_2
-\chi_1^\dagger\partial_t\zeta_2
+i\chi_1^\dagger\hat\nabla\chi_2
-2i\Epsilon\zeta_1^\dagger\chi_2.
\]
Thus Eq.~(2.20) of Ref.~\cite{Bergshoeff:2023vfd} and the free action
\eqref{act-2-bm1-CE-real} agree up to the overall normalization and boundary
terms at $m_{\rm B}=-\Epsilon$.

\paragraph{The opposite endpoint.}
This case follows from the duality \eqref{dual-U}. For clarity, the
same substitution using \eqref{Bergshoeff-colored-field-map-plus} gives
the channel identities
\begin{align*}
-i\xi_+^{{\rm B}\dagger}\partial_t\psi_+^{\rm B}
-m_{\rm B}\xi_+^{{\rm B}\dagger}\psi_+^{\rm B}
&=-\chi_1^\dagger\partial_t\zeta_2
+i\left(\Epsilon+m_{\rm B}\right)\chi_1^\dagger\zeta_2,\\
-i\xi_-^{{\rm B}\dagger}\partial_t\psi_-^{\rm B}
+m_{\rm B}\xi_-^{{\rm B}\dagger}\psi_-^{\rm B}
&=\zeta_1^\dagger\partial_t\chi_2
+i\left(m_{\rm B}-\Epsilon\right)\zeta_1^\dagger\chi_2,\\
-\xi_-^{{\rm B}\dagger}\hat\nabla\psi_+^{\rm B}
&=i\zeta_1^\dagger\hat\nabla\zeta_2.
\end{align*}
At $m_{\rm B}=-\Epsilon$ these are precisely the last four terms of
\eqref{act-2-b1-CE-real}; their Hermitian conjugates, after integration by
parts, are its first four terms.  Hence both colored free endpoints are
related to \eqref{Bergshoeff-magnetic-components} by
different field dictionaries.

\paragraph{Field equations.}
For an independent check at the level of field equations, variation of
\eqref{Bergshoeff-magnetic-components} with respect to the daggered Pauli
blocks, after integrating the conjugate derivatives by parts, yields
\begin{equation}
\label{Bergshoeff-contracted-eom}
\begin{aligned}
(\partial_t-i m_{\rm B})\psi_+^{\rm B}&=0,
&(\partial_t+i m_{\rm B})\psi_-^{\rm B}
-i\hat\nabla\psi_+^{\rm B}&=0,\\
(\partial_t-i m_{\rm B})\xi_+^{\rm B}
+i\hat\nabla\xi_-^{\rm B}&=0,
&(\partial_t+i m_{\rm B})\xi_-^{\rm B}&=0.
\end{aligned}
\end{equation}
Here $m_{\rm B}$ is real, as in
\eqref{Bergshoeff-colored-field-map} and
\eqref{Bergshoeff-colored-field-map-plus}. Dividing out
their common phase, \eqref{Bergshoeff-colored-field-map} turns these four
equations, in order, into the second, first, fourth and third equations of
\eqref{eom-2-bm1-CE-real}.  The different dictionary
\eqref{Bergshoeff-colored-field-map-plus} turns them, in order, into the
fourth, third, second and first equations of \eqref{eom-2-b1-CE-real}.


\subsection{Interacting extensions of the magnetic Carroll models}
\label{app:Bergshoeff-interacting}

We now express our interactions in the field conventions of
Bergshoeff et al.~\cite{Bergshoeff:2023vfd}, obtaining explicit
extensions of their minimal and non-minimal magnetic Carroll models.

\paragraph{Minimal magnetic models.}
Denote the nonzero Pauli blocks of the projected spinors in
\eqref{Bergshoeff-endpoint-map} by $\psi_+^{\rm B}$ and
$\psi_-^{\rm B}$. Thus $\psi_+^{\rm B}=\chi$ and
$\psi_-^{\rm B}=\zeta$. Define
\begin{equation}
\label{Bergshoeff-minimal-interaction-bilinears}
\begin{aligned}
N_{\rm B}&=\psi_+^{{\rm B}\dagger}\psi_+^{\rm B},&
M_{\rm B}&=\psi_+^{{\rm B}\dagger}\psi_-^{\rm B}
+\psi_-^{{\rm B}\dagger}\psi_+^{\rm B},\\
P_{\rm B}&=\psi_+^{{\rm B}T}i\sigma_2\psi_+^{\rm B},&
Q_{\rm B}&=\psi_+^{{\rm B}T}i\sigma_2\psi_-^{\rm B}.
\end{aligned}
\end{equation}
Here $P_{\rm B}$ and $Q_{\rm B}$ are the spin-singlet pairing
bilinears. The five-channel potential becomes
\begin{equation}
\label{Bergshoeff-minimal-quartic}
\begin{split}
V_{4,\rm B}={}&g_1N_{\rm B}^2+g_2N_{\rm B}M_{\rm B}
+g_3M_{\rm B}^2+g_4Q_{\rm B}^\dagger Q_{\rm B}\\
&+i g_5(P_{\rm B}^\dagger Q_{\rm B}-Q_{\rm B}^\dagger P_{\rm B}),
\qquad g_j\in\mathbb R.
\end{split}
\end{equation}
With the normalization of \eqref{Bergshoeff-endpoint-equivalence},
the two minimal models admit the interacting densities
\begin{equation}
\label{Bergshoeff-minimal-interacting-explicit}
\mathcal L_{{\rm B},j}^{\rm int}
:=-\frac12\mathcal L_{\mathrm{magnetic\ Carroll},j}+V_{4,\rm B},
\qquad j=1,2.
\end{equation}
Their mass channels remain
$\operatorname{Im}(m_{\rm B})M_{\rm B}$ and
$\operatorname{Re}(m_{\rm B})N_{\rm B}$, respectively.
The completeness of \eqref{Bergshoeff-minimal-quartic} is the
single-color result of Section~\ref{sec:quartic-basis}.
Higher-order potentials in $N_{\rm B}$ and $M_{\rm B}$ give the
extensions studied in Section~\ref{sec:interacting-Bergshoeff}.

\paragraph{Non-minimal magnetic model.}
For the four independent Pauli blocks of
\eqref{Bergshoeff-magnetic-components}, set
\begin{equation}
\label{Bergshoeff-interaction-blocks}
(u_1^{\rm B},v_1^{\rm B})=(\psi_+^{\rm B},\psi_-^{\rm B}),\qquad
(u_2^{\rm B},v_2^{\rm B})=(-i\xi_-^{\rm B},i\xi_+^{\rm B}).
\end{equation}
Both pairs have the same triangular boost representation. Under
\eqref{Bergshoeff-colored-field-map}, they equal
$e^{-i\Epsilon t}(\chi_a,\zeta_a)$, so the common phase cancels
from the invariant bilinears
\begin{equation}
\label{Bergshoeff-colored-interaction-bilinears}
N_{ab}^{\rm B}:=u_a^{{\rm B}\dagger}u_b^{\rm B},\qquad
M_{ab}^{\rm B}:=u_a^{{\rm B}\dagger}v_b^{\rm B}
+v_a^{{\rm B}\dagger}u_b^{\rm B}.
\end{equation}
In particular,
\[
N_{12}^{\rm B}=-i\psi_+^{{\rm B}\dagger}\xi_-^{\rm B},\qquad
M_{12}^{\rm B}=i(\psi_+^{{\rm B}\dagger}\xi_+^{\rm B}
-\psi_-^{{\rm B}\dagger}\xi_-^{\rm B}),
\]
with $N_{21}^{\rm B}=(N_{12}^{\rm B})^\dagger$ and
$M_{21}^{\rm B}=(M_{12}^{\rm B})^\dagger$.
Consequently, the real-mass non-minimal density has the interacting
extension
\begin{equation}
\label{Bergshoeff-nonminimal-interacting-explicit}
\mathcal L_{\rm B}^{\rm int}
=\mathcal L_{\rm B}+V_{\rm col}(N_{ab}^{\rm B},M_{ab}^{\rm B}),
\qquad V_{\rm col}^\dagger=V_{\rm col},\qquad m_{\rm B}=-\Epsilon,
\end{equation}
where the potential starts at fourth order in the fields.
The pairing channels of \eqref{endpoint-quartic-complete}, evaluated
on either pair in \eqref{Bergshoeff-interaction-blocks}, may also be
added. These constructions supply families of colored interactions;
the five-operator basis classifies the single-pair sector.

For an explicit color-mixed example, the potential
\eqref{colored-Bergshoeff-potential} reads
\begin{equation}
\label{Bergshoeff-colored-potential-explicit}
\begin{aligned}
\mathcal I_\chi&=(\psi_+^{{\rm B}\dagger}\xi_-^{\rm B})
(\xi_-^{{\rm B}\dagger}\psi_+^{\rm B})=N_{12}^{\rm B}N_{21}^{\rm B},\\
\mathcal L_{\rm B}^{\rm int}
&=\mathcal L_{\rm B}+\mathcal U_{\rm col}(\mathcal I_\chi)\\
&=\mathcal L_{\rm B}+g_4\mathcal I_\chi+\frac{g_8}{2}\mathcal I_\chi^2,
\qquad g_4,g_8\in\mathbb R.
\end{aligned}
\end{equation}
The common phase cancels from $\mathcal I_\chi$, leaving the couplings
unchanged. The exact solutions of the colored model therefore map
to those of \eqref{Bergshoeff-colored-potential-explicit} through
\eqref{Bergshoeff-colored-field-map}, with the same density
normalization used in the free dictionary.

\subsection{Spatial field redefinition between endpoint and central actions}
\label{app:spatial-field-redefinition}
For $\Epsilon\neq0$, a spatial field redefinition relates the free
endpoint action \eqref{act-2-bm1-CE-real} to the central action
\eqref{act-2-C-real}. With $\chi_1$ and $\chi_2$ unchanged, set
\begin{equation}
\label{Carroll-differential-map}
\eta_1=\zeta_1+\frac{\hat\nabla\chi_1}{2\Epsilon},\qquad
\eta_2=\zeta_2-\frac{\hat\nabla\chi_2}{2\Epsilon}.
\end{equation}
The non-conjugate part of the density obeys
\begin{align}
&-\chi_2^\dagger\partial_t\zeta_1
+\zeta_2^\dagger\partial_t\chi_1
+i\chi_2^\dagger\hat\nabla\chi_1+2i\Epsilon\chi_2^\dagger\zeta_1
\nonumber\\
&\qquad=
-\chi_2^\dagger(\partial_t-2i\Epsilon)\eta_1
+\eta_2^\dagger\partial_t\chi_1
+\frac1{2\Epsilon}\partial_i
 \left(\chi_2^\dagger\sigma^i\partial_t\chi_1\right).
\label{Carroll-differential-density}
\end{align}
After adding the conjugate, the bulk density has the form
\eqref{act-2-C-real}, with $\zeta_a$ replaced by $\eta_a$; the component
rephasing then yields \eqref{act-2-C-real-kinetic}. The map changes
the boost realization and requires
compatible field domains and boundary conditions. Applied to generic
algebraic potentials, it produces derivative interactions. The free-mode
analysis in Section~\ref{sec:colored-Carroll-free} and the interacting
model of Section~\ref{sec:colored-Carroll} use the original fields.

\section{Completeness of the single-color quartic basis}
\label{app:quartic-completeness}

We prove completeness under the assumptions stated in
Section~\ref{sec:quartic-basis}: one pair of Grassmann Pauli fields,
ordinary Hermitian conjugation, rotations, a common $U(1)$ phase,
endpoint boosts, and no derivatives or imposed discrete symmetries.
Phase invariance requires two daggered and two undaggered components
in each quartic. For this proof, abbreviate the spin-singlet and
spin-triplet pairing bilinears as
\[
P=u^Ti\sigma_2u,\quad Q=u^Ti\sigma_2v,\quad
R=v^Ti\sigma_2v,\quad T_i=u^Ti\sigma_2\sigma_i v .
\]
The antisymmetric product of the four undaggered components decomposes
under rotations as
\[
\Lambda^2(\mathbf2\oplus\mathbf2)
=\mathbf1\oplus\mathbf1\oplus\mathbf1\oplus\mathbf3 .
\]
Its three singlets are $P,Q,R$ and its triplet is $T_i$. Every neutral
rotationally invariant quartic therefore has the unique form
\[
V_4=\sum_{A,B\in\{P,Q,R\}}c_{AB}A^\dagger B
+d\,T_i^\dagger T_i.
\]
There are ten complex coefficients before Hermiticity and boosts
are imposed. Hermiticity requires $c_{AB}=c_{BA}^*$ and
$d\in\mathbb R$, leaving ten real parameters before the boost condition.
For the plus spin-mixing sign,
\[
\delta^{\rm mix}u=0,\qquad
\delta^{\rm mix}v=\frac{i}{2}b^i\sigma_i u .
\]
The Pauli and Grassmann identities then give
\[
\delta^{\rm mix}P=\delta^{\rm mix}Q=0,\qquad
\delta^{\rm mix}R=-i b^iT_i,\qquad
\delta^{\rm mix}T_i=\frac i2 b_iP .
\]
All variations reverse sign for the other endpoint; their kernel is
unchanged. It follows that
\begin{align*}
\delta^{\rm mix}V_4=i b^i\bigg[
&-\left(c_{PR}+\frac d2\right)P^\dagger T_i
+\left(c_{RP}+\frac d2\right)T_i^\dagger P\\
&-c_{QR}Q^\dagger T_i+c_{RQ}T_i^\dagger Q
+c_{RR}\left(T_i^\dagger R-R^\dagger T_i\right)
\bigg].
\end{align*}
The displayed vector structures have distinct degrees in
$(u,v,u^\dagger,v^\dagger)$ except $Q^\dagger T_i$ and $T_i^\dagger Q$.
These two are also independent: with $a=u_1v_2$ and $b=u_2v_1$,
$Q=a-b$ and $T_3=-a-b$; their sum and difference separate diagonal
and crossed monomials. Thus
\[
c_{QR}=c_{RQ}=c_{RR}=0,\qquad c_{PR}=c_{RP}=-\frac d2 .
\]
The kernel has the five independent elements
\[
P^\dagger P,\quad P^\dagger Q,\quad Q^\dagger P,\quad Q^\dagger Q,
\quad K:=P^\dagger R+R^\dagger P-2T_i^\dagger T_i .
\]
Their independence follows from their component monomials: the first
four have distinct field degrees, and only $K$ contains $P^\dagger R$
and $R^\dagger P$. Finally,
\[
P^\dagger P=2N_u^2,\qquad
P^\dagger Q+Q^\dagger P=2N_uM_{uv},\qquad
K=2M_{uv}^2-2Q^\dagger Q .
\]
Imposing Hermiticity leaves five real couplings, precisely the
basis \eqref{endpoint-quartic-complete}. The orbital boost terms
are total derivatives, and the central phases cancel in every neutral
product. No further restriction arises from these parts of the boost.

\section{Boost invariance of the contracted actions}
\label{app:boost-checks}

We verify invariance of the action up to boundary terms, organizing
the calculation by boost class so that each identity covers several
catalog entries. Models related by parameter limits, field relabeling,
the phase map \eqref{boost-C-phase-sim}, or the dualities
\eqref{boost-G-duality}, \eqref{boost-C-duality}, and \eqref{dual-U}
need no independent check. The symmetry generators are transported
together with the fields.

Covariance and Hermiticity are separate properties. In a complex seed,
daggered fields are independent dual variables with the contragredient
transformation; in a Hermitian action this is ordinary conjugation.
The bilinear identities below also hold between two colors carrying
the same representation. Adding the conjugate term then establishes
covariance of the completions constructed from \eqref{L-D12}.

\subsection{Diagonal boost transport}
\label{app:boost-diagonal}

Two fields with the same Bargmann mass transform under a diagonal
Galilean boost as
\[
\delta_b u=-b^iD_i^{(m)}u,\qquad
\delta_b v=-b^iD_i^{(m)}v,
\]
where $D_i^{(m)}$ is defined in \eqref{eqn:Di}. The dual
fields carry the contragredient operator, so for any constant spin
matrix $\Gamma$,
\begin{equation}
\label{app:G-diagonal-bilinear}
\delta_b(u^\dagger\Gamma v)
=-t\,b^i\partial_i(u^\dagger\Gamma v)
=\partial_i\!\left[-t\,b^i(u^\dagger\Gamma v)\right].
\end{equation}
If $m=0$, the same identity holds when a constant-coefficient spatial
differential operator is inserted between the fields, since it commutes with
$t\partial_i$. This includes the massless endpoint densities built only from
the component without spin mixing.

Equation \eqref{app:G-diagonal-bilinear} covers
T.\ref{tab:S1-Galilei-coefficients}.i) and
T.\ref{tab:S1-Galilei-coefficients}.iv), as well as
T.\ref{tab:S2-Galilei-coefficients}.ii) and
T.\ref{tab:S2-Galilei-coefficients}.iv)--vi). In particular, the spatial
fixed-$\Epsilon$ rows of Table~\ref{tab:S2-Galilei-coefficients} are
invariant because their active field is diagonal even when the unused
component has endpoint spin mixing.

For Carroll boosts, use \eqref{eqn:Dt} and its coefficientwise
complex conjugate $D_t^{(\Epsilon)*}:=\partial_t+i\Epsilon$:
\[
\delta_bu=-b^ix_iD_t^{(\Epsilon)}u,\qquad
\delta_bu^\dagger=-b^ix_iD_t^{(\Epsilon)*}u^\dagger .
\]
Since $b^ix_i$ is independent of time, the phases cancel in every
daggered--undaggered bilinear built from fields with the same $\Epsilon$.
Any density assembled from such algebraic bilinears and
constant-coefficient time derivatives, but no spatial derivatives, is
transported according to
\begin{equation}
\label{app:C-diagonal-density}
\delta_b\mathcal L
=-b^ix_i\partial_t\mathcal L
=\partial_t(-b^ix_i\mathcal L).
\end{equation}
This proves all six rows of Table~\ref{tab:S1-Carroll-coefficients}: at an
endpoint the displayed Pauli component is precisely the one without spin
mixing, and the other cases are diagonal sums or specializations. It also
proves T.\ref{tab:S2-Carroll-coefficients}.ii) and
T.\ref{tab:S2-Carroll-coefficients}.v), including the combination
$A_2+2\Epsilon G_2$, and the hermitian central action
\eqref{act-2-C-real}.

\subsection{Endpoint Galilean cancellations}
\label{app:boost-G-endpoints}

The L\'evy--Leblond density \eqref{act-1-b1-NR} is the representative
$C_1$ endpoint. After extracting the common total spatial derivative, the
diagonal part of its $\beta_\star=1$ boost leaves
\[
-i\,b^i\chi^\dagger\partial_i\chi
-im\,b^i\chi^\dagger\sigma_i\zeta
+im\,b^i\zeta^\dagger\sigma_i\chi,
\]
At this endpoint the nonzero spin-mixing variations are
$\delta_b^{\rm mix}\zeta=-\frac{i}{2}b^i\sigma_i\chi$ and
$\delta_b^{\rm mix}\zeta^\dagger
=\frac{i}{2}b^i\chi^\dagger\sigma_i$. Therefore,
using $\{\sigma_i,\sigma_j\}=2\delta_{ij}$,
\begin{align*}
\delta_b^{\rm mix}
\left(\zeta^\dagger\hat\nabla\chi-\chi^\dagger\hat\nabla\zeta\right)
&=\frac{i}{2}b^i\chi^\dagger
\{\sigma^j,\sigma_i\}\partial_j\chi
=i\,b^i\chi^\dagger\partial_i\chi,\\
\delta_b^{\rm mix}\left(2m\zeta^\dagger\zeta\right)
&=im\,b^i\chi^\dagger\sigma_i\zeta
-im\,b^i\zeta^\dagger\sigma_i\chi .
\end{align*}
The derivative term cancels the first diagonal contribution, while
the mass variation cancels the remaining two. Thus
\begin{equation}
\delta_b\mathcal L^{(1,1)}_{\infty,m}
=\partial_i\!\left[-t\,b^i
\mathcal L^{(1,1)}_{\infty,m}\right].
\end{equation}
This establishes T.\ref{tab:S1-Galilei-coefficients}.ii).
T.\ref{tab:S1-Galilei-coefficients}.v) is its $m\to0$ specialization, while
T.\ref{tab:S1-Galilei-coefficients}.iii) follows by field relabeling together
with a common time-dependent phase redefinition. The diagonal cases were already
covered by \eqref{app:G-diagonal-bilinear}.

For the two $C_2$ endpoint seeds define
\[
(u_-,Q_-)=(\zeta,\chi^\dagger\zeta),\qquad
(u_+,Q_+)=(\chi,\chi^\dagger\zeta).
\]
The signs label $\beta_\star=\pm1$, and the corresponding densities are
\[
\mathcal L_\pm=i\,u_\pm^\dagger\hat\nabla u_\pm+2imQ_\pm .
\]
In either class, the diagonal orbital term and the endpoint spin mixing give
\begin{align}
\delta_b(iu_\pm^\dagger\hat\nabla u_\pm)
&=\partial_i[-t\,b^i(iu_\pm^\dagger\hat\nabla u_\pm)]
-m\,b^iu_\pm^\dagger\sigma_i u_\pm,\nonumber\\
\delta_b(2imQ_\pm)
&=\partial_i[-t\,b^i(2imQ_\pm)]
+m\,b^iu_\pm^\dagger\sigma_i u_\pm .
\label{app:G-endpoint-cancellation}
\end{align}
The last terms cancel, proving the invariance of
T.\ref{tab:S2-Galilei-coefficients}.i) and
T.\ref{tab:S2-Galilei-coefficients}.iii). 

\subsection{Endpoint Carrollian cancellations}
\label{app:boost-C-endpoints}

The fixed-$\Epsilon$ $C_2$ endpoints provide the most general Carrollian
check; the fixed-$m$ cases follow by setting $\Epsilon=0$. For convenience,
we restate the relevant seed blocks,
\[
A_2=\zeta^\dagger\partial_t\chi-\chi^\dagger\partial_t\zeta,\quad
B_2=i\chi^\dagger\hat\nabla\chi,\quad
D_2=i\zeta^\dagger\hat\nabla\zeta,\quad
G_2=i\chi^\dagger\zeta .
\]
Let
\[
(u_-,X_-)=(\chi,B_2),\qquad
(u_+,X_+)=(\zeta,D_2),\qquad
\mathcal L_\pm=A_2+X_\pm+2\Epsilon G_2 .
\]
Using the phase-shifted boost with $\beta_\star=\pm1$ and separating the
common Carroll transport, one obtains for both signs
\begin{align}
\delta_bX_\pm
&=\partial_t(-b^ix_iX_\pm)
-i\,b^iu_\pm^\dagger\sigma_i\partial_tu_\pm
-\Epsilon b^iu_\pm^\dagger\sigma_i u_\pm,\nonumber\\
\left.\delta_bA_2\right|_{\rm mix}
&=i\,b^iu_\pm^\dagger\sigma_i\partial_tu_\pm,\nonumber\\
\left.\delta_b(2\Epsilon G_2)\right|_{\rm mix}
&=\Epsilon b^iu_\pm^\dagger\sigma_i u_\pm .
\label{app:C-endpoint-cancellation}
\end{align}
The spin variation of $A_2$ cancels the derivative term in
$\delta_bX_\pm$, and the variation of $2\Epsilon G_2$ cancels its
remaining algebraic term. Restoring the common transport gives
\[
\delta_b\mathcal L_\pm
=\partial_t(-b^ix_i\mathcal L_\pm).
\]
Equation \eqref{app:C-endpoint-cancellation} proves
T.\ref{tab:S2-Carroll-coefficients}.iv) and
T.\ref{tab:S2-Carroll-coefficients}.vi); its $\Epsilon=0$ specialization
proves T.\ref{tab:S2-Carroll-coefficients}.i) and
T.\ref{tab:S2-Carroll-coefficients}.iii).

\bibliographystyle{utphys}
\bibliography{ref}

\end{document}